\documentclass[aps,prc,floatfix,twocolumn,showpacs,superscriptaddress,amssymb,amsmath]{revtex4}

\usepackage{graphicx}

\newcommand{\nggn}{$(n,\gamma)\rightleftarrows(\gamma,n)$}

\begin{document}

\title{Dynamical r-process studies within the neutrino-driven wind
  scenario and its sensitivity to the nuclear physics input}

\author{A.~Arcones} 
\altaffiliation{Present address: Department of Physics, University of Basel,
  Klingelbergstra{\ss}e 82, 4056 Basel,  Switzerland}
\affiliation{Institut f{\"u}r Kernphysik, Technische Universit{\"a}t
  Darmstadt, 64289 Darmstadt, Germany} 
\affiliation{GSI Helmholtzzentrum f\"ur Schwerionenforschung,
  Planckstr. 1, 64291 Darmstadt, Germany}
\author{G.~Mart\'inez-Pinedo} 
\affiliation{GSI Helmholtzzentrum f\"ur Schwerionenforschung,
  Planckstr. 1, 64291 Darmstadt, Germany}

\date{\today}

\begin{abstract}
  We use results from long-time core-collapse supernovae simulations
  to investigate the impact of the late time evolution of the ejecta
  and of the nuclear physics input on the calculated r-process
  abundances. Based on the latest hydrodynamical simulations, heavy
  r-process elements cannot be synthesized in the neutrino-driven
  winds that follow the supernova explosion. However, by artificially
  increasing the wind entropy, elements up to $A=195$ can be made. In
  this way one can reproduce the typical behavior of high-entropy
  ejecta where the r-process is expected to occur. We identify which
  nuclear physics input is more important depending on the dynamical
  evolution of the ejecta. When the evolution proceeds at high
  temperatures (hot r-process), an \nggn\ equilibrium is
  reached. While at low temperature (cold r-process) there is a
  competition between neutron captures and beta decays. In the first
  phase of the r-process, while enough neutrons are available, the
  most relevant nuclear physics input are the nuclear masses for the
  hot r-process and the neutron capture and beta-decay rates for the
  cold r-process. At the end of this phase, the abundances follow a
  steady beta flow for the hot r-process and a steady flow of neutron
  captures and beta decays for the cold r-process. After neutrons are
  almost exhausted, matter decays to stability and our results show
  that in both cases neutron captures are key for determining the
  final abundances, the position of the r-process peaks, and the
  formation of the rare-earth peak. In all the cases studied, we find
  that the freeze out occurs in a timescale of several seconds.
\end{abstract}
\maketitle

\section{Introduction}
\label{sed:intro}
The synthesis of heavy elements by the rapid neutron capture process
(r-process) is a fascinating, long-standing problem which involves
challenges in nuclear physics (experiment and theory), astrophysical
simulations of explosive environments, and observations of metal-poor
stellar atmospheres.  The astrophysical site where the r-process
produces half of the heavy elements has not yet been identified (see
Ref.~\cite{arnould.goriely.takahashi:2007} for a recent
review). However, even if the astrophysical scenario were discovered,
one still has to deal with nuclei far from stability for which no
experimental data are available and theoretical predictions are quite
uncertain~\cite{Grawe.Langanke.Martinez-Pinedo:2007}.

Galactic chemical evolution models \cite{Wanajo.Ishimaru:2006,
  Qian.Wasserburg:2007} indicate that heavy elements are most likely
produced in core-collapse supernova outflows. After a core-collapse
supernova explosion, a proto-neutron star forms and a baryonic
neutrino-driven wind develops \cite{duncan.shapiro.wasserman:1986}.
The matter expands at high velocity, which can become supersonic, and
eventually collides with the slow, early supernova ejecta resulting in
a wind termination shock or reverse shock
\cite{Burrows.Hayes.Fryxell:1995, Janka.Mueller:1996,
  Buras.Rampp.ea:2006, arcones.janka.scheck:2007,Fischer.etal:2010}.
There are several nucleosynthesis process that occur or might occur in
this environment: $\alpha$-process \cite{Woosley.Hoffman:1992,
  Witti.Janka.Takahashi:1994}, $\nu$p-process
\cite{Froehlich.Martinez-Pinedo.ea:2006, Pruet.Hoffman.ea:2006,
  Wanajo:2006}, and
r-process~\cite{Woosley.Wilson.ea:1994,Takahashi.Witti.Janka:1994}.
The production of heavy r-process elements ($A>130$), requires a high
neutron-to-seed ratio. This can be achieved by the following
conditions \cite{Qian.Woosley:1996, hoffman.woosley.qian:1997,
  Otsuki.Tagoshi.ea:2000, Thompson.Burrows.Meyer:2001}: high entropy,
fast expansions, or low electron fraction.  However, these conditions
are not yet realized in hydrodynamical simulations that follow the
outflow evolution during the first seconds of the wind phase after the
explosion \cite{arcones.janka.scheck:2007, Huedepohl.ea:2010,
  Fischer.etal:2010}.

This manuscript aims to explore the sensitivity of the calculated
r-process abundances to the combined effects of the long-time
dynamical evolution and nuclear physics input providing a link between
the behavior of nuclear masses far from stability and features in the
final abundances.  The impact of the nuclear physics input on the
production of heavy elements has been explored in classical r-process
calculations in Ref.~\cite{Kratz.Bitouzet.ea:1993,
  Kratz.Pfeiffer.Thielemann:1998} and with more dynamical but still
parametric calculations (e.g.,
\cite{Howard.Goriely.ea:1993,Freiburghaus.Rembges.ea:1999,
  Meyer.Brown:1997}). However, most of the r-process studies
\cite{arnould.goriely.takahashi:2007} aimed to find the astrophysical
conditions (density, temperature, electron fraction) that reproduce
the solar abundances for a given nuclear physics input. There have
been only few works exploring the impact of different mass models
(e.g., \cite{Wanajo.Goriely.ea:2004,Farouqi.etal:2010}), the effect of
neutron captures when matter decays to stability
\cite{Surman.Engel.ea:1997,Surman.Engel:2001,Surman.Beun.ea:2009,
  beun.etal:2009}, and the importance of beta decays
\cite{Borzov.Cuenca-Garcia.ea:2008, Dillmann.Kratz.ea:2003}. Our
results contribute to improve the present understanding of how nuclear
masses and neutron-capture cross sections determine the r-process
abundances.  The influence of the beta-decay rates will be studied in
future work, here we only explore the effect of beta-delayed neutron
emission.

In this paper we use hydrodynamical trajectories from the
neutrino-driven wind simulations of
Ref.~\cite{arcones.janka.scheck:2007}. As the conditions found in
these trajectories do not allow for the synthesis of heavy r-process
elements \cite{Arcones.Montes:2010}, we artificially increase the
entropy by a factor two in order to produce the third r-process
peak. This allows us to study the nucleosynthesis of heavy elements in
a typical high-entropy neutrino-driven wind in a more consistent way
than with fully parametric expansions
\cite{Freiburghaus.Rembges.ea:1999, Farouqi.etal:2010} or with
steady-state wind models (e.g. \cite{Otsuki.Tagoshi.ea:2000,
  Thompson.Burrows.Meyer:2001}), which cannot consistently explore the
interaction of the wind with the slow supernova ejecta.  The possible
influence of the wind termination shock on the nucleosynthesis was
already suggested in Refs.~\cite{Janka.Mueller:1996,
  Qian.Woosley:1996} and analyzed in different works
\cite{Terasawa.Sumiyoshi.ea:2002, Wanajo.Itoh.ea:2002,Wanajo:2007,
  Kuroda.Wanajo.Nomoto:2008, Panov.Janka:2009}. For the first time, we
use here trajectories obtained in hydrodynamical simulations to
perform a detailed investigation of the effect on the r-process of the
long-time dynamical evolution. We choose two different astrophysical
evolutions, that cover the broad range of conditions found in the
hydrodynamical simulations.  Four different mass models are used to
study the impact of the nuclear physics input on the calculated
abundances.

Our astrophysical and nuclear physics inputs are introduced in
Sect.~\ref{sec:methods}. We study the impact on the final abundances
of the reverse shock and long-time dynamical evolution
(Sect.~\ref{sec:rs_effect}). The sensitivity of the abundances and of
the r-process evolution to the mass model is explored in
Sect.~\ref{sec:mm}. In Sect.~\ref{sec:back}, we discuss the evolution
of the abundances after the r-process freeze-out, that we define as
the moment when the neutron-to-seed ratio drops below one.  Our
conclusions are summarized in Sect.~\ref{sec:conclusions}.

\section{Methods}
\label{sec:methods}
\subsection{Neutrino-driven wind and termination shock}
\label{sec:sim}
Our nucleosynthesis studies are based on a trajectory ejected at 8~s
after bounce in an explosion of a 15~$M_\odot$ progenitor (model
M15-l2-r1 in \cite{arcones.janka.scheck:2007}). This trajectory
represents a typical neutrino-driven wind, whose wind phase can be
described by steady-state models \cite{Qian.Woosley:1996,
  Otsuki.Tagoshi.ea:2000, Thompson.Burrows.Meyer:2001,
  Wanajo.Kajino.ea:2001}. However, hydrodynamical simulations are
required to study the long-time evolution when the supersonic wind
collides with the slow-moving supernova ejecta resulting in a wind
termination shock.

Detailed description of the hydrodynamical simulations can be found in
Refs.~\cite{arcones.janka.scheck:2007,Scheck.Kifonidis.Janka.Mueller:2006}. In
these simulations, Newtonian hydrodynamics
\cite{Scheck.Kifonidis.Janka.Mueller:2006, Kifonidis.Plewa.ea:2006}
with general relativistic corrections for the gravitational potential
\cite{Marek.Dimmelmeier.ea:2006} is combined with a simplified
neutrino transport approximation assuming Fermi-Dirac neutrino spectra
\cite{Scheck.Kifonidis.Janka.Mueller:2006}. This is computationally
very efficient and reproduces the results of Boltzmann transport
simulations qualitatively.  The equation of state used in the
simulations includes neutrons, protons, alpha particles, and a
representative nucleus ($^{54}\text{Mn}$) treated as non-relativistic
Boltzmann gases in nuclear statistical equilibrium
\cite{Janka.Mueller:1996}. The central part ($\rho \gtrsim 10^{13} \,
\text{g}/\text{cm}^3$) of the proto-neutron star is removed from the
computational domain and a Lagrangian inner boundary (placed below the
neutrinosphere) describes the neutron star evolution. The neutrino
cooling (i.e., neutrino energies and luminosities) and contraction of
the proto-neutron star are parametrized at the inner boundary to
account for possible uncertainties in the high density equation of
state.

The neutron-to-seed ratio found in the simulations of
Ref.~\cite{arcones.janka.scheck:2007} after freeze-out of
charged-particle reactions is too low ($Y_n/Y_{\rm{seed}}\approx
10^{-2}$) to permit the formation of heavy r-process nuclei. Only
elements with $Z<48$ are produced, i.e. light element primary process
(LEPP) elements \cite{Arcones.Montes:2010}. However, we can still use
this trajectory for r-process studies, if the neutron-to-seed ratio is
artificially increased by assuming a smaller initial electron fraction
or by rising the entropy. The electron fraction is determined by
electron neutrino and antineutrino energies and luminosities (see
e.g., Ref.~\cite{Qian.Woosley:1996}) and we keep it as given by the
simulations ($Y_e \approx 0.47$). Hereafter, all calculations are
performed on the same trajectory with the density decreased by a
factor of two~\footnote{Notice that we need only a factor two
  increased in the entropy compared to the factor five required in
  Ref.~\cite{Takahashi.Witti.Janka:1994} because the expansion is
  faster in the simulations of \cite{arcones.janka.scheck:2007}.}
overall to get also a factor two higher entropies ($S\propto T^3/\rho
\approx 200\ k_B / \mathrm{nuc}$) and thus high neutron-to-seed ratio
($Y_n/Y_{\mathrm{seed}}\approx 70$).  This is enough to produce nuclei
around the $A=195$ peak and mimics the hydrodynamical conditions of a
neutrino-driven wind where the r-process does occur and of other
astrophysical environments that involve ejection of high entropy
matter. Therefore, it can be used as basis to study the combined
influence on the abundances of the long-time dynamical evolution and
of the nuclear physics input.

Variations in the late evolution of the ejecta are expected as shown
in multidimensional simulations \cite{Arcones.Janka:2010}. Although
the neutrino-driven wind stays spherically symmetric in absence of
rotation, the interaction of the wind with the slow, early supernova
ejecta and thus the resulting reverse shock depends on the progenitor
structure \cite{arcones.janka.scheck:2007} and on the anisotropic
pressure distribution of the supernova ejecta, where the wind
propagates through~\cite{Arcones.Janka:2010}. Since our trajectory
corresponds to a spherically symmetric simulation, we have modified it
to account for the possible variation of the reverse shock radius and
of the long-time evolution. The changes are done only after
charged-particle reactions freeze out to assure same initial
conditions for the r-process.

As we will show, the reverse shock has a non-negligible influence on
nucleosynthesis. When the wind collides with the slow-moving ejecta, the
expansion velocity drops as kinetic energy is transformed into
internal energy, with the consequent increase in temperature and
density. The density ($\rho_{\text{rs}}$), temperature
($T_{\text{rs}}$), and velocity ($u_{\text{rs}}$) of the shocked
matter can be calculated with the Rankine-Hugoniot conditions,
corresponding to mass, momentum, and energy conservation:
\begin{subequations}
  \label{eq:rh}
\begin{align}
  \rho_{\text{w}}u_{\text{w}} &=
  \rho_{\text{rs}}u_{\text{rs}} \, ,\label{eq:rh1}\\
  P_{\text{w}} + \rho_{\text{w}}u_{\text{w}}^2 & =
  P_{\text{rs}} + \rho_{\text{rs}}u_{\text{rs}}^2 \, ,  \label{eq:rh2}\\
  \frac{1}{2}u_{\text{w}}^2 + \epsilon_{\text{w}}+
  \frac{P_{\text{w}}}{\rho_{\text{w}}} &= \frac{1}{2}u_{\text{rs}}^2 +
  \epsilon_{\text{rs}}+ \frac{P_{\text{rs}}}{\rho_{\text{rs}}} \,
  , \label{eq:rh3}
\end{align}
\end{subequations}
here $\rho$, $u$, $P$, and $\epsilon$ are the density, velocity,
pressure, and specific internal energy, respectively. Quantities
in the wind have the subscript ``w'' and in the shocked material,
above the reverse shock, the subscript ``rs''. In
addition to these equations, one needs an equation of state that
relates pressure and energy. In the last equation we take $\epsilon
\approx 3P/\rho$ which is a good approximations because the
environment is radiation-dominated.  The lhs of these equations is
known from the wind, therefore we combine them to obtain two possible
solutions for the matter velocity after the shock: 1) $u_{\text{rs}} =
u_{\text{w}}$, no shock; 2) $u_{\text{rs}}= u_{\text{w}}/7 + 8/7 \,
P_{\text{w}}/(\rho_{\text{w}}u_{\text{w}})$. Once the velocity is
known, the density and pressure of the shocked matter are calculated
with Eqs.~(\ref{eq:rh1}),~(\ref{eq:rh2}). Any other thermodynamical
variable, including temperature and entropy, can be obtained from an
equation of state (EoS). Here, we use the Timmes
EoS~\cite{Timmes.Arnett:1999}.

Once the conditions of the shocked matter are known, we need to
describe their evolution. The density and velocity evolutions have to
fulfill the condition of constant mass outflow ($\dot{M} = 4\pi r^2 v
\rho$). Two extreme expansions can be identified: 1) the velocity is
constant, and the density thus decreases as $r^{-2}$; 2) the density
is constant and then it is the velocity that decreases as
$r^{-2}$. The latter expansion was used in
Ref.~\cite{Kuroda.Wanajo.Nomoto:2008} but it implies a decrease of the
velocity down to a few m~s$^{-1}$ in about a second, while in full
hydrodynamical simulations \cite{arcones.janka.scheck:2007,
  Fischer.etal:2010}, the velocities stay around
$10^3$--$10^4$~km~s$^{-1}$. Following these simulations, we prescribe
an extrapolation for the evolution of matter which is between these
two extreme cases.  The density of the shocked matter stays constant
and the velocity decreases as $r^{-2}$ during one second. Afterwards,
we keep the velocity constant and the density decreases as $r^{-2}$. A
similar extrapolation is obtained with the prescription used in
Ref.~\cite{Wanajo.etal:2010}.

\subsection{Nucleosynthesis network and nuclear physics input}
\label{sec:netw}

We start our nucleosynthesis calculations at a temperature $T=10$~GK
where the composition is given by nuclear statistical equilibrium and
is dominated by free neutrons and protons. The evolution of the
composition is followed with two reactions networks. During the seed
formation we use an extended network that includes all possible
charged-particle reactions. While for the r-process we take advantage
of a faster network that only considers neutron capture, beta decay,
photodissociation, alpha decay, and fission. However, fission
reactions play a negligible role in the present calculations and will
not be further discussed.

The extended nuclear reaction network consists of 3347 nuclei from
neutron and protons to Europium. Neutral and charged-particle
reactions are the same as in the REACLIB compilation used
by~\citep{Froehlich.Hauser.ea:2006}. For the weak-interaction rates
(electron/positron captures and $\beta$-decays) we use the rates of
Refs.~\cite{Fuller.Fowler.Newman:1982b,Fuller.Fowler.Newman:1982a} for
nuclei with $A\leq 45$ and those of
Refs.~\cite{Langanke.Martinez-Pinedo:2000,Langanke.Martinez-Pinedo:2001}
for $45<A\leq65$. Neutrino interactions are important during the seed
formation since they control the amount of free neutrons and are taken
from Ref.~\cite{Zinner:2007}. When the temperature drops below $T\sim
3$~GK charged-particle reactions freeze out, i.e end of the
$\alpha$-process \cite{Woosley.Hoffman:1992,
  Witti.Janka.Takahashi:1994}, and the r-process phase, characterized
by a domination of neutron capture, begins. The subsequent evolution
is followed by a r-process network of 5300 nuclei between $Z=14$ and
$Z=110$. Since we are interested in the detailed evolution of matter
when decays to stability, we had to improve the original network of
D.~Mocelj PhD~\cite{Mocelj:2007}. This network was based on the
algorithm suggested by Ref.~\cite{Cowan.Cameron.Truran:1983} which is
also used in Refs.~\cite{Freiburghaus.Rembges.ea:1999,
  Farouqi.etal:2010}. Such algorithm assumes that the neutron
abundance ($Y_n$) stays constant during a time step (see Appendix).
This is very efficient during the early r-process phase, but it
becomes numerically unstable when the neutron-to-seed ratio drops
below one and $Y_n$ decreases very fast. This problem can be cured
using smaller time steps, which increases however the computational
time without completely removing the numerical instabilities.  In our
updated network, the equation for the neutron abundance evolution (see
Eq.~\ref{eq:dyn}) is included and the resulting system of equations is
solved using a fully implicit scheme based on the Newton-Raphson
method~\cite{Hix.Thielemann:1999b,Hix.Meyer:2006} and the sparse
matrix solver package PARDISO~\cite{Schenk.Gaertner:2004}.

Our nucleosynthesis calculations are based on four different mass
models and their consistently calculated neutron capture rates: the
Finite Range Droplet Model (FRDM)~\citep{Moeller.Nix.ea:1995}, the
quenched version of the Extended Thomas-Fermi with Strutinsky Integral
(ETFSI-Q)~\citep{Pearson.Nayak.Goriely:1996}, the version 17 of the
Hartree-Fock-Bogoliubov masses
(HFB-17)~\citep{Goriely.Chamel.Pearson:2009}, and the Duflo-Zuker mass
formula~\citep{Duflo.Zuker:1995}. For the FRDM and ETFSI-Q mass models
the neutron-capture rates are taken from
Ref.~\cite{Rauscher.Thielemann:2000} that uses the statistical model
code NON-SMOKER~\cite{Rauscher.Thielemann:1998}. The HFB-17 neutron
capture rates are taken from the Bruslib
database~\footnote{\protect\url{http://www-astro.ulb.ac.be/Html/talys.html}}
and were computed with the statistical model code
TALYS~\citep{Goriely.Hilaire.Koning:2008}. For the Duflo-Zuker mass
formula, we have evaluated the neutron-capture rates using the
analytic approximation suggested in Ref.~\cite{Michaud.Fowler:1970}
that reproduces the results of more sophisticated statistical model
calculations~\citep{Woosley.Fowler.ea:1975,mathews.mengoni.ea:1983}
(see also Sect.~\ref{sec:back}). For the range of temperatures we are
considering, the temperature dependence of the $(n,\gamma)$ rates can
be neglected and we use thus values that correspond to a temperature
of 30~keV. The inverse $(\gamma,n)$ rates are obtained from detailed
balance (see Eq.~\ref{eq:photonv}).

We use theoretical beta-decay rates from
Ref.~\cite{Moeller.Pfeiffer.Kratz:2003} supplemented by experimental
data whenever available (NuDat 2 database~\citep{NuDat2}). We realize
that the beta decays should be calculated consistently with the mass
model, however such calculations have not been performed so far.  We
plan to explore the sensitivity of r-process calculations to different
beta-decay rates in future work.

\section{Results}
\label{sec:results}
\subsection{Wind termination shock and long-time evolution}
\label{sec:rs_effect}

Here we analyze the impact of the reverse shock on the r-process
abundances and dynamics. All calculations presented in this section
are based on the ETFSI-Q mass model.

\begin{figure}[!ht]
  \centering
    \includegraphics[width=\linewidth,angle=0]{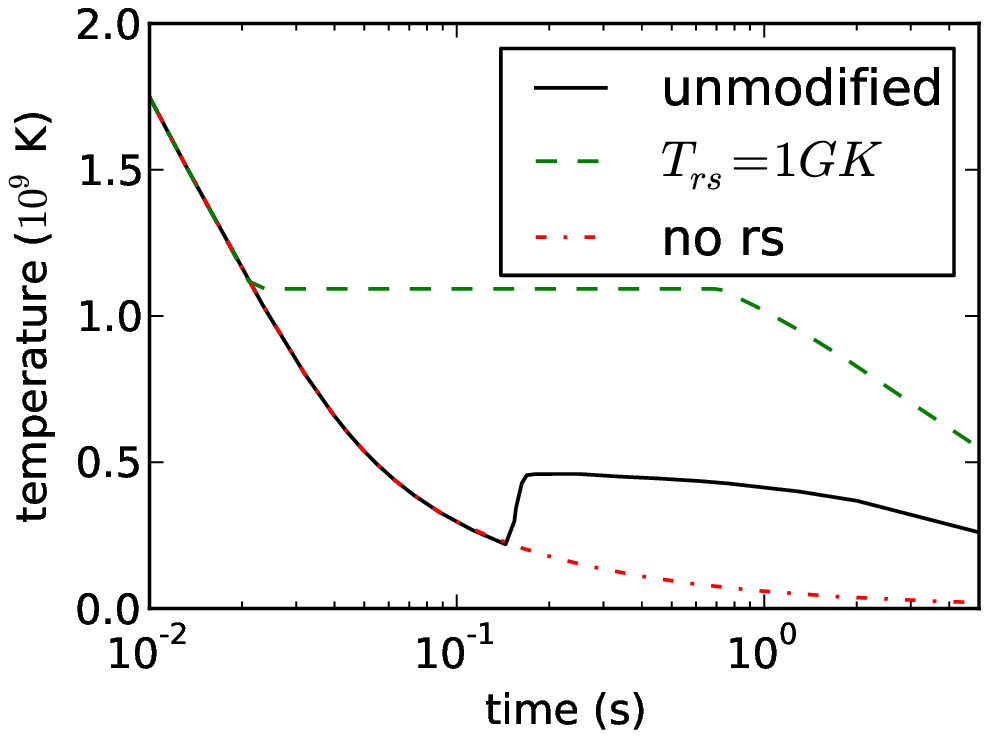}\\
    \includegraphics[width=\linewidth,angle=0]{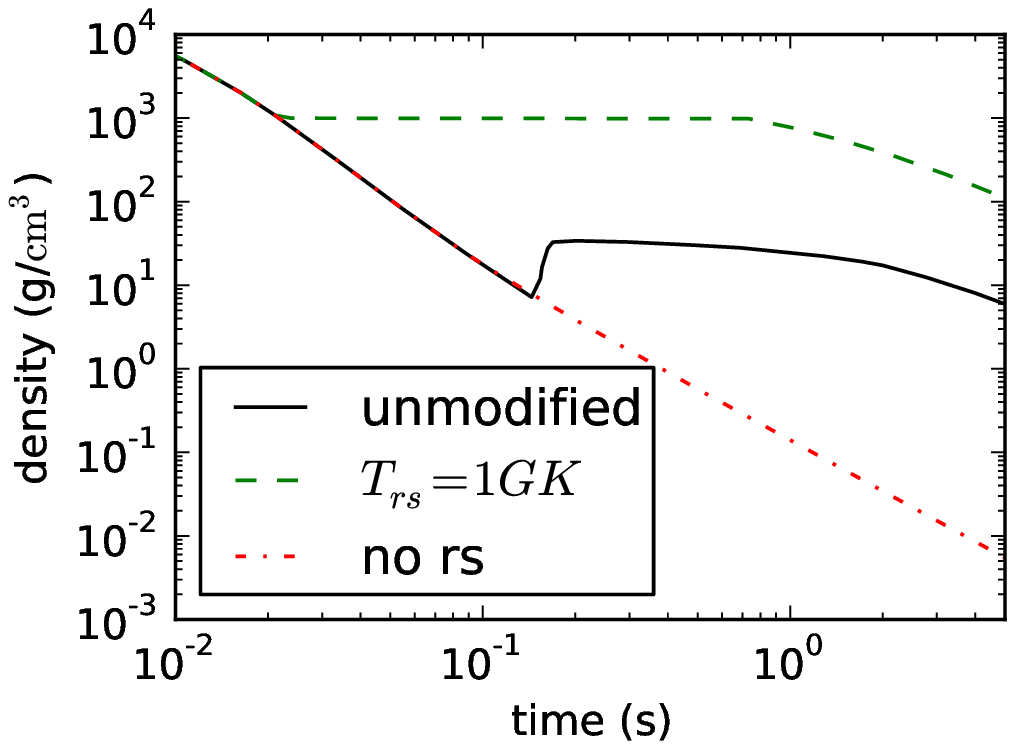}
    \caption{(Color online) Temperature and density evolution of a
      mass element ejected at 8~s after bounce of the supernova
      explosion based on model M15-l2-r1 of
      Ref.~\cite{arcones.janka.scheck:2007}, see discussion in
      Sect.~\ref{sec:methods}.  The solid black line corresponds to
      the original trajectory from the supernova simulation labeled as
      ``unmodified''. Notice, that density have been divided by a
      factor 2 to get higher neutron-to-seed ratio. The green dashed
      line represents a evolution with the reverse shock at
      temperature of 1~GK\@. For the evolution shown by the red
      dashed-dotted line the reverse shock was removed.}
  \label{fig:rs_td}
\end{figure}

Our nucleosynthesis study correspond to the trajectories shown in
Fig.~\ref{fig:rs_td}. The solid black line, labeled as ``unmodified'',
represents the trajectory from Ref.~\cite{arcones.janka.scheck:2007}
introduced in Sect.~\ref{sec:sim}.  The position of the reverse shock
and the evolution after it are not modified, but the density is
overall reduced by a factor of two. In the dashed green line the
reverse shock is assumed to be at temperature of 1~GK and the
subsequent evolution is calculated as described in
Sect.~\ref{sec:sim}.  The dashed-dotted red line, labeled as ``no
rs'', reproduces a case without reverse shock, where matter expands
without colliding with the slow, early supernova ejecta. In this case,
we assume an adiabatic (constant entropy) expansion with constant
velocity starting at the position of the reverse shock in the
simulation. As the mass outflow is constant, the density decreases
with radius as $\rho \propto r^{-2}$. 

\begin{figure}[!ht]
  \centering
  \includegraphics[width=\linewidth,angle=0]{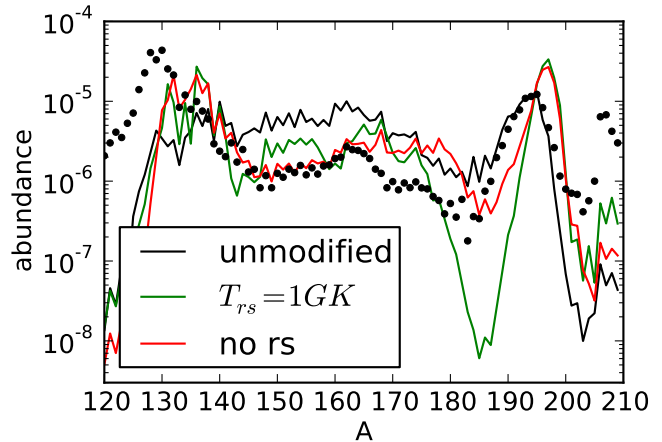}
  \caption{(Color online) Final abundances for the different
    evolutions of Fig.~\ref{fig:rs_td} compared to solar r-process
    abundances shown by dots.}
  \label{fig:rs_abund}
\end{figure}

\begin{figure*}[!ht]
  \centering
    \includegraphics[width=0.5\linewidth,angle=0]{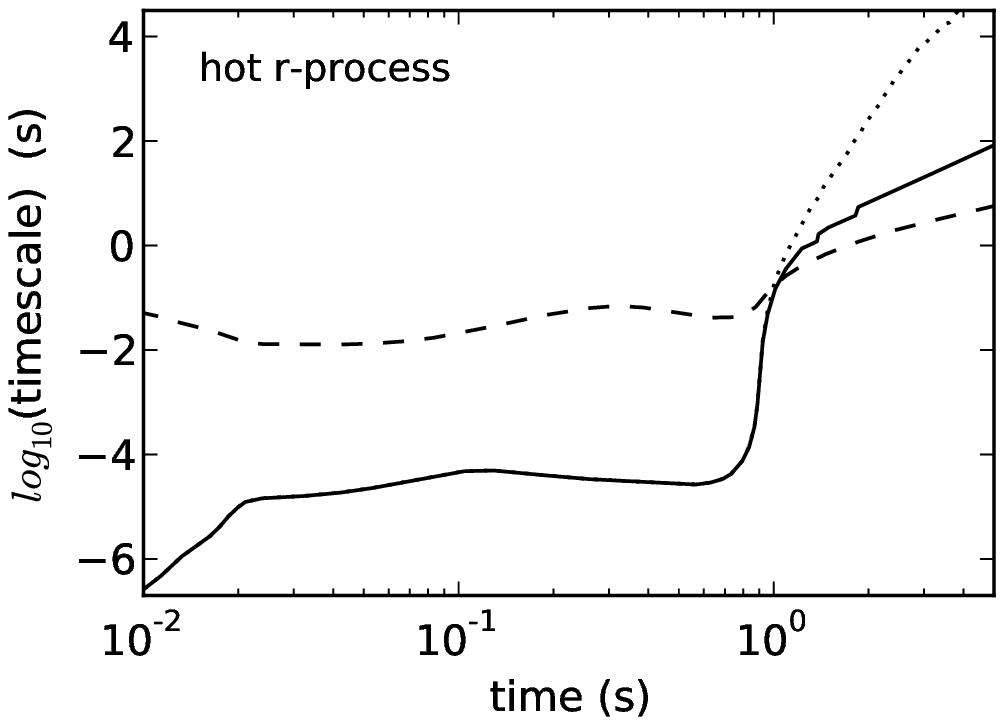}%
    \includegraphics[width=0.5\linewidth,angle=0]{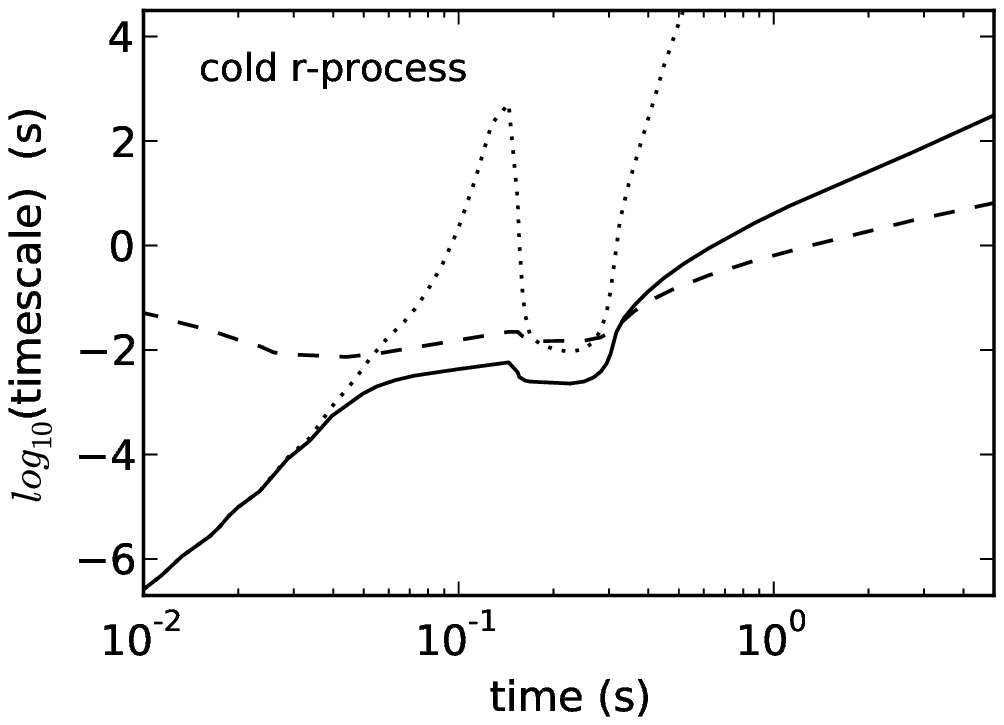}
    \caption{Evolution of relevant timescales for the trajectories
      shown in Fig.~\ref{fig:rs_td} and labeled as
      ``$T_{\mathrm{rs}}=1$~GK'' (left panel) and ``unmodified''
      (right panel). Solid, dashed, and dotted lines represent the
      neutron capture, beta decay, and photodissociation timescales,
      respectively.}
  \label{fig:rs_tau}
\end{figure*}

Figure \ref{fig:rs_abund} shows the final r-process abundances for the
different trajectories together with solar r-process abundances
(dots)~\cite{Sneden.etal:2008}. None of the calculations reproduce the
solar abundances around the second peak ($A\sim 130$), since we have
chosen the conditions which produce mainly the third r-process peak
($A\sim 195$). A pattern similar to solar is expected from a
superposition of different
trajectories~\cite{Freiburghaus.Rembges.ea:1999,Wanajo.Kajino.ea:2001}.
Some of the deficiencies seen in Fig.~\ref{fig:rs_abund} are due to
the mass model and will be discussed in the next section. However,
there are features that depend on the dynamical evolution. In order to
understand the abundances under different late time evolutions
(Fig.~\ref{fig:rs_td}), we look at the characteristic timescales for
the r-process: neutron capture, photodissociation, and beta decay that
are defined, respectively, as:
\begin{subequations}
  \label{eq:timescales}
  \begin{eqnarray}
    \frac{1}{\tau_{(n,\gamma)}} & = & \frac{\sum_{Z,A} N_n \langle \sigma v
      \rangle_{(Z,A)} Y(Z,A)}{\sum_{Z,A}  Y(Z,A)}, \label{eq:tng}\\  
    \frac{1}{\tau_{(\gamma,n)}} & =&
    \frac{\sum_{Z,A} \lambda_{\gamma} (Z,A) Y(Z,A)}{\sum_{Z,A}
      Y(Z,A)}, \label{eq:tgn} \\    
    \frac{1}{\tau_{\beta}} & = & \frac{\sum_{Z,A}\lambda_{\beta}(Z,A)
Y(Z,A)}
    {\sum_{Z,A}  Y(Z,A)}, \label{eq:tbeta}
  \end{eqnarray}
\end{subequations}
where $N_n$ is the neutron number density, $\langle\sigma
v\rangle_{(Z,A)}$ the neutron capture or $(n,\gamma)$ rate,
$\lambda_{\gamma}(Z,A)$ the photodissociations or $(\gamma,n)$ rate,
and $\lambda_{\beta}(Z,A)$ the beta-decay rate. The evolution of these
timescales is shown in Fig.~\ref{fig:rs_tau} for the trajectories
labeled as ``$T_{rs} = 1$~GK'' (left panel) and ``unmodified'' (right
panel). The trajectory labeled ``no rs'' follows a behavior very
similar to the unmodified one.

For the trajectory with the reverse shock at $T=1$~GK
(Fig.~\ref{fig:rs_tau}, left panel), there is a competition between
neutron captures and photodissociation that last until neutrons are
exhausted at around 1~s. During this evolution under \nggn\
equilibrium, the beta-decay timescale is longer than the other two,
i.e.\ $\tau_{(n,\gamma)} = \tau_{(\gamma,n)} \ll \tau_{\beta} $.  In
the following, this kind of evolution is called \emph{hot
  r-process}. When the reverse shock is at lower temperatures
(Fig.~\ref{fig:rs_tau}, right panel), the photodissociation timescale
becomes longer than the other two timescales once the temperature
drops below $\sim 0.5$~GK. The subsequent evolution proceeds by a
competition between beta decay and neutron capture, i.e.\
$\tau_{(n,\gamma)} \approx \tau_{\beta} \ll \tau_{(\gamma,n)}$. This
evolution was already studied in Ref.~\cite{Blake.Schramm:1976} and
has been recently named as cold r-process~\cite{Wanajo:2007} and
rn-process~\cite{Panov.Janka:2009}. We call this evolution \emph{cold
  r-process}.

The relevant nuclear physics input depends on the dynamical evolution.
The hot r-process proceeds initially in \nggn\ equilibrium, therefore
the neutron separation energy is the key quantity that determines the
location of the r-process path, see Eq.~(\ref{eq:yequiv}). For the
cold r-process the evolution proceeds by a competition between neutron
captures and beta decays. Therefore, the most relevant nuclear physics
inputs are beta-decay and neutron-capture rates. Nuclear masses are
also important because they enter in the calculation of both.  The
evolution after freeze out is dominated by beta decays and neutron
captures for both cold and hot r-process.

The final abundances shown in Fig.~\ref{fig:rs_abund} indicate that
cold r-process calculations lead to broader peaks because the
r-process proceeds farther away from stability. The hot r-process,
which evolves in \nggn\ equilibrium, results in a huge trough in the
abundances before the third r-process peak (green line in
Fig.~\ref{fig:rs_abund}). This is due to the behavior of the neutron
separation energy just before the $N=126$ shell closure (see
Fig.~\ref{fig:s2npath}) and will be discussed in the next section.

\begin{figure}
  \centering
  \begin{tabular}{c}  
    \includegraphics[width=\linewidth,angle=0]{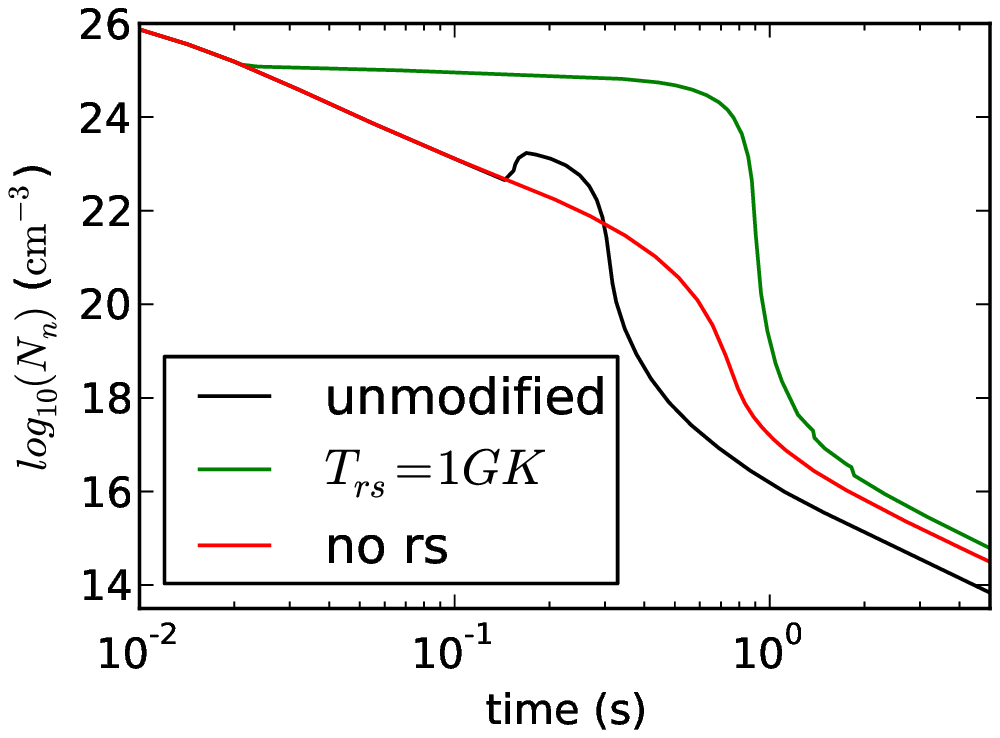}    \\
    \includegraphics[width=\linewidth,angle=0]{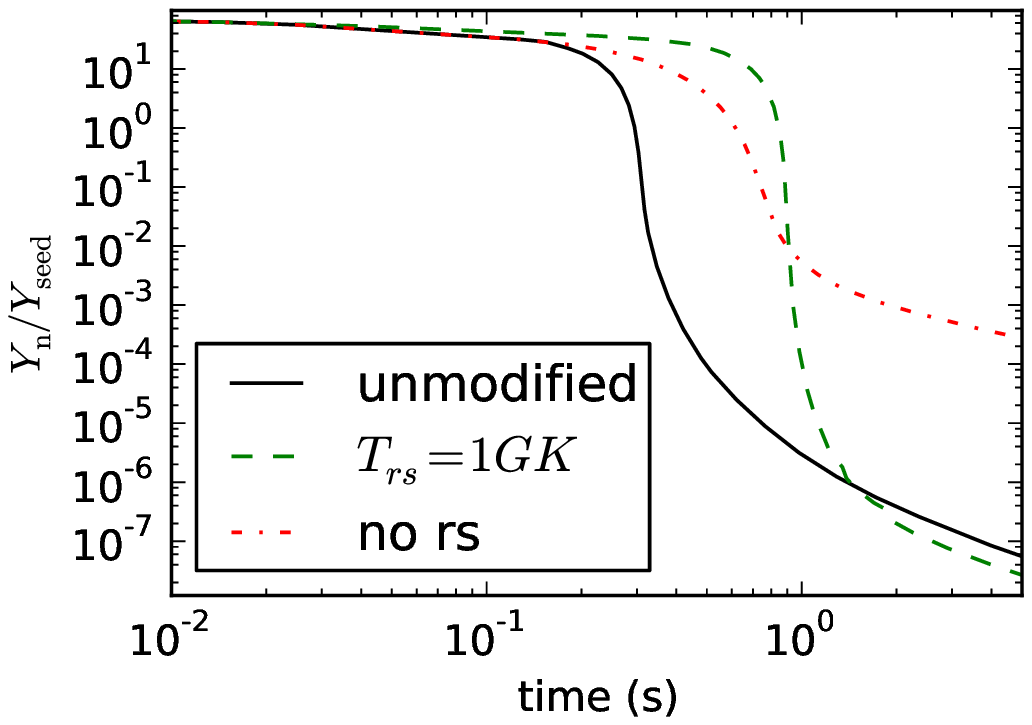} \\
    \includegraphics[width=\linewidth,angle=0]{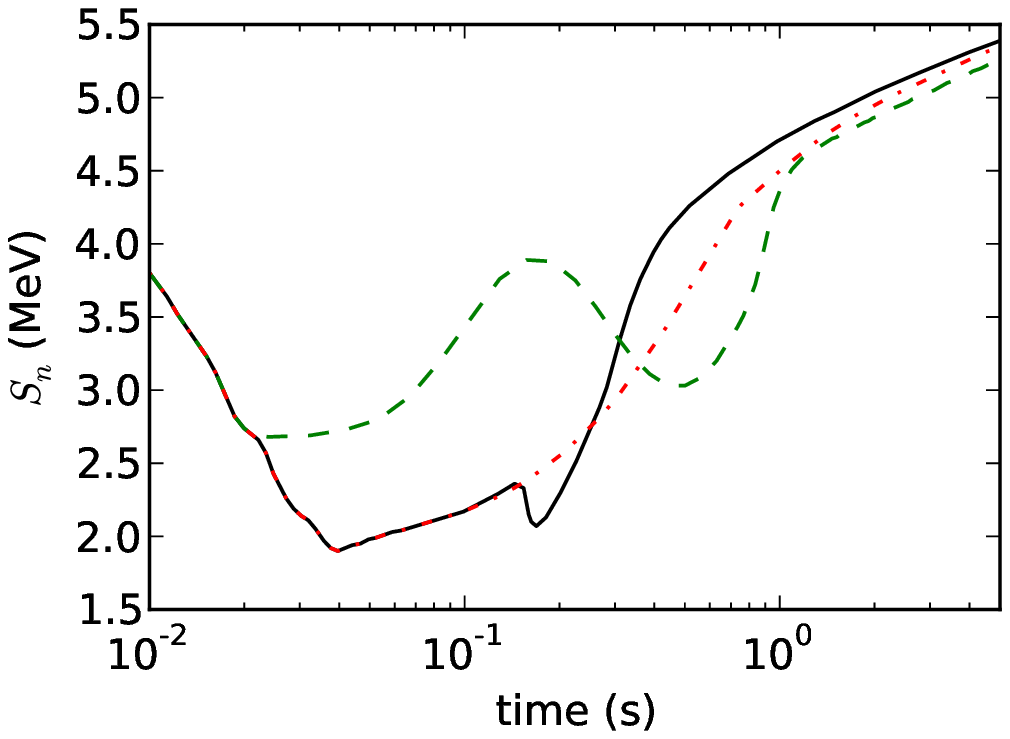}
  \end{tabular}
  \caption{(Color online) Neutron density, neutron-to-seed ratio, and
    average neutron separation energy evolution
    (Eq.~(\ref{eq:average_sn})) for the same trajectories shown in
    Fig.~\ref{fig:rs_td}.}
  \label{fig:rs_nn}
\end{figure}

After we have introduced the two possibilities for the r-process: hot
and cold, the evolution of relevant quantities (neutron density,
neutron-to-seed ratio, and average neutron separation energy) will be
explained.  The average neutron separation energy (shown in the bottom
panel of Fig.~\ref{fig:rs_nn}) is defined as:
\begin{equation}
  \langle S_n \rangle = \frac{\sum_{Z,A} S_{n}(Z,A) Y(Z,A)}
  {\sum_{Z,A} Y(Z,A)},
  \label{eq:average_sn}
\end{equation}
where $S_n (Z,A)$ is the neutron separation energy of a nucleus with
mass number $A$ and charge number $Z$, and $Y(Z,A)$ is its
abundance. The neutron separation energy is smaller for nuclei far
from stability, since their neutrons become less bound. Therefore, we
can use this quantity to study the evolution of the r-process.

We consider first the hot r-process which is labeled as
``$T_{\mathrm{rs}}=1$~GK" in
Figs.~\ref{fig:rs_td},~\ref{fig:rs_abund}, and~\ref{fig:rs_nn}. The
average neutron separation energy (Fig.~\ref{fig:rs_nn}, bottom panel)
decreases very fast initially, as matter moves away from stability
after charged-particle reactions freeze out.  $\langle S_n \rangle$
comes to a minimum after 30~ms, because the matter flow has reached
$N=82$ shell closure. Here the abrupt drop of individual neutron
separation energies and the high photodissociation rates prevent
matter to move farther away from stability. Therefore, the r-process
path moves to higher Z by beta decays and successive neutron captures,
while the neutron number stays constant at $N=82$. This is shown in
Fig.~\ref{fig:s2npath} by the dots that mark the r-process path. Once
the neutron separation energy for nuclei beyond $N=82$ becomes large
enough, the flow of matter can continue moving towards heavier
nuclei. The r-process path reaches the $N=126$ shell closure at around
500~ms as indicated by the second minimum in $\langle S_n
\rangle$. When matter starts to pass through the $N=126$ shell
closure, neutrons are exhausted in our calculations and the matter
decays to stability.

The evolution of the average neutron separation energy for the other
two trajectories (cold r-process) is very different because
photodissociation becomes negligible once the temperature is $\lesssim
0.5$~GK. The evolution proceeds by a competition of neutron captures
and beta decays and this allows matter to move farther from stability
compared to the hot r-process. Therefore, the average beta-decay
lifetime becomes shorter which speeds up the flow of matter towards
heavier nuclei and broadens the minimum in $\langle S_n \rangle$. The
faster evolution leads to a more rapid decrease of the neutron-to-seed
ratio (Fig.~\ref{fig:rs_nn}, middle panel) and therefore the r-process
ends earlier than in the hot r-process.

The case without reverse shock (red line in Fig.~\ref{fig:rs_td}) is
extreme because the neutron density decreases initially very fast
(Fig.~\ref{fig:rs_nn}, upper panel). This leads to a drop of the
neutron captures, which explains the high values of the neutron
density and neutron-to-seed ratio that are maintained at later times.
Having a large neutron-to-seed ratio at late times allows a
continuation of neutron captures, even after several
seconds. Therefore, the peak at $A=195$ is shifted towards higher mass
numbers as shown in Fig.~\ref{fig:rs_abund} (see also discussion in
Ref.~\cite{Surman.Engel:2001}). In the hot r-process, neutron captures
after freeze-out also lead to the shift of the third peak, even when
the r-process path at freeze-out is located at a neutron separation
energy of $S_n = 2.8$~MeV, similar to the one used in the classical
r-process calculations of Ref.~\cite{Kratz.Bitouzet.ea:1993}, where
the peak in the final abundances is obtained at the right position.

\subsection{Influence of the mass model}
\label{sec:mm}
\begin{figure*}[htb]
  \centering
  \includegraphics[width=0.45\linewidth]{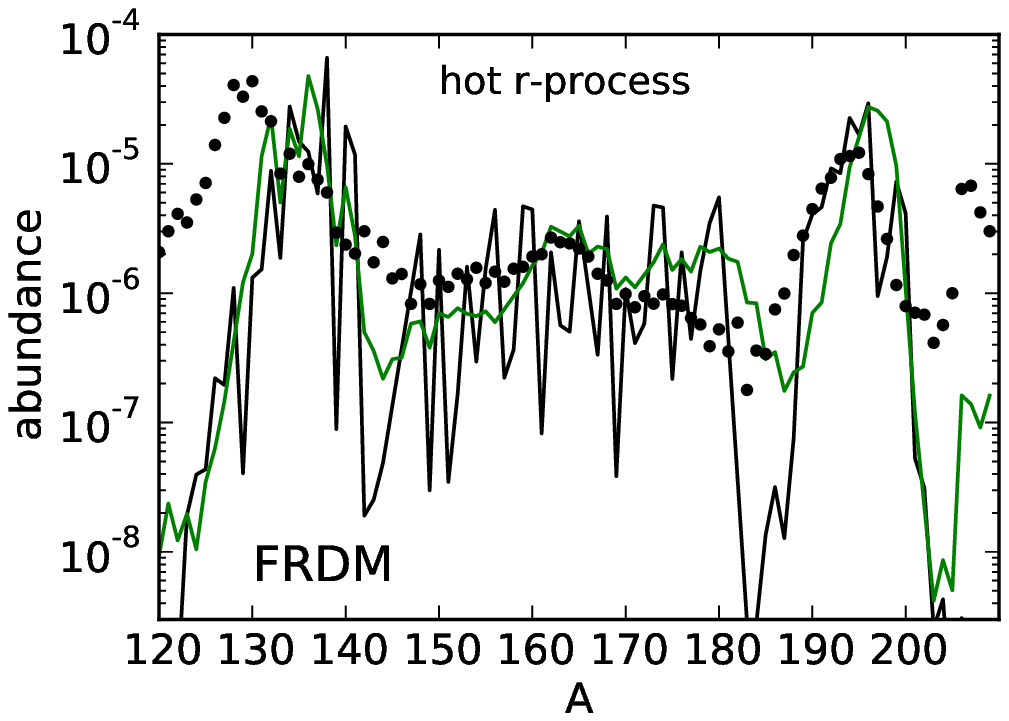}%
  \includegraphics[width=0.45\linewidth]{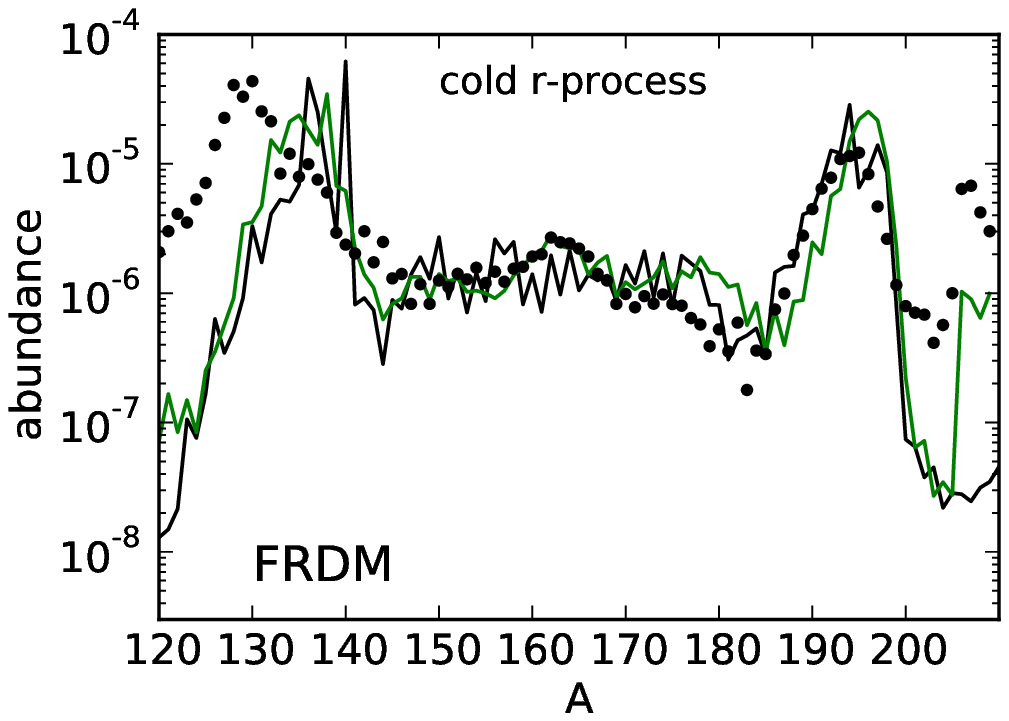}\\
  \includegraphics[width=0.45\linewidth]{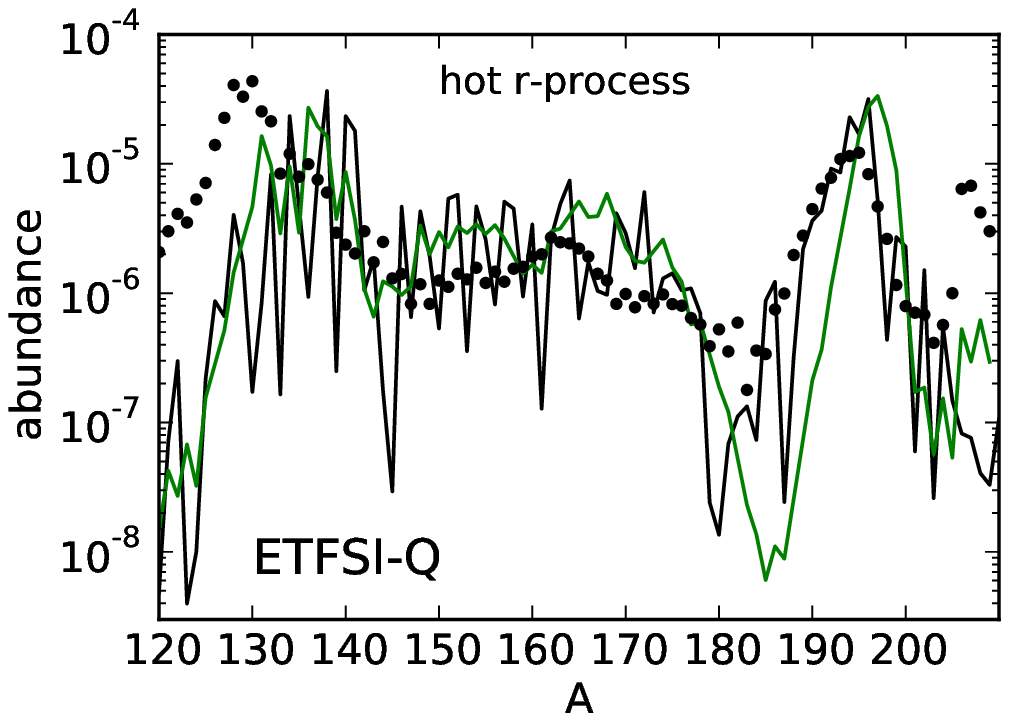}%
  \includegraphics[width=0.45\linewidth]{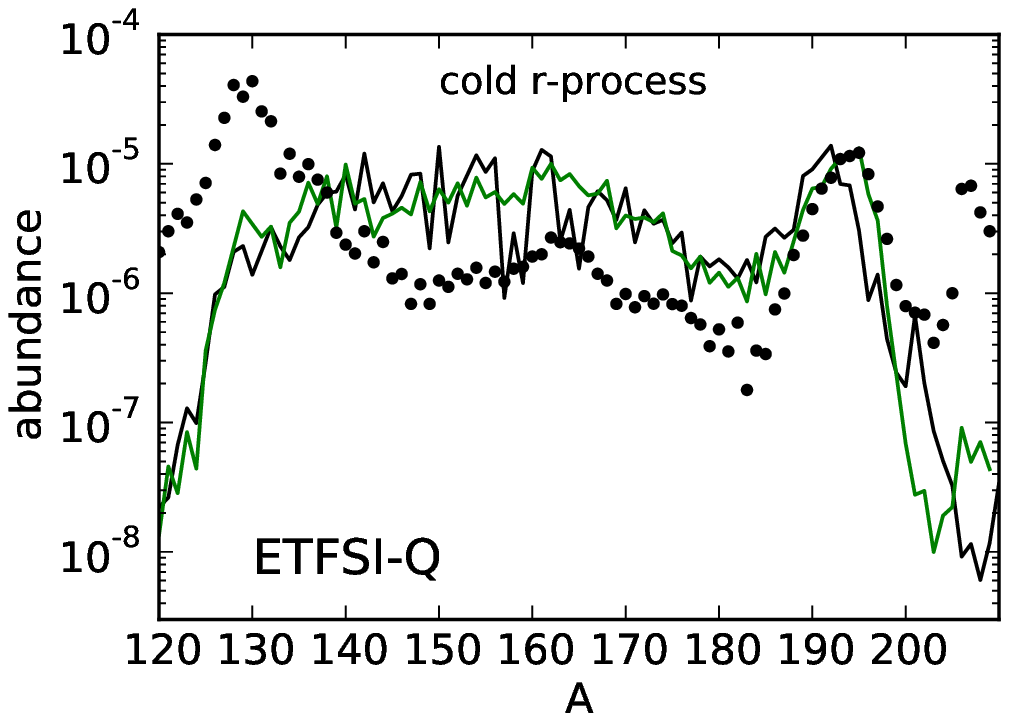}\\
  \includegraphics[width=0.45\linewidth]{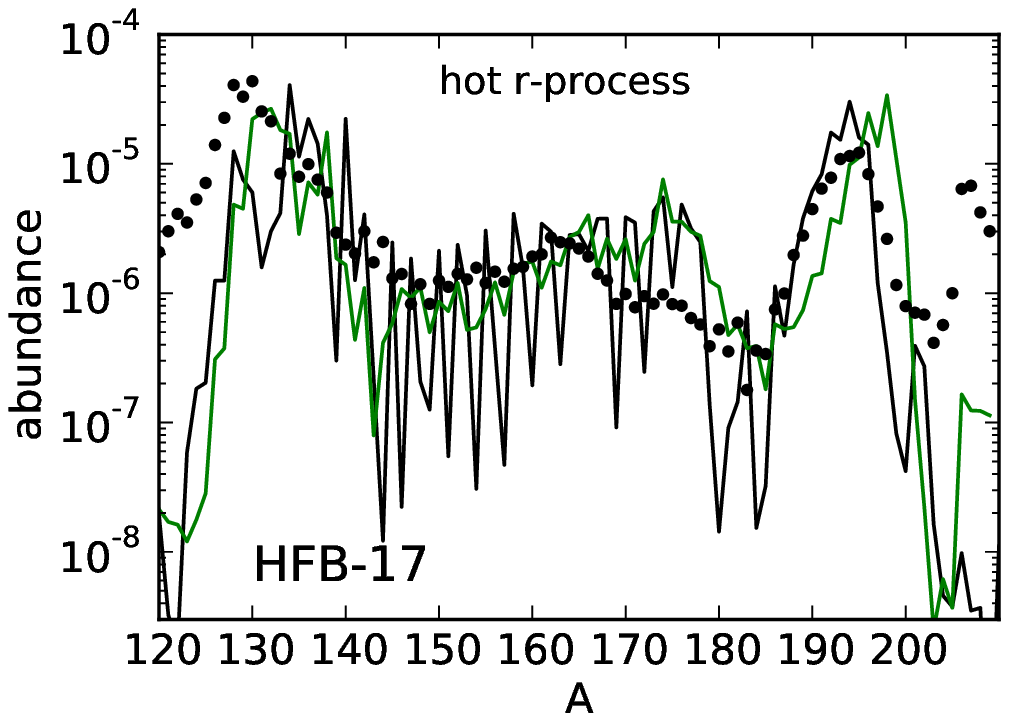}%
  \includegraphics[width=0.45\linewidth]{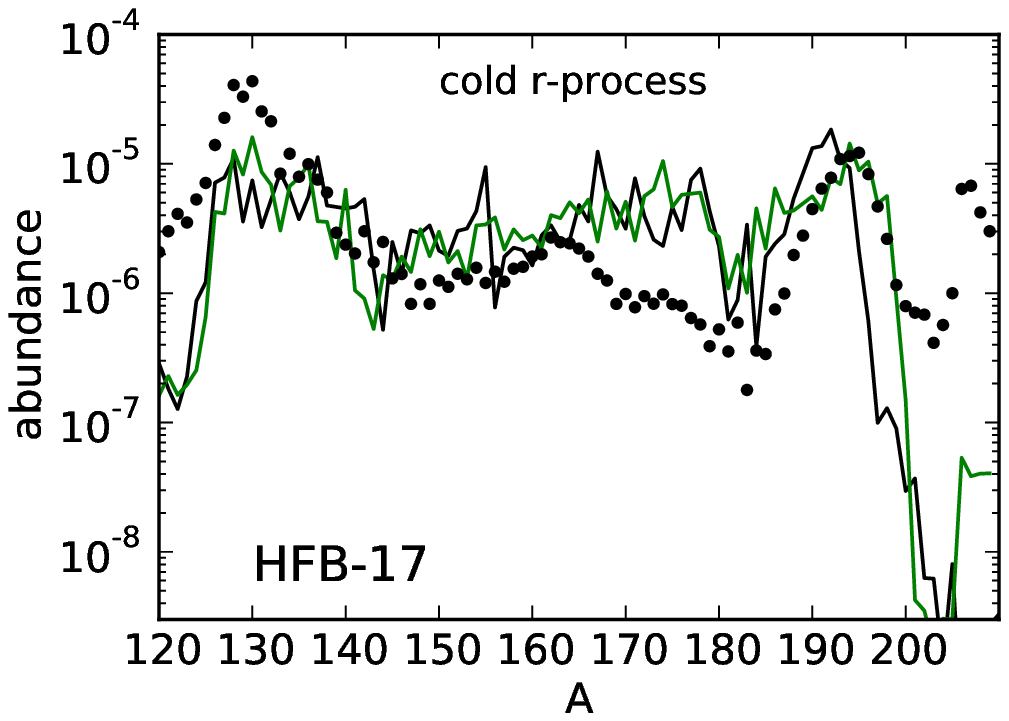}\\
  \includegraphics[width=0.45\linewidth]{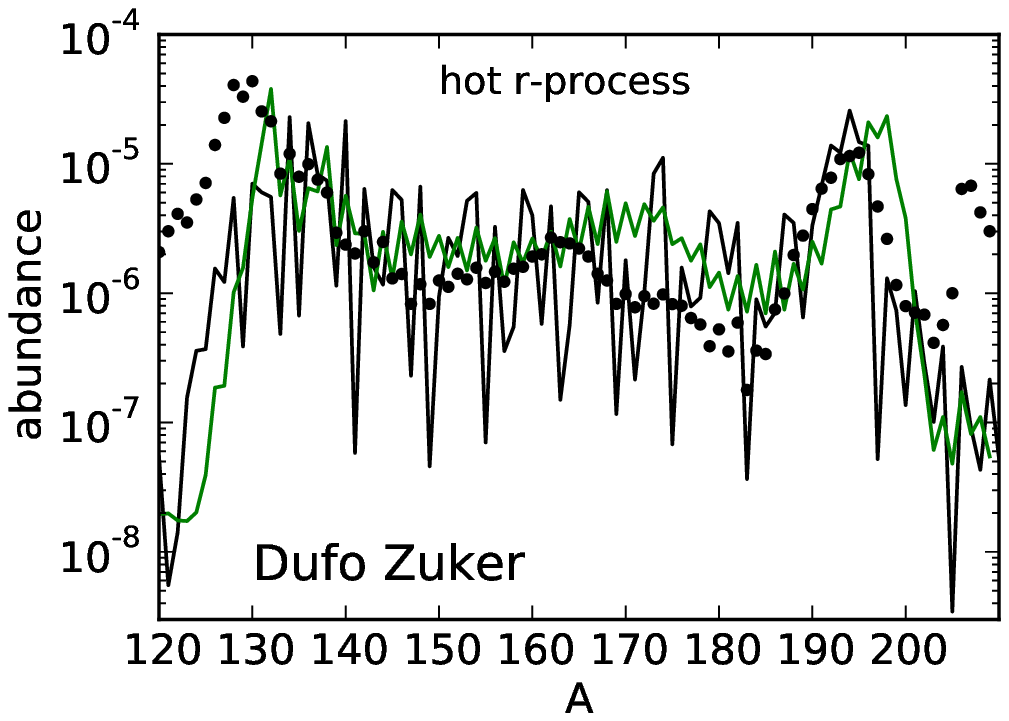}%
  \includegraphics[width=0.45\linewidth]{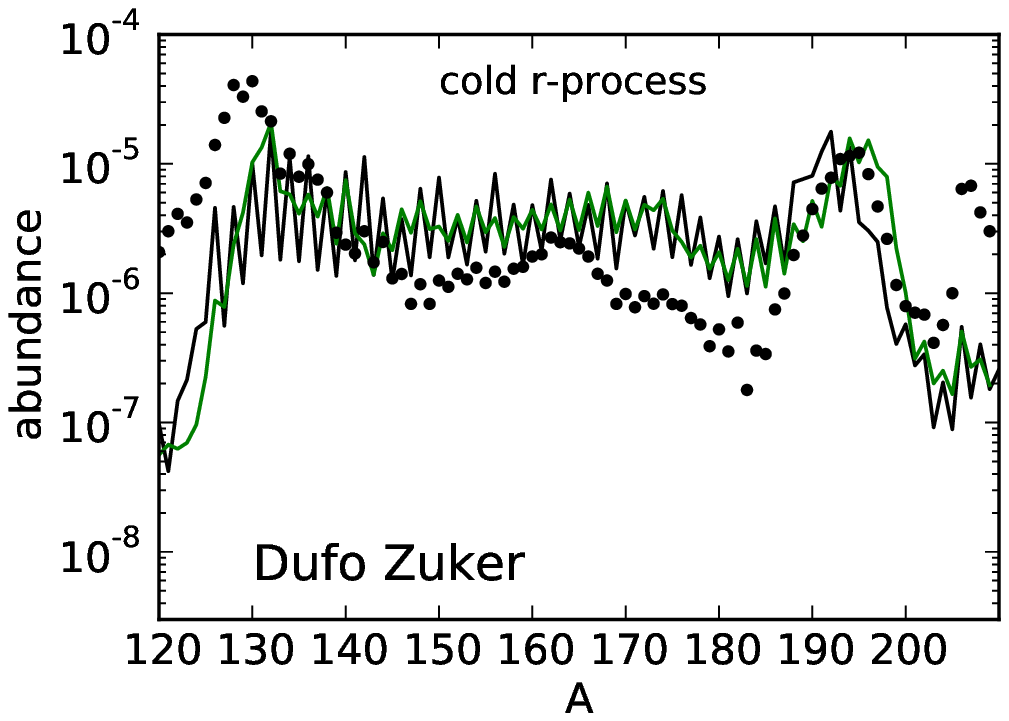}\\
  \caption{(Color online) Freeze-out (black lines) and final (green
    lines) abundances for the hot (left column) and cold (right
    column) r-process. The calculations are based on the mass model
    that is indicated in the label.}
  \label{fig:mm_abund}
\end{figure*}

Nuclear masses are a key nuclear physics input for r-process
calculations as they determine the energy thresholds for all relevant
reactions: neutron capture, photodissociation, and beta decay. In this
section we present results based on the four mass models introduced in
Sect.~\ref{sec:netw} and link features in the abundances with the
behaviour of the two neutron separation energy ($S_{2n}$).  In the
r-process evolution, two phases can be distinguished depending on
whether the neutron-to-seed ratio is larger or smaller than one, i.e.
before and after freeze-out. The abundances at freeze-out (shown in
Fig.~\ref{fig:mm_abund} by black lines) contain the information of the
pre-freeze-out phase when there are enough neutrons to be captured by
each individual nucleus.  After freeze-out, there are two important
facts to be considered: 1) nuclei compete for capturing the few
neutrons available, 2) the neutron-capture and beta-decay rates are
comparable. These two facts are common to the hot and cold r-process
and are important for determining the final abundances shown in
Fig.~\ref{fig:mm_abund} by the green lines. Here we focus on the
difference among mass models and general features will be described in
the next section.

\begin{figure*}[ht]
  \includegraphics[width=0.5\linewidth,angle=0]{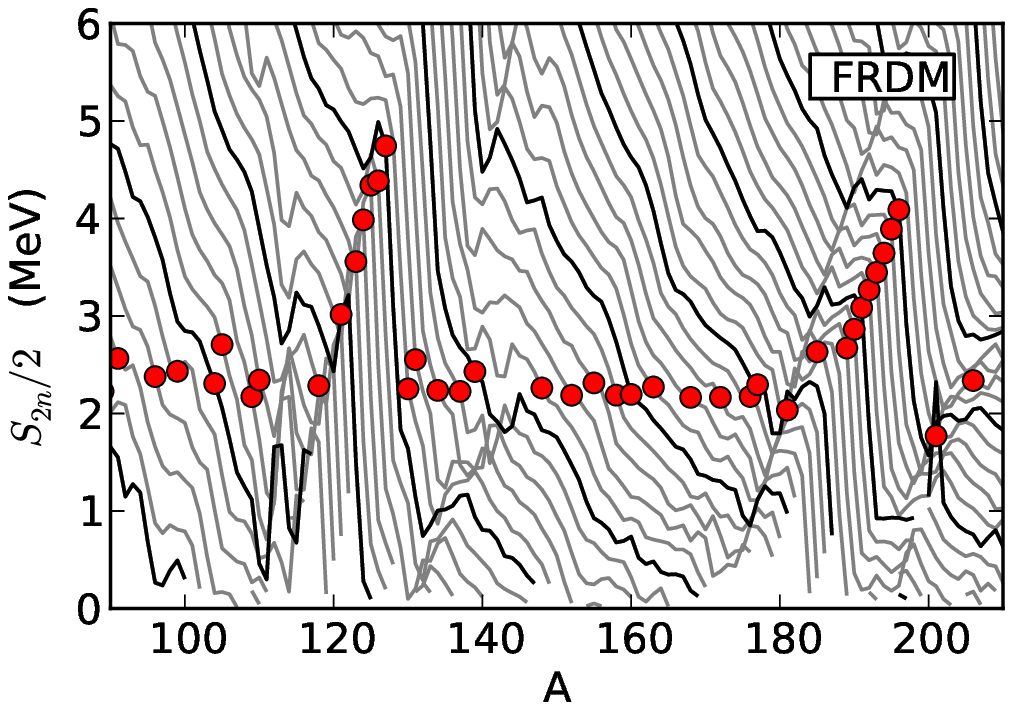}%
  \includegraphics[width=0.5\linewidth,angle=0]{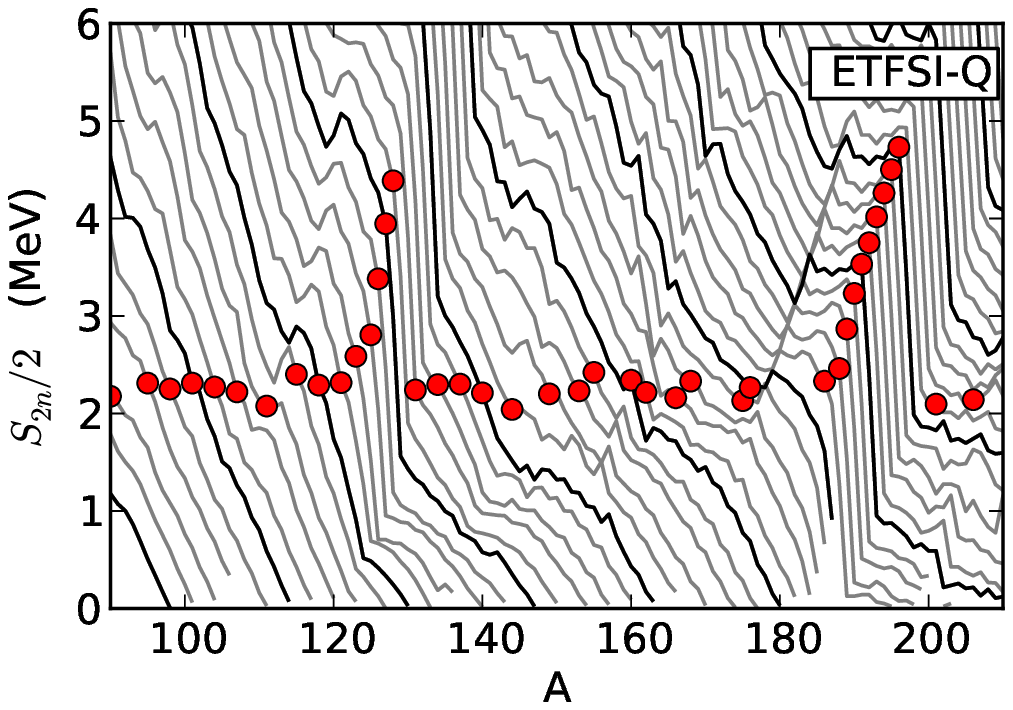}\\
  \includegraphics[width=0.5\linewidth,angle=0]{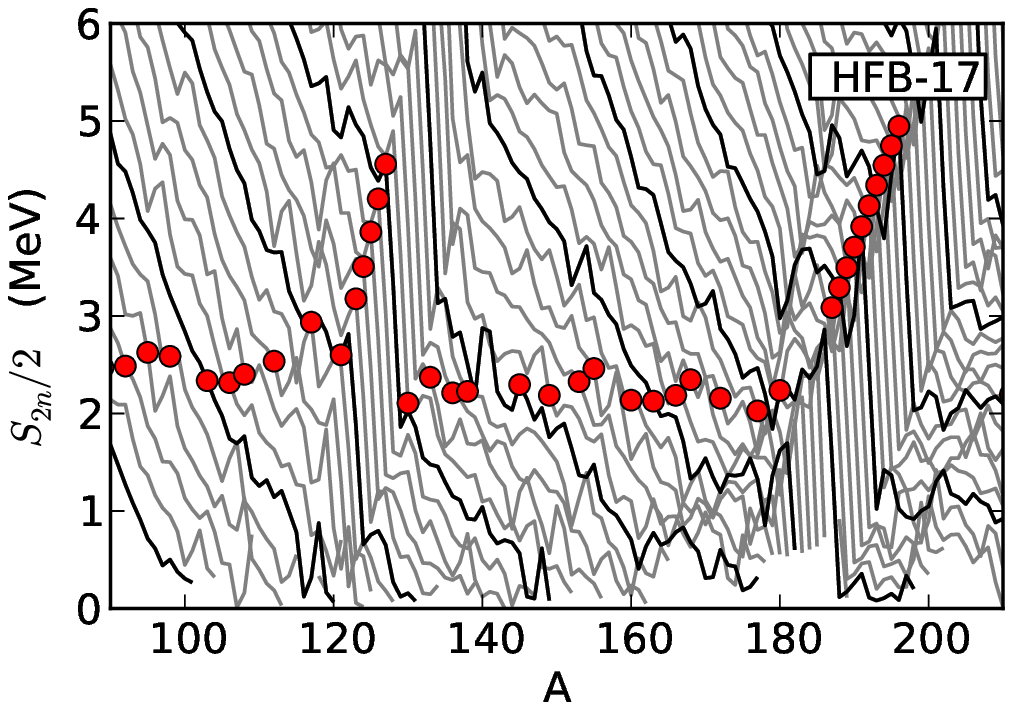}%
  \includegraphics[width=0.5\linewidth,angle=0]{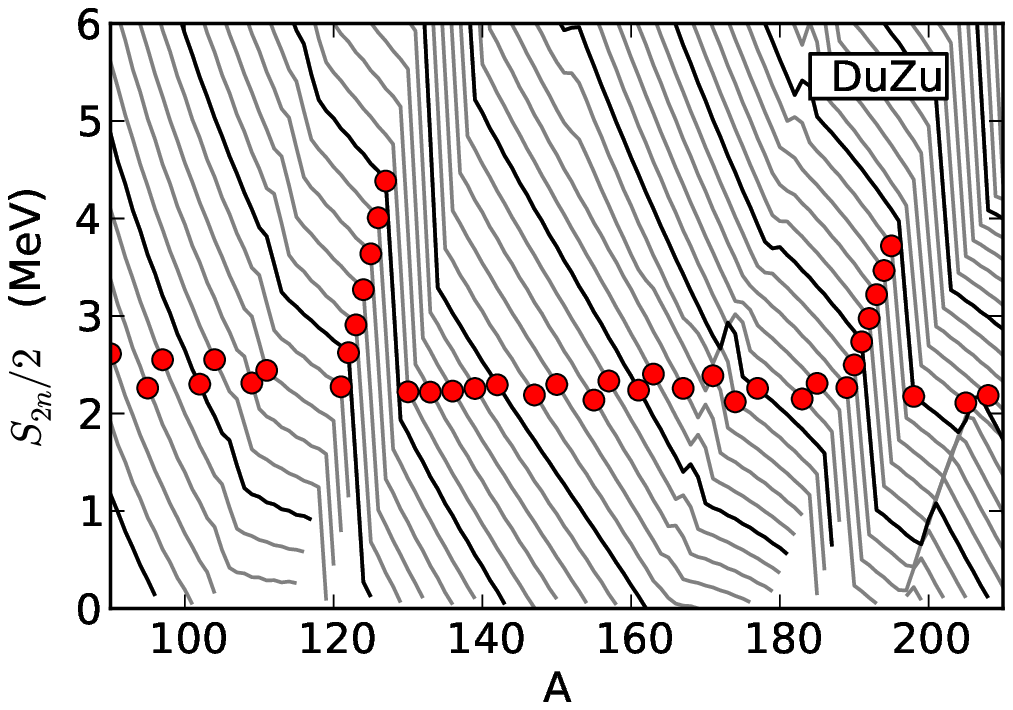}
  \caption{(Color online) Two neutron separation energy for constant
    proton number as a function of mass number. Black lines correspond
    to proton numbers starting at Z~=~30 on the left and increasing in
    steps of 5 and grey lines are shown in steps of 1. The hot
    r-process path is shown at freeze-out ($Y_n/Y_{\mathrm{seed}}=1$)
    by dots.}
  \label{fig:s2npath}
\end{figure*}

Figure~\ref{fig:mm_abund} shows the abundances obtained using the four
mass models for the hot (left column) and cold (right column)
r-process.  The freeze-out abundances (black lines) are characterized,
especially in the hot r-process, by the presence of strong
fluctuations that have almost disappeared in the final abundances
(green lines), as expected from solar r-process abundances. These
fluctuations are due to the fact that $(n,\gamma)$ and $(\gamma,n)$
reactions favor nuclei with an even neutron number. Consequently, as Z
increases in moving from one isotopic chain to the next, N increases
by at least two units (except at the magic numbers where it stays
constant). Therefore, some mass numbers are not present in the
r-process path as shown by the dots in Fig.~\ref{fig:s2npath}.

The freeze-out abundances can be understood looking at the two neutron
separation energy in Fig.~\ref{fig:s2npath}. Two kind of features in
$S_{2n}$ leave a fingerprint on the abundances: 1) The abrupt drop in
$S_{2n}$ at the magic numbers $N=82$ and $N=126$, leads to
accumulation of matter at these neutron numbers and to the formation
of peaks in the abundance distribution.  2) In regions where $S_{2n}$
is constant or presents a saddle point behaviour, an equilibrium can
not be achieved between neutron captures and photodissociation. In the
hot r-process, this leads to troughs in the abundances around $A \sim
110, 140$ for the FRDM mass model and around $A \sim 185$ for FRDM,
ETFSI-Q, and HFB-17 mass models.  These features of the two-neutron
separation energies have less impact for the cold r-process because
photodissociation reactions are suppressed due to the low
temperatures.

\begin{figure*}[htb]
  \centering
  \includegraphics[width=0.5\linewidth]{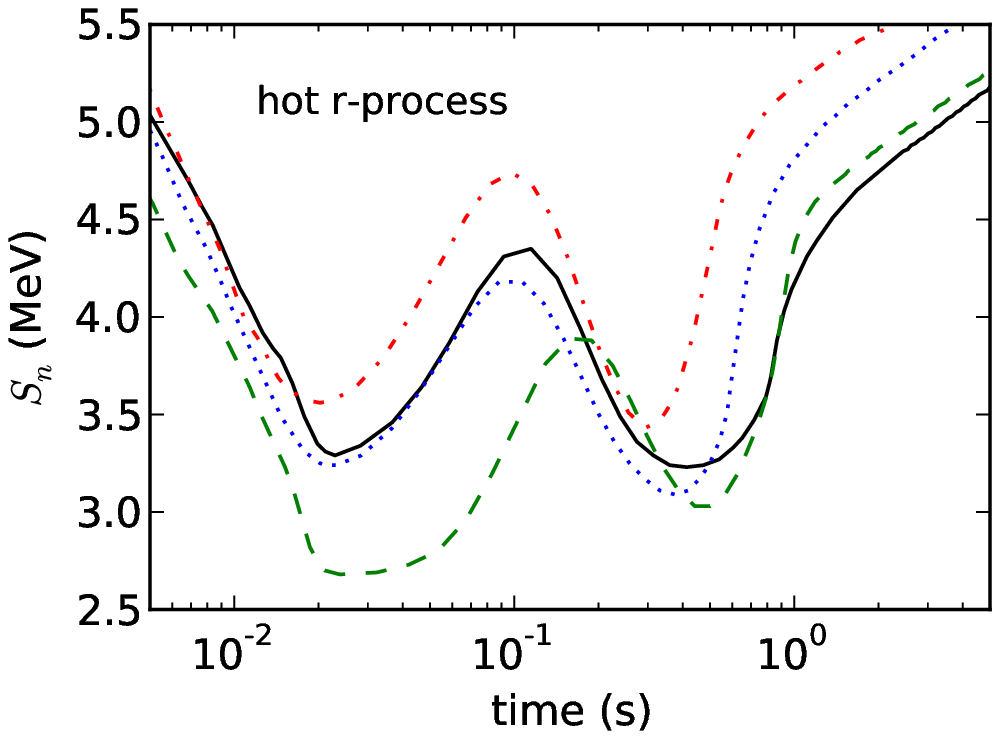}%
  \includegraphics[width=0.5\linewidth]{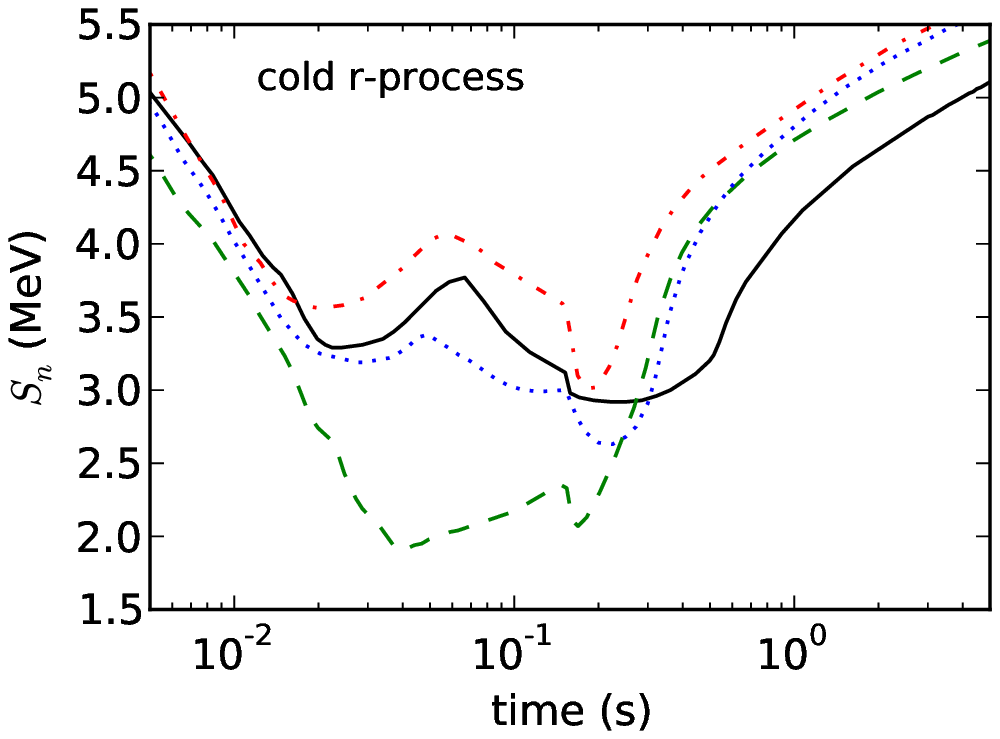}\\
  \includegraphics[width=0.5\linewidth]{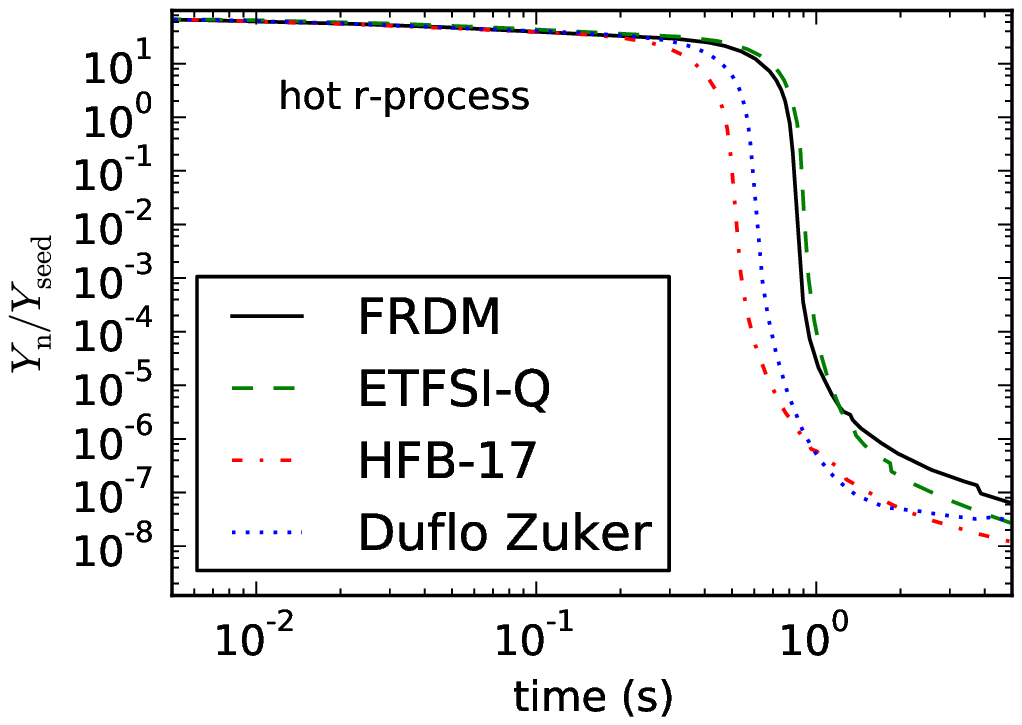}%
  \includegraphics[width=0.5\linewidth]{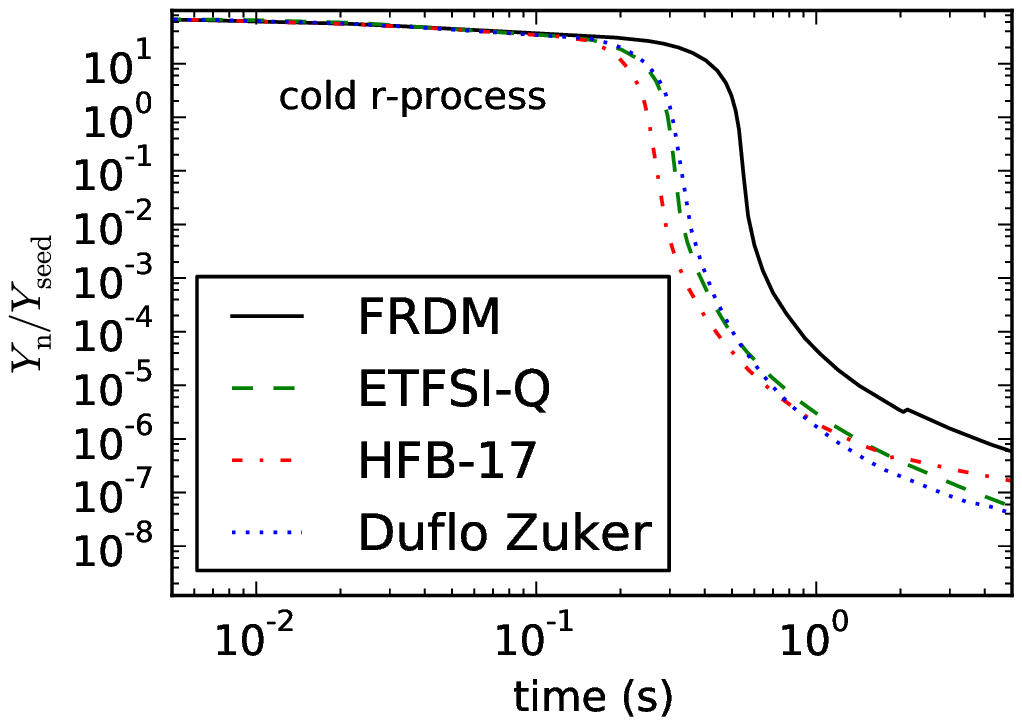}\\
  \caption{(Color online) Evolution of the average neutron separation
    energy and neutron-to-seed ratio for the hot (left column) and
    cold (right column) r-process and based on the four mass models
    discussed in the text.}
  \label{fig:mm_all}
\end{figure*}

We use the average neutron separation energy and the neutron-to-seed
ratio (see Fig.~\ref{fig:mm_all}) to discuss the evolution of matter
during the r-process for different mass models in the hot r-process.
The average neutron separation energy, $\langle S_n \rangle$, shows
two minima and one maximum for all mass models. However, $\langle S_n
\rangle$ significantly differs during the early evolution in the
calculation based on ETFSI-Q nuclear masses.  For $t \approx 30$~ms,
when matter flow approaches $N=82$ shell closure, $\langle S_n
\rangle$ is smaller and the minimum is broader.  Due to the quenching
of the $N=82$ shell gap introduced in the ETFSI-Q mass
model~\cite{Pearson.Nayak.Goriely:1996}, the abrupt drop in $S_{2n}$
at $N=82$ dissapears for $Z<43$ (see Fig.~\ref{fig:s2npath}).
Consequently, with this mass model the r-process proceeds through
nuclei with smaller neutron separation energy before reaching the
$N=82$ shell closure. This occurs at $Z=43$ for ETFSI-Q, while for the
other mass models it is reached already for $Z=40$.  The width of the
first minimum is related to the sum of half-lives of the nuclei on the
r-process path with $N<82$. This is substantially larger in the
calculations with ETFSI-Q mass model. There is thus a delay in the
time required to overcome the $N=82$ shell closure and a slow down of
the speed at which neutrons are captured. Therefore, the r-process
freeze-out occurs at later times for ETFSI-Q than for HFB-17 and
Duflo-Zuker. The situation is different for FRDM. Here the neutron
separation energy drops abruptly just before $N=90$ and even becomes
negative for nuclei like $^{133}$Pd, $^{134}$Ag, and $^{137}$Cd. This
region is reached when the r-process breaks out of the $N=82$ shell
closure. As the beta-decay half-lives of these nuclei are relatively
long (around 100~ms for the rates used in the present
calculation~\cite{Moeller.Pfeiffer.Kratz:2003}) matter accumulates
producing peaks at $^{138}$Sn in the hot r-process calculation and at
$^{134}$Cd and $^{140}$Sn in the cold r-process (see upper panels of
Fig.~\ref{fig:mm_abund}).  These nuclei represent a barrier to the
flow of neutron captures to more neutron-rich isotopes and
consequently the matter has to wait for their beta-decay before
heavier nuclei can be reached. This effect results in a broader second
minima for the FRDM average neutron separation energy in
Fig.~\ref{fig:mm_all} and in a longer duration for the r-process.

In the phase after freeze-out, the few available neutrons are not
equally captured in all regions. This leads to different position of
troughs and peaks depending on the mass model used. The most visible
feature in the final abundances is the trough in the ETFSI-Q
abundances before the third peak. This is also present in the
freeze-out abundances based on FRDM and HFB-17, but not on
Duflo-Zuker. During the decay to stability the trough is filled when
using the FRDM and HFB-17 mass models while for the ETFSI-Q becomes
even larger. The two neutron separation energies in
Fig.~\ref{fig:s2npath} show that for FRDM there is a drop of $S_{2n}$
for $N=122$ followed by a rise before the $N=126$ magic shell. This
leads to the formation of the trough at $A\sim 184$. For ETFSI-Q the
two neutron separation energies are almost constant for nuclei in the
region $A=180$--190 and $Z\sim 60$. This produces two troughs in the
freeze-out abundances at $A=180$ and $A=187$. The situation is more
complicated for HFB-17 (as the neutron separation energies show larger
fluctuations) with the net result of troughs at $A=180$ and $A=184$ in
the freeze-out abundances. \ During the decay to stability, in the
calculations based on FRDM, HFB-17, and Duflo-Zuker neutron captures
move matter from the region before the trough to higher mass numbers
and the trough is partially filled. In contrast, ETFSI-Q presents
higher $S_{2n}$ (and thus higher $(n,\gamma)$ rates) in the region
just before $N=126$. This leads to a shift of matter from the trough
towards the peak that produces an enhancement of the first one.

Notice that Duflo-Zuker abundances present stronger odd-even effects
than the abundances calculated with the other models as shown in
bottom panels of Fig.~\ref{fig:mm_abund}.  However, this effect is not
due to the mass model itself but to the computed neutron-capture
rates, which here are based on the simple approximation suggested in
Ref.~\cite{Michaud.Fowler:1970}. The importance of neutron-capture
rates on the final abundance will be discussed in detail in
Sects.~\ref{sec:back} and \ref{sec:ncap}.

\subsection{Decay to stability}
\label{sec:back}
As we have shown in the previous section, there are still reactions
occurring after freeze out that contribute to the redistribution of
matter and to the production of the final abundances. Classical
r-process studies (see for example Ref.~\cite{Kratz.Bitouzet.ea:1993},
but still amply used for r-process chronometers
studies~\cite{Schatz.Toenjes.ea:2002, Roederer.Kratz.ea:2009}) neglect
the neutron captures after freeze out and consider that beta-delayed
neutron emission is the only mechanism to redistribute and smooth the
abundances after freeze-out. Dynamical r-process
calculations~\cite{Freiburghaus.Rembges.ea:1999} have shown that
neutron captures during freeze-out can reduce odd-even effects but
also shift the peaks~\cite{Surman.Engel:2001} and produce the small
rare-earth peak around
$A\approx160$~\cite{Surman.Engel.ea:1997}. However, in some of these
studies (e.g., \cite{Freiburghaus.Rembges.ea:1999,Farouqi.etal:2010})
neutron captures were suppressed once the $Y_n/Y_{\text{seed}}$ ratio
was small and only beta-decays were considered. Here, we show that
even when $Y_n/Y_{\text{seed}} \approx 10^{-5}$, neutron captures are
still key for determining the final abundances.  For these conditions
the neutron density is around $10^{18}$~cm$^{-3}$, leading to a
typical time between neutron captures of 100~ms, that is comparable
with the beta-decay half-lives.

\begin{figure*}[htb]
  \centering
  \includegraphics[width=0.5\linewidth]{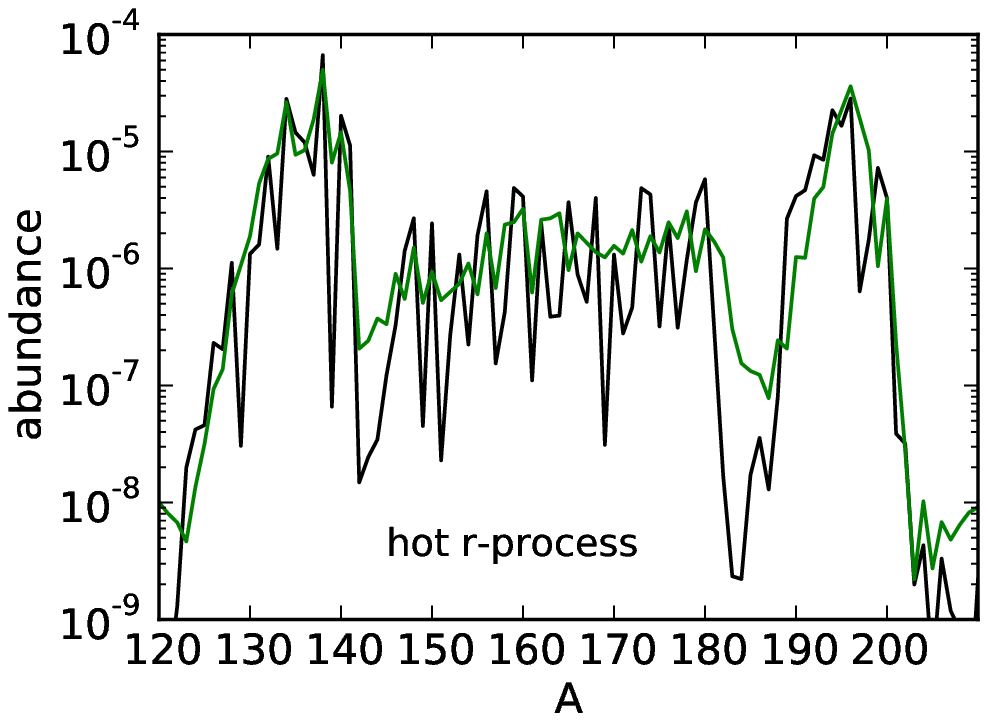}%
  \includegraphics[width=0.5\linewidth]{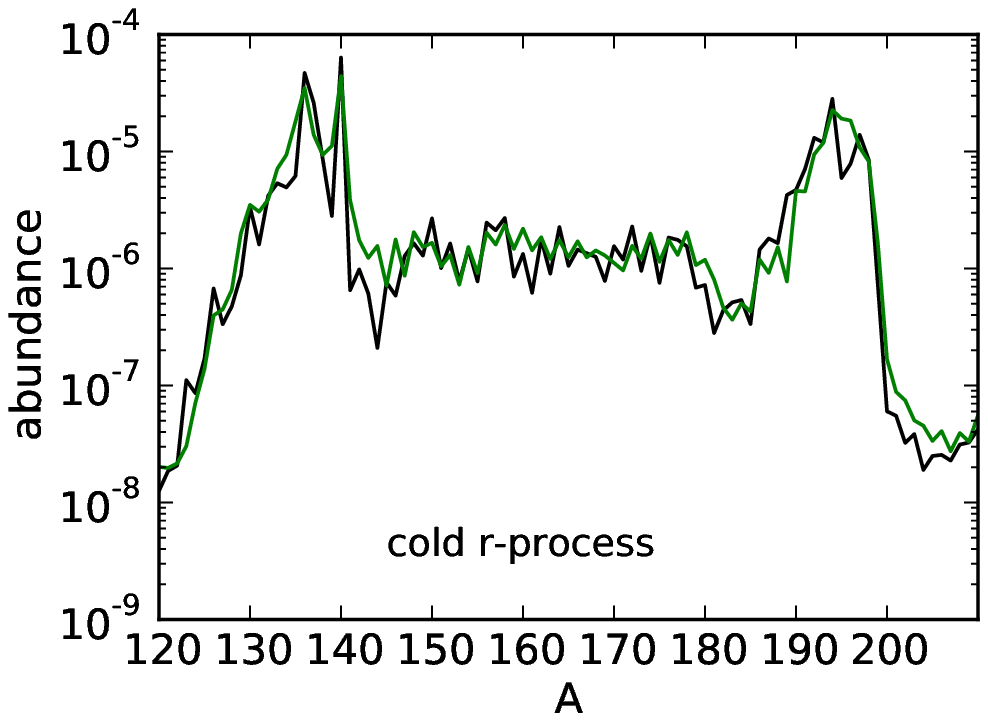}\\
  \includegraphics[width=0.5\linewidth]{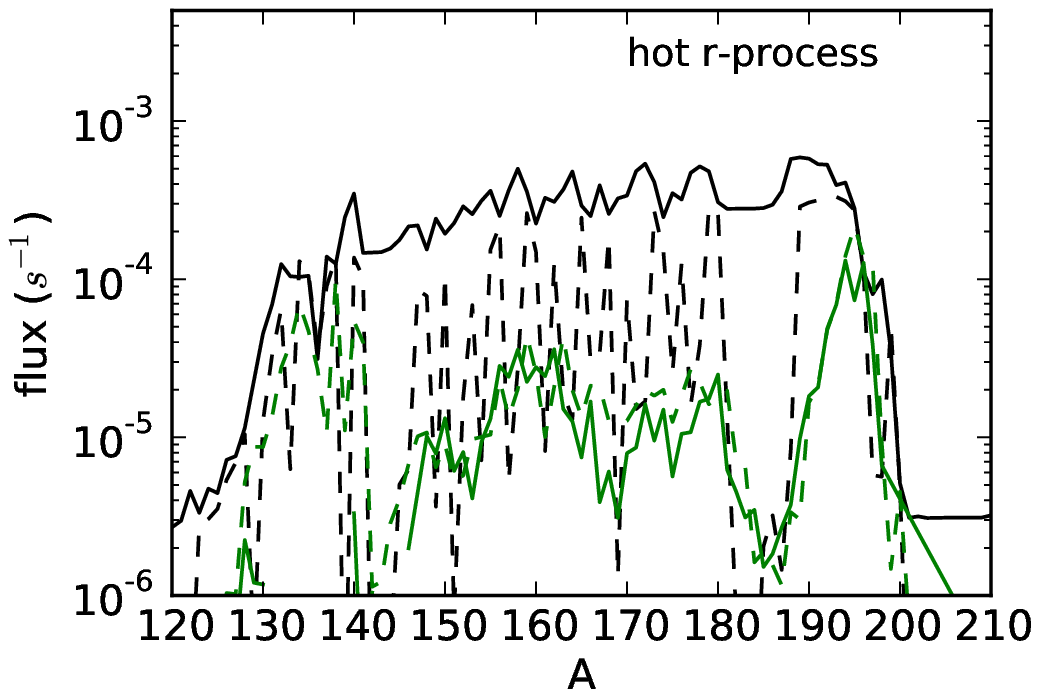}%
  \includegraphics[width=0.5\linewidth]{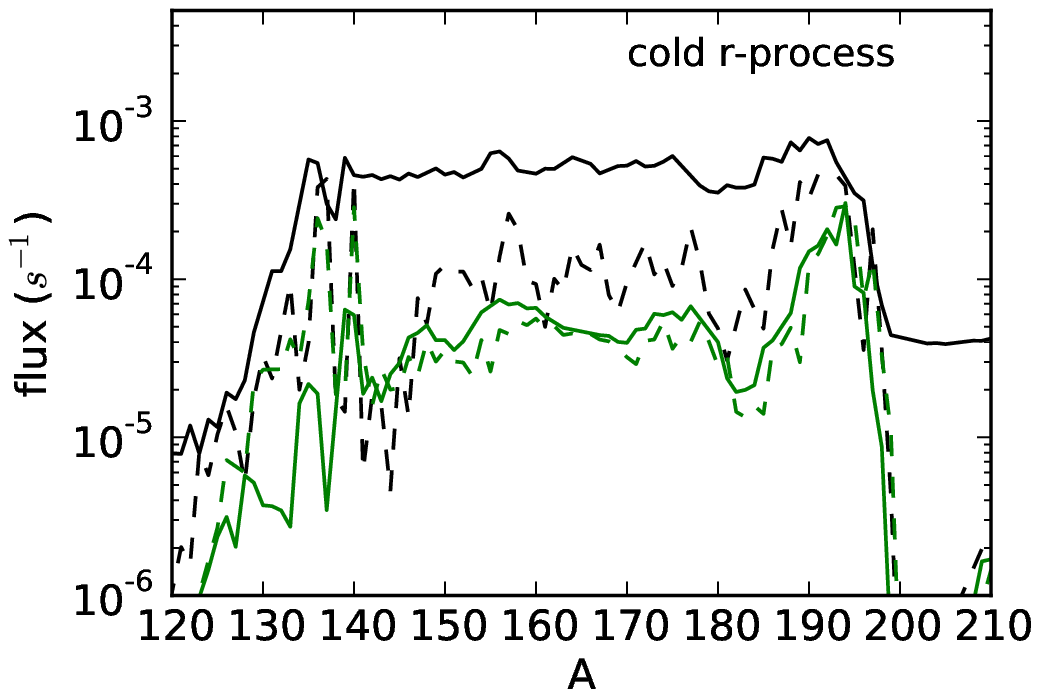}\\
  \includegraphics[width=0.5\linewidth]{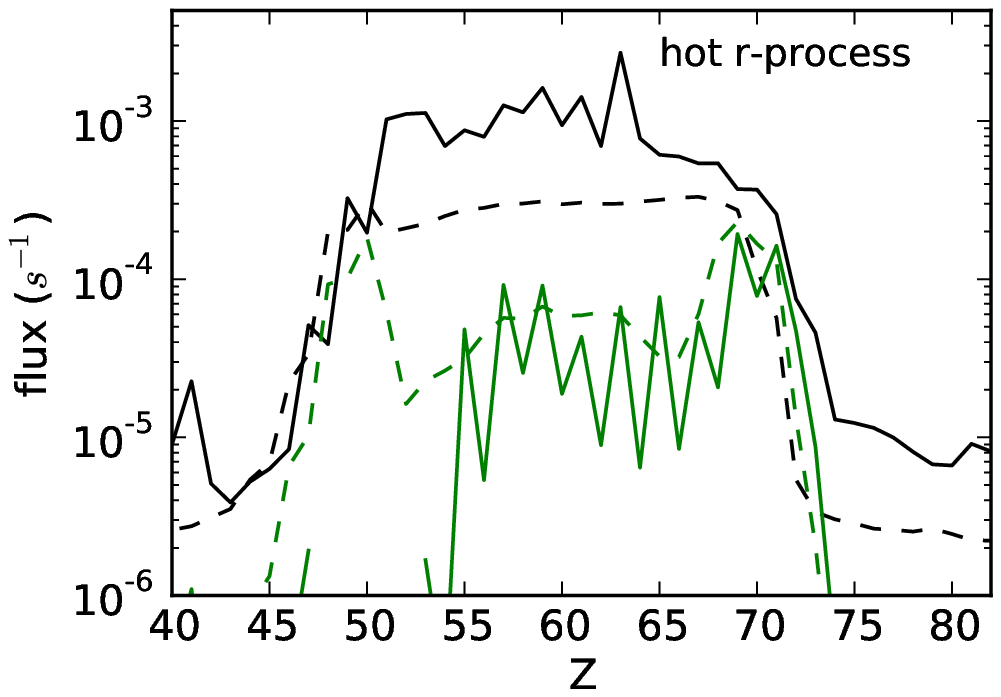}%
  \includegraphics[width=0.5\linewidth]{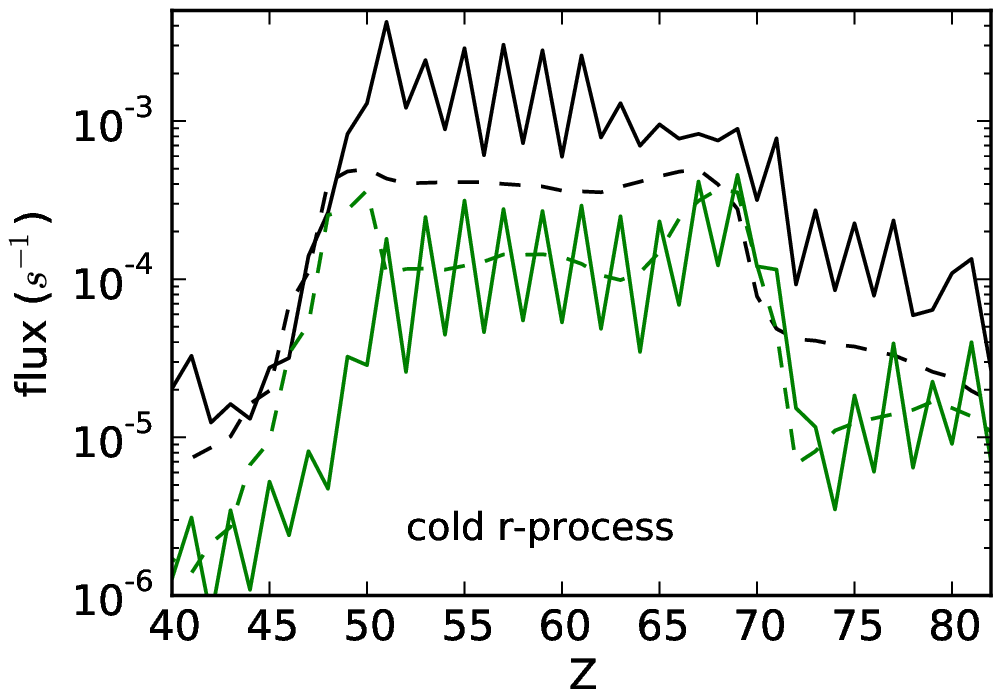}
  \caption{(Color online) Abundances and fluxes
    (Eqs.~\ref{eq:fbeta}--\ref{eq:fn}) vs. A and Z are shown for the
    hot (left column) and cold (right column) r-process. The black
    lines correspond to freeze out ($Y_n/Y_{\mathrm{seed}}=1$) and the
    green lines to the time when $\tau_{\beta}=\tau_{(n,\gamma)}$.
    The beta decay flux is represent by dashed lines and the net
    neutron capture flux by solid lines.}
  \label{fig:back_flux}
\end{figure*}

The processes producing the smoothing of the r-process abundances act
mainly between freeze out ($Y_n/Y_{\mathrm{seed}}=1$) and the moment
when $\tau_{(n,\gamma)}=\tau_{\beta}$.  The upper panels in
Fig.~\ref{fig:back_flux} show the abundances at freeze-out (black
lines) and when the average neutron capture (Eq.~\eqref{eq:tng}) and
beta-decay (Eq.~\eqref{eq:tbeta}) timescales are identical (green
lines) for hot (left column) and cold (right column) r-process.  The
abundances at freeze-out present large fluctuations (particularly in
the hot r-process) which have almost completely disappeared at the
later time as indicated by the green line.  There is still some
redistribution of matter at even later times that leads to the final
abundances shown in Fig.~\ref{fig:mm_abund} and to the formation of
the rare-earth peak around $A\approx160$.

The competition between beta decay (with delayed-neutron emission) and
neutron capture when matter decays to stability is important to
understand the differences in the final abundances between the hot and
cold r-process.  Let us consider a nucleus with charge $Z$ and mass
number $A$. Its abundance can change by beta-decay, neutron capture,
and photodissociation.  The competition among these processes can be
quantified by the beta decay flux:
\begin{equation}
  \label{eq:fbeta}
  F_{\beta}(Z,A)  =  \lambda_{\beta}(Z,A) Y(Z,A) \, ,
\end{equation}
and by the net neutron capture flux:
\begin{equation}
  \label{eq:fn}
  F_n(Z,A)  =   Y(Z,A) N_n \langle \sigma v \rangle_{Z,A} -
  \lambda_\gamma (Z,A+1) Y(Z,A+1).
\end{equation}

In order to visualize these quantities, it is convenient to define the
fluxes for an isotopic chain, $F_\beta(Z) =\sum_A F_\beta(Z,A)$ and
$F_n(Z)=\sum_A F_n(Z,A)$, and the fluxes for an isobaric chain,
$F_\beta(A)=\sum_Z F_\beta(Z,A)$ and $F_n(A)=\sum_Z F_n(Z,A)$.  As
discussed in the Appendix, we expect that for the hot r-process a
beta-flow equilibrium is achieved \cite{Kratz.Bitouzet.ea:1993}. This
implies that the $F_\beta(Z)$ reaches a constant value independent of
$Z$. In the cold r-process one expects that both $F_\beta(Z)$ and
$F_n(A)$ become constant. This is confirmed in
Fig.~\ref{fig:back_flux} that shows the net neutron capture and
beta-decay fluxes versus mass number (middle panels) and versus proton
number (bottom panels). Notice, that in the
s-process~\cite{Kaeppeler:1999} $F_n(A)$ is also constant and there
are strong odd-even effects in the abundances. This is due to the
large odd-even effects present in the neutron capture rates and the
fact that beta-decay can be assumed instantaneous compared to
s-process timescales. In contrast, in our case these strong odd-even
effects are not present because beta-decay rates become similar to
neutron-capture rates as matter decays to stability.

The fluxes $F_n(A)$ and $F_{\beta}(A)$ present several features when
$\tau_{\beta}=\tau_{(n,\gamma)}$ that can explain how matter is
redistributed.  In the regions where beta-decay dominates over neutron
captures, nuclei will beta decay without substantially changing the
mass number, e.g., see Fig.~\ref{fig:back_flux} for $A>195$.  On the
other hand, nuclei in regions where neutron capture dominates over
beta-decay will predominantly capture neutrons and consequently the
abundances will shift to higher mass numbers, as shown in
Fig.~\ref{fig:back_flux} for $A=182$--195.  The formation of the
rare-earth peak (not yet present in the abundances shown in upper
panels of Fig.~\ref{fig:back_flux}) is also due to neutron capture as
matter decays to stability.  In the region $A=162$--168 the beta-decay
and neutron-capture fluxes are very similar, while in the region $A<
162$ the latter dominates (Fig.~\ref{fig:back_flux}).  This produces a
net movement of matter from nuclei with $A<162$ to nuclei with $A
\approx 162$ that will result in the formation of the rare-earth peak.
The formation of this peak has to wait until the moment when the beta
decay fluxes become larger than the neutron capture fluxes in that
region. The reason why the neutron capture fluxes can become locally
smaller for these nuclei (when compared with slightly heavier or
lighter nuclei) is the presence of a deformed sub-shell closure around
$A \approx 162$ (see Fig.~\ref{fig:s2npath}) that results in a sudden
drop of neutron-capture rates.

In the hot r-process, the fluxes (Fig.~\ref{fig:back_flux}, right
column) present more fluctuations due to photodissociation
reactions. These reactions can result in negative net neutron capture
fluxes in regions where the photodissociation is still important.
This leads to an extra supply of neutrons that can be important at
later times. Negative net neutron capture fluxes appear in the hot
r-process for $A\lesssim145$ and $Z\lesssim 53$ once
$\tau_{\beta}=\tau_{(n,\gamma)}$. They are not shown in
Fig.~\ref{fig:back_flux} as we use a logarithmic scale.

\subsubsection{The role of neutron capture}
\label{sec:ncap}

\begin{figure}[htb]
  \centering
  \includegraphics[width=\linewidth]{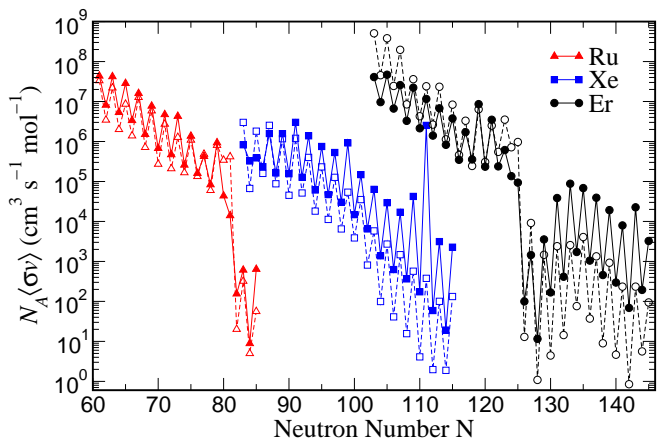}
  \caption{(Color online) Neutron-capture rates as computed in
    Ref.~\cite{Rauscher.Thielemann:2000} (solid lines and filled
    symbols) compared to rates based on the analytical approximation
    of Ref.~\cite{Michaud.Fowler:1970} (dashed lines and empty
    symbols). The rates are shown for 3 different isotopic chains in
    the region of relevance for r-process
    nucleosynthesis.\label{fig:compncap}}
\end{figure}

In order to explore the impact of neutron capture we compare two
different sets of rates based both on the FRDM mass model. The first
set corresponds to statistical model calculations of
Ref.~\cite{Rauscher.Thielemann:2000} and it was used in previous
sections. The second set is computed with the analytic approximation
suggested in Ref.~\cite{Michaud.Fowler:1970}. These two sets of
neutron-capture rates are compared in Fig.~\ref{fig:compncap} for
three different isotopic chains in regions relevant for the
r-process. The Ru isotopes are populated in the region $N\lesssim 82$,
Xe isotopes for $N\sim 100$, and Er isotopes for $N\lesssim 126$. The
largest differences between both sets of neutron-capture rates occur
for nuclei just before neutron shell closures. For these nuclei the
level density around neutron separation energy becomes rather low and
consequently the rates are very sensitive to the treatment of
parity~\cite{Loens.Langanke.ea:2008} and to the dipole strength
distribution~\cite{Litvinova.Loens.ea:2009}. In addition, the
statistical model may not be applicable for some of these nuclei at
r-process temperatures (see Fig.~7 of
Ref.~\cite{Rauscher.Thielemann.Kratz:1997}) and direct capture should
be included~\cite{mathews.mengoni.ea:1983,Goriely:1997}.

\begin{figure*}[htb]
  \centering
  \includegraphics[width=0.5\linewidth]{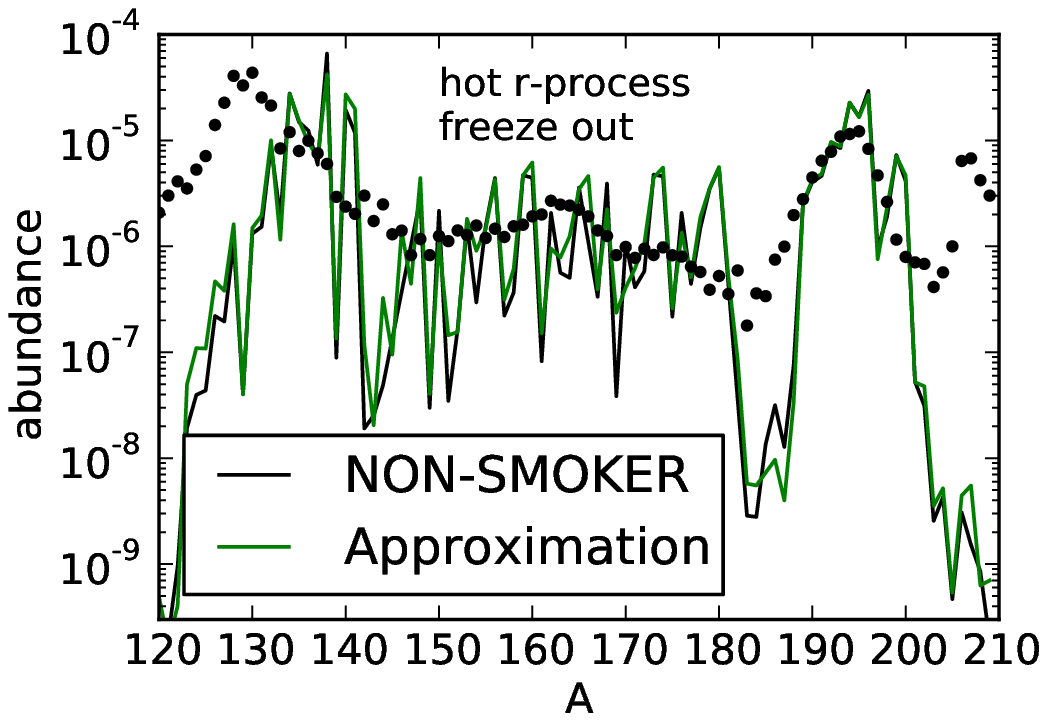}%
  \includegraphics[width=0.5\linewidth]{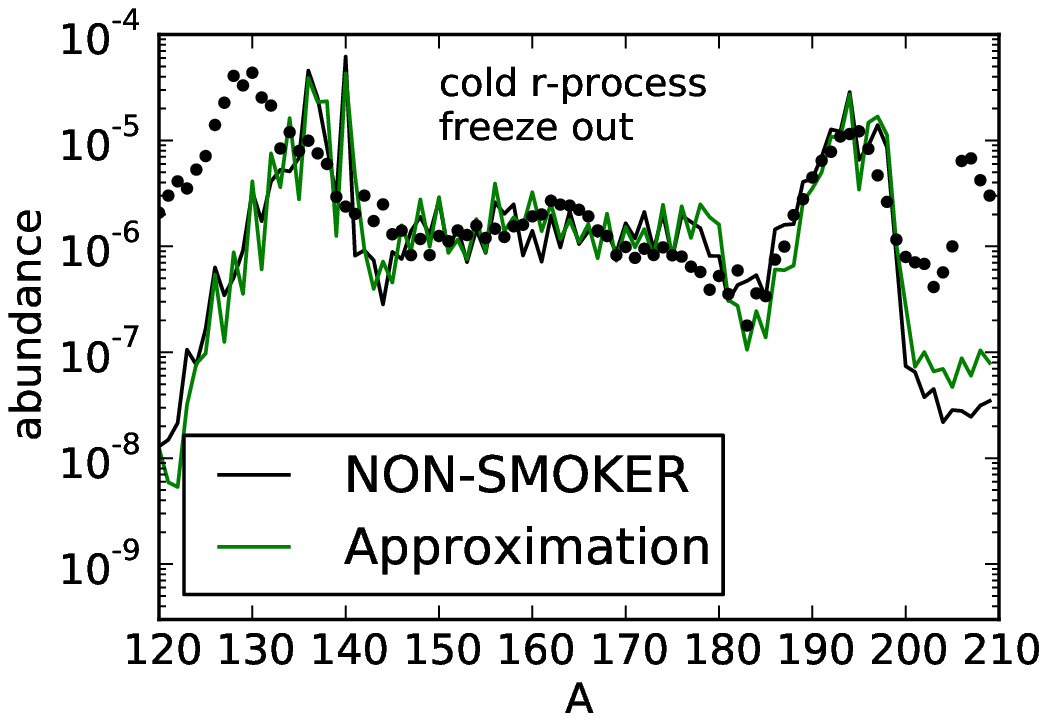}\\
  \includegraphics[width=0.5\linewidth]{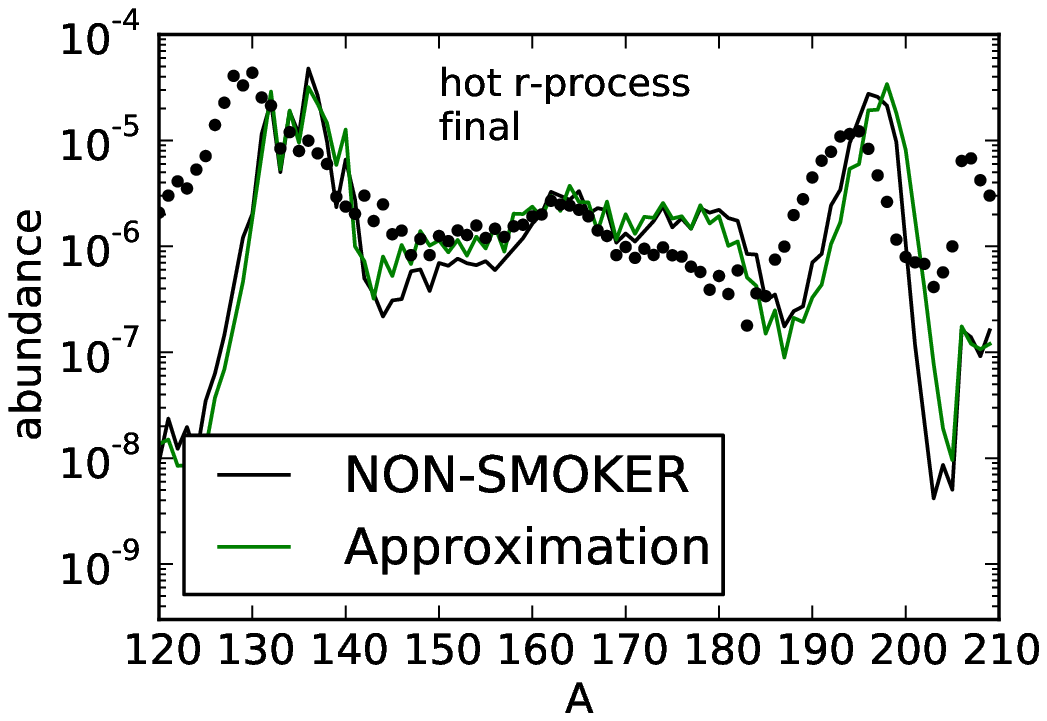}%
  \includegraphics[width=0.5\linewidth]{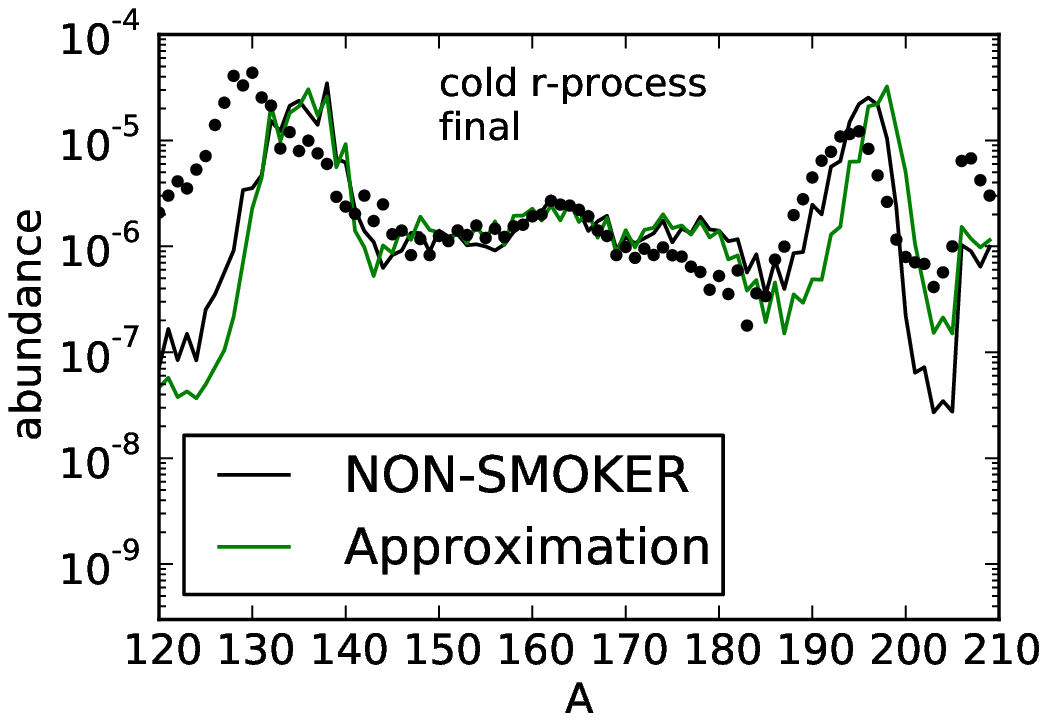}
  \caption{(Color online) Abundances at freeze-out (upper panels) and
    after decay to stability (bottom panels) for the hot (left panels)
    and cold (right panels) r-process. The abundances are obtained
    using two different sets of neutron-capture rates both based in
    the same nuclear mass model (FRDM~\cite{Moeller.Nix.ea:1995}). The
    first set of neutron-capture rates, labeled NON-SMOKER (black
    lines), corresponds to the calculations of
    Ref.~\cite{Rauscher.Thielemann:2000}. The second set, labeled as
    ``Approximation'' (green lines), has been obtained using the
    analytical approximation derived in
    Ref.~\cite{Michaud.Fowler:1970}.}
 \label{fig:back_approxabund}
\end{figure*}

The abundances based on the two sets of neutron-capture rates are
shown in Fig.~\ref{fig:back_approxabund} for the hot and cold
r-process. At freeze-out both sets of neutron-capture rates leads to
very similar abundances. In the hot r-process, the abundances are
independent of the neutron-capture rates since the evolution proceeds
under \nggn\ equilibrium (see Appendix). The cold r-process is
characterized by the competition between neutron capture and beta
decay, however only small changes are present in the freeze-out
abundances when the neutron-capture rates are varied. Notice that the
position and the height of the peaks are the same.  In contrast, the
final abundances exhibit significant differences: The third r-process
peak is more shifted towards higher mass number and the abundances
between peaks are less smooth for the calculations based on the
approximate neutron-capture rates.

The competition of the nuclei to capture the few neutrons available
can be quantified by the net probability of a nucleus for neutron
capture, that we define as
\begin{equation}
    P_{n,\gamma}(Z,A)  =  \frac{F_n(Z,A)}{\sum_{Z,A} F_n(Z,A)}\, ,
  \label{eq:probncap}
\end{equation}
using net neutron capture flux, $F_n(Z,A)$, introduced in
Eq.~(\ref{eq:fn}). The denominator in this expression represents the
change of neutron abundance due to neutron captures and
photodissociations and it is always positive since the neutron
abundance continuously decreases during an r-process calculation. The
numerator is positive for the majority of nuclei, although can become
negative for some of them. The change in abundances after freeze-out
is thus more pronounced in regions with larger $P_{n,\gamma}(Z,A)$.

\begin{figure*}[htb]
  \centering
  \includegraphics[width=0.5\linewidth]{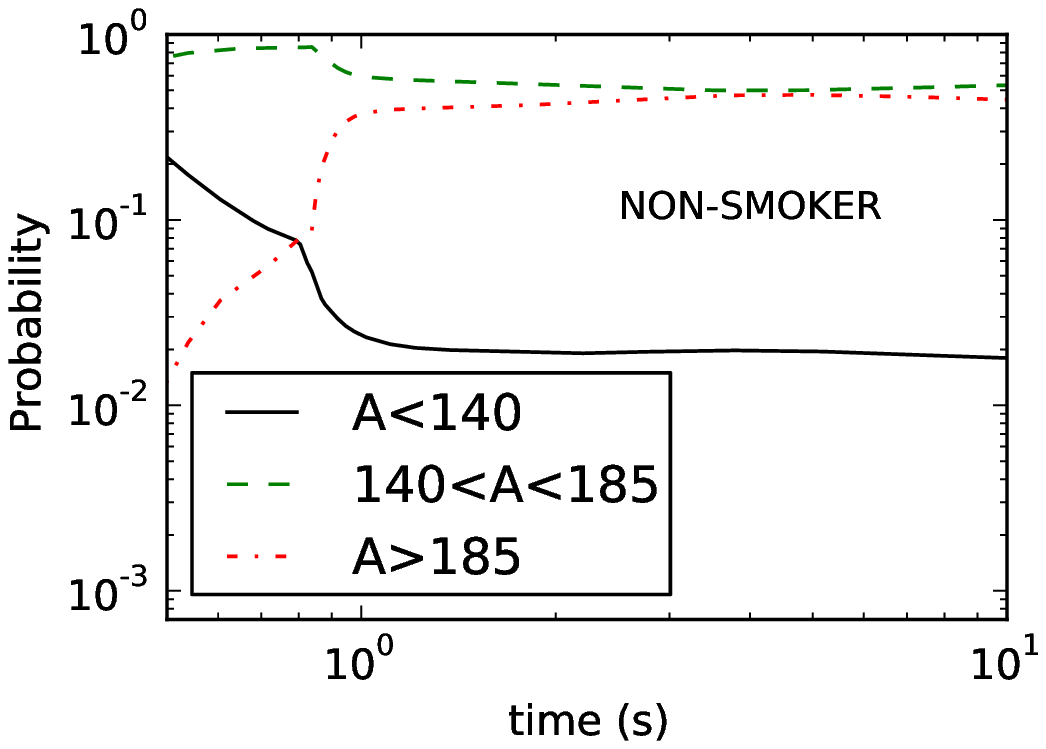}%
  \includegraphics[width=0.5\linewidth]{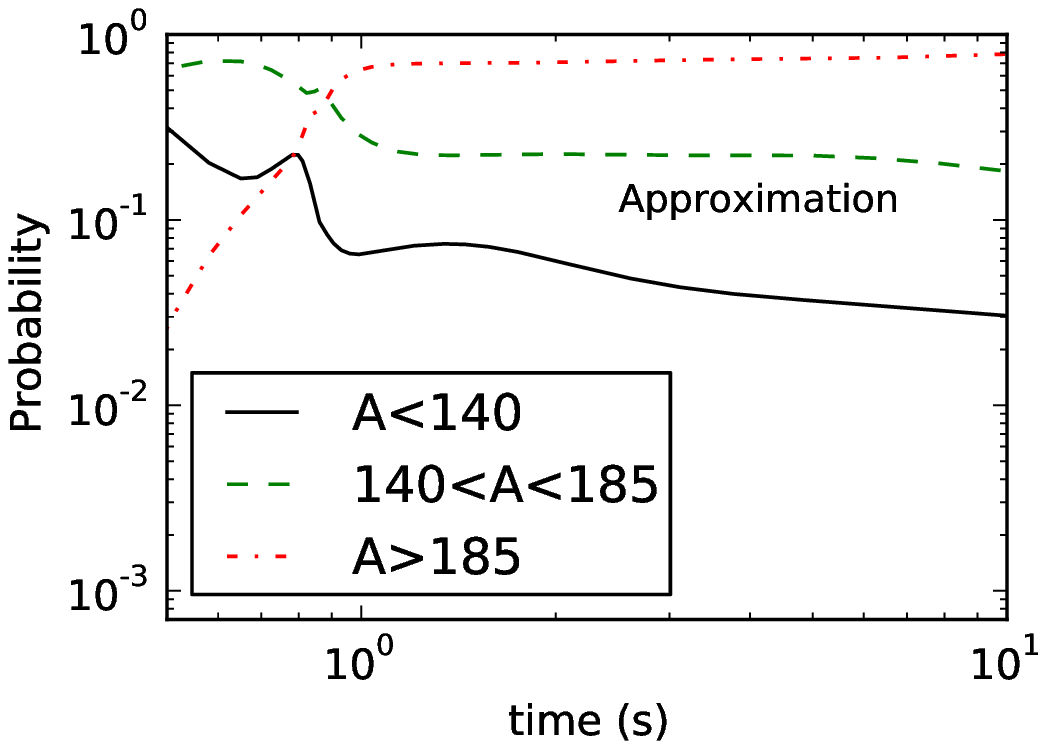}
  \caption{(Color online) Probability of neutron capture
    (Eq.~(\ref{eq:probncap})) vs. time for different mass number
    intervals based on NON-SMOKER rates (left panel) and on the
    approximated rates (right panel).}
  \label{fig:back_approxPn_t}
\end{figure*}

Figure~\ref{fig:back_approxPn_t} shows the neutron-capture probability
for different regions: around the second r-process peak ($A<140$),
between peaks ($140\leq A < 185$), and around the third peak
($A\ge185$) for the hot r-process. The results are qualitatively the
same for the cold r-process. As expected from the agreement between
the freeze-out abundances, the initial evolution of the probabilities
is rather similar and follows the build up of increasingly heavier
nuclei during the r-process. At early times most of the captures takes
place in the region around and below the second r-process peak. Later
as nuclei in the region between peaks are produced, the neutrons are
mainly captured in this region.  Just before freeze-out ($t \approx
1$~s), there is an increase in the capture probability around the
third peak.  The evolution after freeze-out strongly depends on the
neutron captures. In the calculation with NON-SMOKER rates
(Fig.~\ref{fig:back_approxPn_t}, left panel), the neutron-capture
probability is similar for the regions around the third peak and
between peaks, with the latter dominating slightly. This is in
contrast to the neutron-capture probability based on the approximate
rates (Fig.~\ref{fig:back_approxPn_t}, right panel) which is clearly
higher in the region of the third peak than in the region between
peaks. Therefore, neutrons are mainly captured around the third
peak. This leads to the shift of the peak towards even higher $A$ than
with the NON-SMOKER rates and leaves pronounced odd-even features
between peaks. Consequently, when the approximate rates are used,
beta-delayed neutron emission is the only mechanism to smooth the
abundances between $A=130$ and $A=195$ after freeze out. While using
the NON-SMOKER rates, both (neutron capture and beta-delayed neutron
emission) contribute to produce a smother r-process distribution.

Our results demonstrate that neutron-capture rates are key for the
determination of the final r-process abundances, specially around the
third peak and between peaks where a robust pattern is found in old
metal-poor halo stars and solar system (see e.g.,
Ref.~\cite{Sneden.etal:2008}). Similar conclusions were reached in
Ref.~\cite{Rauscher:2005} where individual neutron-capture rates were
changed by an arbitrary factor. However, our results illustrate more
clearly the non-local character of the competition for the few
available neutrons which can produce global changes in the r-process
abundances. In addition to the variation of the third-peak position,
we find also substantial changes in the abundances around $A\sim 150$,
even if both regions are not directly connected by any
reaction. Similar global changes were obtained in
Refs.~\cite{Surman.Beun.ea:2009,Beun.Blackmon.ea:2009,Farouqi.etal:2010}
where the sensitivity of the r-process abundances to changes of the
neutron-capture rates around $A=130$ was studied.

\subsubsection{Beta-delayed neutron emission}
\label{sec:betan}

\begin{figure*}
  \centering
  \includegraphics[width=0.5\linewidth,angle=0]{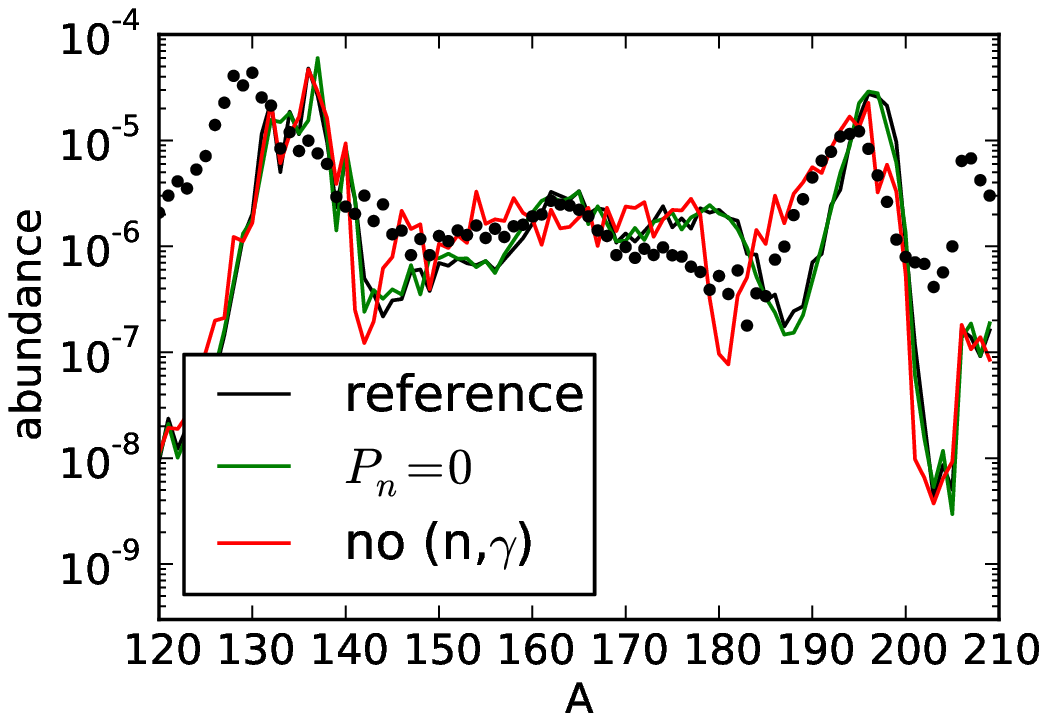}%
  \includegraphics[width=0.5\linewidth,angle=0]{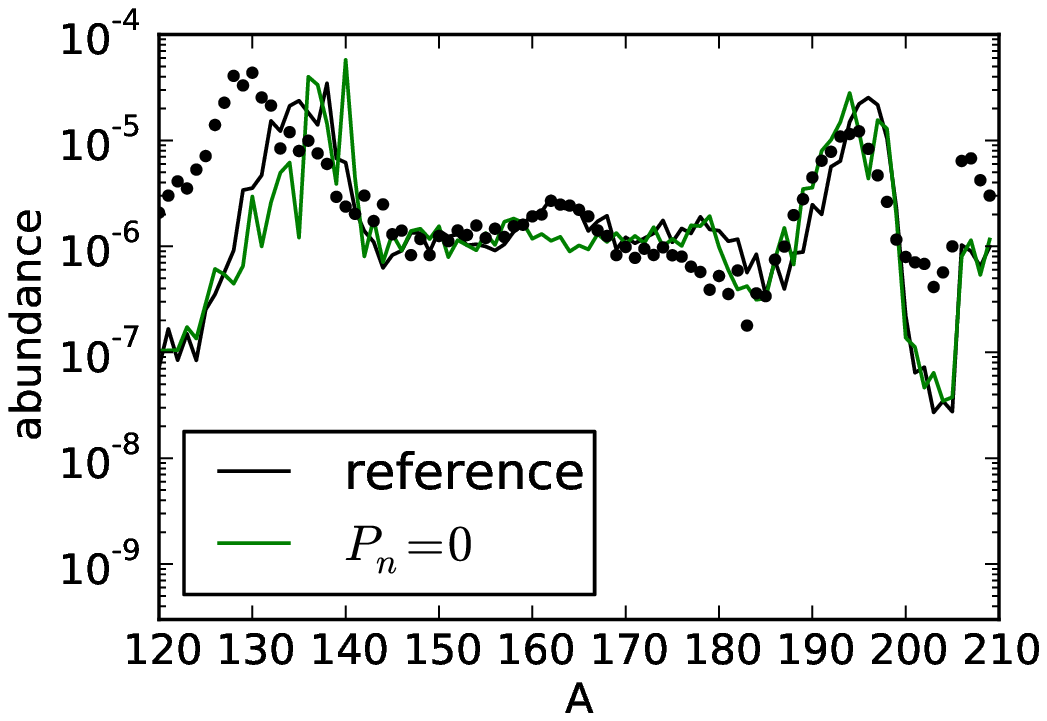}\\
  \includegraphics[width=0.5\linewidth,angle=0]{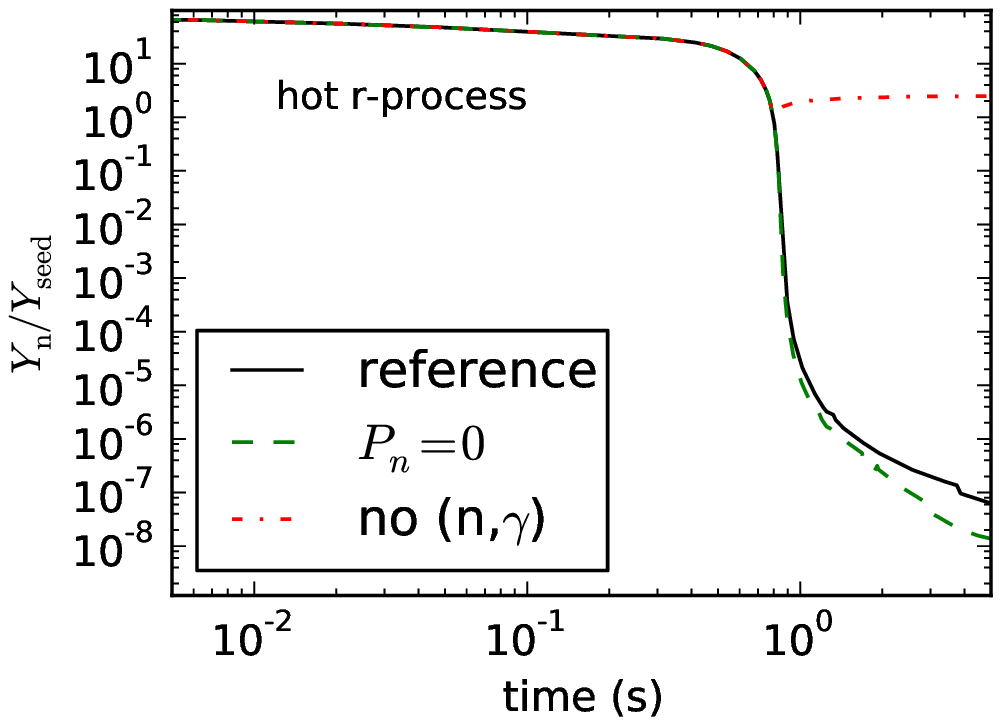}%
  \includegraphics[width=0.5\linewidth,angle=0]{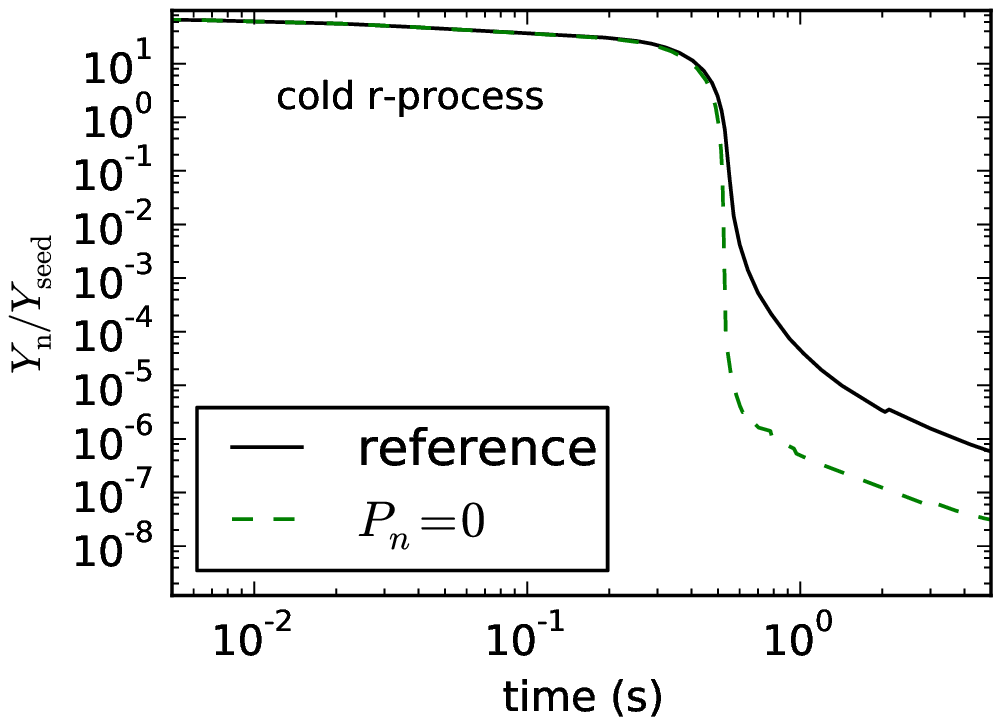}
  \caption{(Color online) Effect of the beta-delayed neutron emission
    in the hot (left column) and cold (right column) r-process, using
    neutron captures and beta decays based on the FRDM mass model.
    The black lines are for the reference case calculated with the
    standard nuclear input: neutrons are emitted with given
    probability ($P_n$) after beta decay. The green lines are for the
    case $P_n=0$, i.e. A is conserved during beta decay. For the red
    dash-dotted line the neutron captures and the photodissociation
    reactions are suppressed after freeze-out.}
  \label{fig:back_betan}
\end{figure*}

After we have shown the importance of beta decay when matter decays to
stability, it is worth to analyze the effect of beta-delayed neutron
emission. In the classical r-process, where no neutron captures are
considered after freeze out, beta-delayed neutron emission is the only
way to redistribute matter and get smooth final abundances.  In
Fig.~\ref{fig:back_betan} we explore the effect of beta-delayed
neutron emission in the hot (left column) and cold (right column)
r-process. The black lines correspond to the standard network
calculations with beta-delayed neutron emission included and the green
lines to calculations where beta-decay takes place without emitting
neutrons, i.e.\ $A$ is conserved.  The effect of beta-delayed neutron
emission and its subsequent capture was also investigated for
different long-time dynamical evolutions in~\cite{Wanajo.Itoh.ea:2002,
  Farouqi.etal:2010}.

After freeze-out, more neutrons are present when beta-delayed neutron
emission is considered and thus the neutron-to-seed ratio is higher as
shown in the bottom panels of Fig.~\ref{fig:back_betan}. However, in
our hot r-process such differences are too small to have an impact on
the final abundances. In this case, photodissociation produces also
neutrons in addition to beta decay. Our results seem to contradict
classical r-process calculations, where the redistribution of matter
is due only to beta-delayed neutron emission. One should notice that
our freeze-out path is in agreement with classical r-process
calculations (see e.g.~\cite{Kratz.Bitouzet.ea:1993}). Moreover, if we
neglect all neutron captures after freeze-out (as it is done in the
classical r-process) and only consider beta-delayed neutron emission,
we obtain the abundances shown by the red line in
Fig.~\ref{fig:back_betan}. This calculation reproduces the position
and width of the third peak but present larger oscillations in the
final abundances. Similar results were also found in the classical
r-process calculations of Ref.~\cite{Kratz.Bitouzet.ea:1993}
demonstrating that beta-delayed neutron emission cannot completely
remove the fluctuations in the freeze-out abundances. Once neutron
captures are considered the abundance distribution becomes smoother
like in the solar system and the rare-earth peak forms (black line in
the upper panels of Fig.~\ref{fig:back_betan}). However, the third
peak becomes narrower and shifts to mass number values larger than
$A\sim195$.

In our cold r-process, the beta-delayed neutron emission has more
impact on the final abundances (Fig.~\ref{fig:back_betan}).  Since
photodissociation is negligible, the r-process path can move farther
away from stability reaching nuclei with higher probability of
emitting neutrons after beta-decay. This leads to significant
differences in the neutron-to-seed ratio shown in the bottom panel of
Fig.~\ref{fig:back_betan}. The freeze-out is more instantaneous
without beta-delayed neutron emission and this has two main effects:
the third peak is less shifted and the rare earth peak is not
produced.  Therefore, we can conclude that beta-delayed neutron
emission is very important to supply the neutrons that through several
captures will determine the final abundances. Moreover, the freeze-out
cannot be totally instantaneous because neutron capture are required
to form the rare earth peak at late times.

\subsubsection{Non-instantaneous freeze-out}
\label{sec:freezeout}
Finally, we want to discuss a general feature present in the evolution
of the neutron-to-seed ratio. After an initial slow decrease, the
neutron-to-seed reaches a phase of fast decline once its value becomes
around one. However, this fast decline, that will correspond to an
instantaneous freeze-out as assumed in classical r-process
calculations, is always interrupted and the neutron-to-seed ratio
follows a more moderate decrease afterward. This is a generic feature
found in all dynamical calculations~(see
e.g. Refs.~\cite{Meyer.Mathews.ea:1992,Howard.Goriely.ea:1993,%
  Woosley.Wilson.ea:1994}) and indicates that the freeze-out effects
discussed here will always be important. We found that the sudden
change in the evolution of the neutron-to-seed ratio, mathematically
corresponding to the appearance of an inflexion point, occurs when the
average neutron-capture rate (Eq.~\eqref{eq:tng}) and the
average beta-decay rate (Eq.~\eqref{eq:tbeta}) are identical,
i.e. $\tau_{(n,\gamma)}=\tau_{\beta}$. Interestingly, this happens
even in the calculations where beta-delayed neutron emission is
artificially switched off, see Fig.~\ref{fig:back_betan}. This phase
of moderate decline of the neutron-to-seed ratio is the so-called
s-process phase of the r-process~\cite{Meyer:1993b} in which
beta-decay dominates over neutron-capture. During this phase, the
rare-earth peak is formed in our calculations based on the FRDM mass
model. However, in these calculations the third r-process peak is
shifted to mass number values larger than $A\sim 195$.

\section{Summary and conclusions}
\label{sec:conclusions}

We have studied the impact of the long-time dynamical evolution and
nuclear physics input on the r-process abundances. Our calculations
are based on hydrodynamical trajectories from core-collapse supernova
simulations of Ref.~\cite{arcones.janka.scheck:2007} with the entropy
increased by a factor two (i.e., density decreased by a factor two) in
order to produce the third r-process peak.  We have chosen two
different evolutions to cover the two possible physical conditions at
which the r-process occurs in high entropy ejecta.  These evolutions
are identical during the seed formation phase and differ only after
the temperature becomes $\lesssim 3$~GK and the nucleosynthesis flow
is dominated by neutron captures, i.e. the r-process phase. This
guarantees that changes in the resulting abundances are due to the
modification of the long-term evolution and/or the nuclear physics
input and not to changes in the initial conditions. The long-time
evolution is varied assuming that the reverse shock is at different
temperatures, which is justified based on two-dimensional
simulations~\cite{Arcones.Janka:2010}. The two typical long-time
evolutions are:

\begin{itemize}
\item \emph{hot r-process} that occurs when the reverse shock is at
  high enough temperatures ($T\gtrsim 0.5$~GK) to reach an \nggn\
  equilibrium that lasts until neutrons are exhausted.
\item \emph{cold r-process} (with the reverse shock at low temperatures) that
  takes place under a competition between neutron capture and beta
  decay.
\end{itemize}

The main difference between these evolutions is that in the cold
r-process the photodissociation is
negligible~\cite{Wanajo:2007}. Therefore, the r-process path can move
farther away from stability reaching nuclei with shorter beta-decay
half-lives and leading to a faster evolution and an earlier freeze out
than in the hot r-process.

We can distinguish two phases during the r-process that are
characterized by different nuclear physics processes. Before
freeze-out, $Y_n/Y_{\text{seed}}>1$, the most relevant nuclear physics
input depends on the type of r-process taking place, hot or cold
r-process:

\begin{itemize}
\item In the hot r-process the most relevant input are the nuclear
  masses as they determine the r-process path via the neutron
  separation energy.  A very good approximation to the freeze-out
  abundances of the hot r-process is obtained assuming that matter
  achieves steady beta-flow, i.e. $\lambda_\beta(Z) Y(Z) =
  \text{constant}$~(see Eq.~\eqref{eq:yequivz}). It is well
  known~\cite{Burbidge.Burbidge.ea:1957,Cameron:1957} that in this
  case the peaks in the r-process abundance distribution at $A=130$
  and $A=195$ are associated with long beta-decay half-lives at the
  magic neutron numbers $N=82$ and $N=126$ where the r-process path is
  closer to the stability.

  The abundances at freeze-out in the hot r-process are characterized
  by the presence of large troughs that occur in regions where the two
  neutron separation energy is almost constant or presents a saddle
  point. This typically occurs before (after) a magic neutron number
  where a transition from deformed (spherical) to spherical (deformed)
  shapes takes place. As experimental data for two neutron separation
  energies do not show this behavior~\cite{Audi.Bersillon.ea:2003},
  this may indicate a drawback in some of the mass models
  used. However, the solar abundances suggest a small trough in the
  region before the third peak, $A=180$--190, that could be related
  with a transition from deformed to spherical nuclei.

\item In the cold r-process the most relevant inputs are beta-decay
  and neutron-capture rates. The nuclei at the r-process path are
  those with similar neutron-capture and beta-decay rates for a given
  neutron density. Our results are the first to show that the
  abundances at freeze-out achieve steady flow for both beta decays
  and neutron captures, i.e.  $\lambda_\beta(Z) Y(Z)$ and
  $\langle\sigma v\rangle_A Y(A)$ are constant (see
  Eq.~\eqref{eq:ydotzdota}). This suggests that the cold r-process is
  more robust than the hot r-process, as the abundances fulfil
  additional constraints.

\end{itemize}

The sensitivity of the mass model has been investigated by
consistently using neutron separation energies and neutron capture
rates based on the mass models: FRDM, ETFSI-Q, HFB-17, and
Duflo-Zuker. Our results show peculiarities coming from each mass
model. In ETFSI-Q the quenching of the $N=82$ shell closure leads to a
slow down of the evolution and to a later freeze-out. Moreover, the
large values of $S_{2n}$ before $N=126$ in this mass model make the
trough in the freeze-out abundances for $A\approx 185$ bigger due to
neutron captures when matter decays to stability. Results based on
FRDM are clearly affected by the anomalous behaviour of $S_{2n}$
before $N=90$, which produces the accumulation of matter and thus the
formation of peaks around $A\approx 135$ even in the cold r-process.

In order to study the evolution after freeze-out we have used the
fluxes for neutron captures and beta decays. They help us to explain
the final features in the abundances, such as the exact position of
the r-process peaks and the formation of the rare earth peak. These
are our most significance outcomes:

\begin{itemize}
\item The abundances at freeze-out can be approximated assuming a
  steady beta-flow (hot r-process) or a steady flow of beta decays and
  neutron captures (cold r-process) for a given neutron density and
  temperature. However, the final abundances are determined by the
  evolution after freeze-out. In all cases considered, the final
  abundances are substantially different and smother than the
  freeze-out abundances. Most of the smoothing takes place just after
  freeze-out when the timescale for neutron captures is still shorter
  than the one for beta-decays, $\tau_{(n,\gamma)} < \tau_{\beta}$, see
  Eq.~\eqref{eq:timescales}. Hence, neutron captures play a dominant
  role in producing a smooth distribution. Nevertheless, the evolution
  once $\tau_{(n,\gamma)} > \tau_{\beta}$ is also important.  During
  this phase the decrease of the neutron-to-seed ratio is rather
  moderate and determined by the timescale at which matter beta decays
  to stability, even in calculations where beta-delayed neutron
  emission is artificially suppressed.

\item The impact of the neutron capture rates has been investigated by
  comparing results based on the same mass model (FRDM) but different
  sets of neutron-capture rates: A set is based in statistical model
  calculations with the code
  NON-SMOKER~\cite{Rauscher.Thielemann:2000} and the other in the
  analytical approximation derived in
  Ref.~\cite{Michaud.Fowler:1970}. We find that the abundances at
  freeze-out for hot and cold r-process are rather similar for both
  sets of neutron capture rates. However, after freeze-out we find
  that most of the neutrons are captured in the region between
  r-process peaks when using the NON-SMOKER rates. In contrast, with
  the approximated rates neutrons are captured more probably in the
  region around the third peak. The end result is a larger shift of
  the third peak and a less smooth abundance distribution between
  peaks with the approximated rates than with the NON-SMOKER
  rates. This emphasizes the important role of neutron captures after
  freeze-out.

\item The small rare-earth peak, observed around $A\sim160$ in the
  solar r-process distribution, must necessarily be formed after the
  freeze-out, since it is not present in any of the freeze-out
  abundances. Furthermore, it is also not present in the abundances
  when $\tau_{(n,\gamma)} \approx \tau_{\beta}$. We find, see also
  Refs.~\cite{Surman.Engel.ea:1997,Surman.Engel:2001}, that the
  rare-earth peak forms by neutron captures when matter decays to
  stability.

\item The main role of beta-delayed neutron emission is the supply of
  neutrons. In the hot r-process we find no difference in the
  abundances calculated with and without beta-delayed neutron
  emission. Since temperature is high, photodissociation prevents the
  path from reaching regions far from stability where the probability
  of emitting neutrons after beta decay is large. Furthermore,
  photodissociation reactions are also a source of neutrons. In
  contrast, in the cold r-process the neutron-to-seed ratio reaches
  significantly smaller values when beta-delayed neutron emission is
  suppressed. This leads to a reduction in the shift of the third peak
  after freeze out but also inhibits the formation of the rare earth
  peak. This confirms the argument given above that the peak forms by
  neutron captures.
\end{itemize}

Our results clearly rise the importance of future experiments to
measure nuclear masses, neutron capture rates and beta-decay
half-lives for nuclei far from stability. This will provide not only
direct input for network calculations, but also important constraints
for the theoretical nuclear models.  We have shown that the r-process
abundances are very sensitive to the set of neutron capture
rates as they determine the regions in which neutrons are capture
predominantly.  More experimental effort is necessary for an
improved determination of neutron capture cross sections. Since these
experiments are difficult, sensitivity studies to determine the most
relevant neutron capture rates will be necessary. Our results show a
strong interplay between the late-time evolution of the ejected matter
and the nuclear physics input. This could constrain the astrophysical
conditions once future radioactive experimental facilities deliver
high quality experimental data for r-process nuclei.


\appendix*
\section{Formalism}
\label{sec:formalism}

During the r-process, and assuming that charged particle reactions,
fission and alpha decays can be neglected, the evolution of the
abundances is mainly determined by neutron capture, photodissociation,
and beta-decays. This results in the following differential equation
that determines the change of the abundance of a nucleus with charge
$Z$ and mass number $A$:

\begin{widetext}
\begin{eqnarray}
  \label{eq:yza}
  \frac{d Y(Z,A)}{dt} & = & \rho N_A \langle \sigma v\rangle_{Z,A-1}
  Y_n Y(Z,A-1)+ 
  \lambda_{\gamma}(Z,A+1) Y(Z,A+1) \nonumber \\
  & & + \sum_{j=0}^J \lambda_{\beta jn}
  (Z-1,A+j) Y(Z-1,A+j) \nonumber \\
   & & - \left(\rho N_A \langle\sigma v\rangle_{Z,A} Y_n  +
    \lambda_{\gamma} (Z,A) + \sum_{j=0}^J \lambda_{\beta
      jn}(Z,A)\right) Y(Z,A)
\end{eqnarray}
\end{widetext}
where $Y_n$ is the neutron abundance, $\langle\sigma v(Z,A)\rangle$ is
the thermal averaged neutron-capture rate, and $\lambda_{\gamma}
(Z,A)$ the photodissociation rate for a nucleus $^AZ$, while
$\lambda_{\beta jn}(Z,A)$ is the $\beta^-$ decay rate of $^AZ$ with
emission of $j$ delayed neutrons (up to a maximum of $J$). The
photodissociation rate is related to the neutron capture rate by
detailed balance:

\begin{widetext}
\begin{equation}
  \label{eq:photonv}
  \lambda_\gamma(Z,A+1) =  \langle\sigma v\rangle_{Z,A}
  \left(\frac{m_u k T}{2\pi\hbar^2}\right)^{3/2} \frac{2
    G(Z,A)}{G(Z,A+1)}\left(\frac{A}{A+1}\right)^{3/2}
  \exp\left[-\frac{S_n(Z,A+1)}{kT}\right], 
\end{equation}
\end{widetext}
where $G$ is the particition function and $S_n = m_n +
M(Z,A-1)-M(Z,A)$ is the neutron separation energy with $m_n$ the
neutron mass and $M(Z,A)$ the mass of the nucleus.

If the assumption is made that the neutron abundance varies slowly
enough, it can be assumed that the neutron density, $N_n = Y_n \rho
N_A$, is constant over a timestep. In this case the network can be
divided into separate pieces for each isotopic chain and solve then
sequentially, beginning with the lowest
$Z$~\cite{Cowan.Cameron.Truran:1983}. However, this approximation
becomes numerically unstable when the neutron abundance becomes small,
$Y_n \lesssim 10^{-5}$. Consequently, it is better to include in the
set of differential equations the one determining the change of the
neutron abundance:

\begin{eqnarray}
  \label{eq:dyn}
  \frac{dY_n}{dt} = & - & \sum_{Z,A} \rho N_A \langle \sigma v
  \rangle_{Z,A} Y_n Y(Z,A) \nonumber\\ 
 & + & \sum_{Z,A} \lambda_{\gamma} (Z,A)
  Y(Z,A) \nonumber \\
  &+&\sum_{Z,A}\left(\sum_{j=1}^J j \lambda_{\beta jn}(Z,A)\right) Y(Z,A)
\end{eqnarray}

The system of differential equations defined by
Eq.~\eqref{eq:yza} and~\eqref{eq:dyn} allows for several
approximations that are valid in different physical regimes. A
commonly used assumption in classical r-process calculations is the
$(n,\gamma)\rightleftarrows(\gamma,n)$ equilibrium. This approximation
is valid whenever the neutron density ($N_n \gtrsim
10^{20}$~cm$^{-3}$) and temperature ($T\gtrsim
1$~GK)~\cite{Cameron.Cowan.Truran:1983} are large enough to warrant
that both the rate of neutron capture ($N_n \langle\sigma v\rangle$
and the photodissociation rate ($\lambda_\gamma$) are much larger than
the beta decay rate ($\lambda_\beta$) for all the nuclei participating
in the network. Under this conditions the evolution of the system is
mainly determined by the beta decay rates as the abundances along an
isotopic chain are inmediatly adjusted to an equilibrium between
neutron captures and photodissociations, i.e. $N_n Y(Z,A)
\langle\sigma v\rangle_{Z,A} = \lambda_\gamma(Z,A+1)
Y(Z,A+1)$. Combining these result with Eq.~\eqref{eq:photonv} one
obtains that the abundances in an isotopic chain are given by the
simple relation:

\begin{widetext}
\begin{equation}
  \label{eq:yequiv}
  \frac{Y(Z,A+1)}{Y(Z,A)} = N_n\left(\frac{2\pi\hbar^2}{m_ukT}\right)^{3/2}
  \left(\frac{A+1}{A}\right)^{3/2}\frac{G(Z,A+1)}{2G(Z,A)}
  \exp\left[\frac{S_n(Z,A+1)}{kT}\right].
\end{equation}
\end{widetext}
For each isotopic chain, the above equation defines a nucleus that has
the maximum abundance and which is normally known as waiting point
nucleus as the flow of neutron captures ``waits'' for this nucleus to
beta-decay. The set of waiting point nuclei constitutes the r-process
path. The maximum of the abundance distribution can be determined
setting the left-hand side of Eq.~(\ref{eq:yequiv}) to 1, which
results in a value of $S_n$ that is the same for all isotopic chains
for a given neutron density and temperature:

\begin{equation}
  \label{eq:sneq}
  S^0_n (\text{MeV}) = \frac{T_9}{5.04} \left(34.075 - \log N_n +
    \frac{3}{2} \log T_9 \right),
\end{equation}
where $T_9$ is the temperature in units of $10^9$~K and $N_n$ is the
neutron density in cm$^{-3}$. Equation~(\ref{eq:sneq}) implies that
the r-process proceeds along lines of constant neutron separation
energies towards heavy nuclei. For typical r-process conditions this
corresponds to $S^0_n \sim 2$--3~MeV. Due to pairing, the most
abundance isotopes have always an even neutron number. For this
reason, it may be more appropriate to characterize the most abundance
isotope in an isotopic chain as having a two-neutron separation energy
$S_{2n} = 2 S^0_n$~\cite{Goriely.Arnould:1996}. The two-neutron
separation energy is not a continuous function of neutron number but
shows large jumps particularly close to magic neutron numbers. For
this reason r-process nuclei near to magic numbers have neutron
separation energies much larger than the typical 2--3~MeV and the
r-process path moves closer to the stability (see
figure~\ref{fig:s2npath}). 

If $(n,\gamma) \rightleftarrows (\gamma,n)$ equilibrium is valid it is
sufficient to consider the time evolution of the total abundance of an
isotopic chain $Y(Z) = \sum_A Y(Z,A)$ as the abundances of different
isotopes are fully determined by Eq.~(\ref{eq:yequiv}). From
Eq.~(\ref{eq:yza}) we can determine the time evolution of $Y(Z)$
obtaining:

\begin{equation}
  \label{eq:ydotz}
  \frac{dY(Z)}{dt} = \lambda_\beta (Z-1) Y(Z-1) - \lambda_\beta (Z) Y(Z)
\end{equation}
where $\lambda_\beta (Z) = \sum_A \lambda_\beta (Z,A) Y(Z,A)/Y(Z)$. In
this case the r-process evolution is independent of the
neutron-capture rates, only beta-decays are necessary for
Eq.~(\ref{eq:ydotz}) and masses via $S_n$ in Eq.~(\ref{eq:yequiv}). If
the r-process proceeds in $(n,\gamma) \rightleftarrows (\gamma,n)$
equilibrium and its duration is larger than the beta decay lifetimes
of the nuclei present, Eq.~(\ref{eq:ydotz}) tries to reach an
equilibrium denoted as steady
$\beta$-flow~\cite{Kratz.Bitouzet.ea:1993} that satisfies for each $Z$
value:

\begin{equation}
  \label{eq:yequivz}
  \lambda_\beta (Z-1) Y(Z-1) = \lambda_\beta (Z) Y(Z)
\end{equation}
In this case the peaks at $A=130$ and 195 in the solar r-process
distribution can be attributed to the long $\beta$-decay lifetimes of the
waiting point nuclei with $N=82$ and 126, where the r-process path
gets closer to the stability (see figure~\ref{fig:s2npath}). This is
the case for the equilibrium calculations discussed in the text before
freeze-out of neutron captures. 

The r-process can also operate under such a low temperatures that the
photodissociation rates in Eq.~\eqref{eq:yza} can be neglected. Under
this conditions the r-process operates under a competition of neutron
captures and beta decays. If one neglects beta-delayed neutron
emission, Eq.~\eqref{eq:yza} can be reduced to two independent
equations that govern the evolution of the total abundance along an
isotopic chain, $Y(Z)$ and along an isobaric chain, $Y(A)=\sum_Z
Y(Z,A)$:

\begin{subequations}
  \label{eq:ydotzdota}
  \begin{equation}
    \label{eq:ydotz2}
    \frac{dY(Z)}{dt} = \lambda_\beta (Z-1) Y(Z-1) - \lambda_\beta (Z) Y(Z)
  \end{equation}
  \begin{equation}
    \label{eq:ydota}
    \frac{dY(A)}{dt} = N_n \langle\sigma v\rangle_{A-1} Y(A-1) - N_n
    \langle\sigma v\rangle_A Y(A)    
  \end{equation}
\end{subequations}
were $\langle\sigma v\rangle_A = \sum_Z \langle\sigma v\rangle_{Z,A}
Y(Z,A)/Y(A)$. If the r-process duration is longer than the beta-decay
and neutron capture lifetimes, Eq.~\eqref{eq:ydotzdota} reaches an
equilibrium that we will denote as steady flow that satisfies for each
$Z$ and $A$:

\begin{subequations}
  \label{eq:equivza}
  \begin{equation}
    \label{eq:yequivz2}
    \lambda_\beta (Z-1) Y(Z-1) = \lambda_\beta (Z) Y(Z)
  \end{equation}
  \begin{equation}
    \label{eq:yequiva}
    \langle\sigma v\rangle_{A-1} Y(A-1) = \langle\sigma v\rangle_A Y(A)
  \end{equation}
\end{subequations}
In addition, as the r-process occurs under a competition of
beta-decays and neutron captures one obtains that $N_n \langle \sigma
v\rangle_A Y(A) \approx \lambda_\beta(Z) Y(Z)$. As the abundances
along an isotopic and isobaric chain are dominated by a single nucleus
this condition determines also the nuclei that participate in the
r-process, i.e.\ the r-process path. Similarly to what happens in the
equilibrium case, the peaks in the abundance distribution correspond to
long beta decay lifetimes. However, in this case the peaks are in
addition associated with long neutron capture lifetimes.

\begin{acknowledgments}
  We thank K.~Langanke, H.~P.~Loens, F.~Montes, K.~Otsuki,
  I.~Petermann, and F.~K.~Thielemann for valuable discussions. We are
  grateful to D.~Mocelj for providing us with the first version of the
  network. This work was supported by the Deutsche
  Forschungsgemeinschaft through contract SFB 634 and by the ExtreMe
  Matter Institute (EMMI). A.~Arcones is supported by the Swiss
  National Science Foundation.
\end{acknowledgments}
%
%


\begin{thebibliography}{89}
\expandafter\ifx\csname natexlab\endcsname\relax\def\natexlab#1{#1}\fi
\expandafter\ifx\csname bibnamefont\endcsname\relax
  \def\bibnamefont#1{#1}\fi
\expandafter\ifx\csname bibfnamefont\endcsname\relax
  \def\bibfnamefont#1{#1}\fi
\expandafter\ifx\csname citenamefont\endcsname\relax
  \def\citenamefont#1{#1}\fi
\expandafter\ifx\csname url\endcsname\relax
  \def\url#1{\texttt{#1}}\fi
\expandafter\ifx\csname urlprefix\endcsname\relax\def\urlprefix{URL }\fi
\providecommand{\bibinfo}[2]{#2}
\providecommand{\eprint}[2][]{\url{#2}}

\bibitem[{\citenamefont{Arnould et~al.}(2007)\citenamefont{Arnould, Goriely,
  and Takahashi}}]{arnould.goriely.takahashi:2007}
\bibinfo{author}{\bibfnamefont{M.}~\bibnamefont{Arnould}},
  \bibinfo{author}{\bibfnamefont{S.}~\bibnamefont{Goriely}}, \bibnamefont{and}
  \bibinfo{author}{\bibfnamefont{K.}~\bibnamefont{Takahashi}},
  \bibinfo{journal}{Phys. Repts.} \textbf{\bibinfo{volume}{450}},
  \bibinfo{pages}{97} (\bibinfo{year}{2007}).

\bibitem[{\citenamefont{{Grawe} et~al.}(2007)\citenamefont{{Grawe}, {Langanke},
  and {Mart{\'{\i}}nez-Pinedo}}}]{Grawe.Langanke.Martinez-Pinedo:2007}
\bibinfo{author}{\bibfnamefont{H.}~\bibnamefont{{Grawe}}},
  \bibinfo{author}{\bibfnamefont{K.}~\bibnamefont{{Langanke}}},
  \bibnamefont{and}
  \bibinfo{author}{\bibfnamefont{G.}~\bibnamefont{{Mart{\'{\i}}nez-Pinedo}}},
  \bibinfo{journal}{Rep. Prog. Phys.} \textbf{\bibinfo{volume}{70}},
  \bibinfo{pages}{1525} (\bibinfo{year}{2007}).

\bibitem[{\citenamefont{Wanajo and Ishimaru}(2006)}]{Wanajo.Ishimaru:2006}
\bibinfo{author}{\bibfnamefont{S.}~\bibnamefont{Wanajo}} \bibnamefont{and}
  \bibinfo{author}{\bibfnamefont{Y.}~\bibnamefont{Ishimaru}},
  \bibinfo{journal}{Nucl. Phys. A} \textbf{\bibinfo{volume}{777}},
  \bibinfo{pages}{676} (\bibinfo{year}{2006}).

\bibitem[{\citenamefont{{Qian} and {Wasserburg}}(2007)}]{Qian.Wasserburg:2007}
\bibinfo{author}{\bibfnamefont{Y.}~\bibnamefont{{Qian}}} \bibnamefont{and}
  \bibinfo{author}{\bibfnamefont{G.~J.} \bibnamefont{{Wasserburg}}},
  \bibinfo{journal}{Phys. Repts.} \textbf{\bibinfo{volume}{442}},
  \bibinfo{pages}{237} (\bibinfo{year}{2007}).

\bibitem[{\citenamefont{{Duncan} et~al.}(1986)\citenamefont{{Duncan},
  {Shapiro}, and {Wasserman}}}]{duncan.shapiro.wasserman:1986}
\bibinfo{author}{\bibfnamefont{R.~C.} \bibnamefont{{Duncan}}},
  \bibinfo{author}{\bibfnamefont{S.~L.} \bibnamefont{{Shapiro}}},
  \bibnamefont{and}
  \bibinfo{author}{\bibfnamefont{I.}~\bibnamefont{{Wasserman}}},
  \bibinfo{journal}{Astrophys. J.} \textbf{\bibinfo{volume}{309}},
  \bibinfo{pages}{141} (\bibinfo{year}{1986}).

\bibitem[{\citenamefont{Burrows et~al.}(1995)\citenamefont{Burrows, Hayes, and
  Fryxell}}]{Burrows.Hayes.Fryxell:1995}
\bibinfo{author}{\bibfnamefont{A.}~\bibnamefont{Burrows}},
  \bibinfo{author}{\bibfnamefont{J.}~\bibnamefont{Hayes}}, \bibnamefont{and}
  \bibinfo{author}{\bibfnamefont{B.~A.} \bibnamefont{Fryxell}},
  \bibinfo{journal}{\apj} \textbf{\bibinfo{volume}{450}}, \bibinfo{pages}{830}
  (\bibinfo{year}{1995}).

\bibitem[{\citenamefont{Janka and M{\"u}ller}(1996)}]{Janka.Mueller:1996}
\bibinfo{author}{\bibfnamefont{H.-T.} \bibnamefont{Janka}} \bibnamefont{and}
  \bibinfo{author}{\bibfnamefont{E.}~\bibnamefont{M{\"u}ller}},
  \bibinfo{journal}{Astron. \& Astrophys.} \textbf{\bibinfo{volume}{306}},
  \bibinfo{pages}{167} (\bibinfo{year}{1996}).

\bibitem[{\citenamefont{Buras et~al.}(2006)\citenamefont{Buras, Rampp, Janka,
  and Kifonidis}}]{Buras.Rampp.ea:2006}
\bibinfo{author}{\bibfnamefont{R.}~\bibnamefont{Buras}},
  \bibinfo{author}{\bibfnamefont{M.}~\bibnamefont{Rampp}},
  \bibinfo{author}{\bibfnamefont{H.-T.} \bibnamefont{Janka}}, \bibnamefont{and}
  \bibinfo{author}{\bibfnamefont{K.}~\bibnamefont{Kifonidis}},
  \bibinfo{journal}{Astron. \& Astrophys.} \textbf{\bibinfo{volume}{447}},
  \bibinfo{pages}{1049} (\bibinfo{year}{2006}).

\bibitem[{\citenamefont{{Arcones} et~al.}(2007)\citenamefont{{Arcones},
  {Janka}, and {Scheck}}}]{arcones.janka.scheck:2007}
\bibinfo{author}{\bibfnamefont{A.}~\bibnamefont{{Arcones}}},
  \bibinfo{author}{\bibfnamefont{H.-T.} \bibnamefont{{Janka}}},
  \bibnamefont{and} \bibinfo{author}{\bibfnamefont{L.}~\bibnamefont{{Scheck}}},
  \bibinfo{journal}{Astron. \& Astrophys.} \textbf{\bibinfo{volume}{467}},
  \bibinfo{pages}{1227} (\bibinfo{year}{2007}).

\bibitem[{\citenamefont{{Fischer, T.} et~al.}(2010)\citenamefont{{Fischer, T.},
  {Whitehouse, S. C.}, {Mezzacappa, A.}, {Thielemann, F.-K.}, and
  {Liebend\"orfer, M.}}}]{Fischer.etal:2010}
\bibinfo{author}{\bibnamefont{{Fischer, T.}}},
  \bibinfo{author}{\bibnamefont{{Whitehouse, S. C.}}},
  \bibinfo{author}{\bibnamefont{{Mezzacappa, A.}}},
  \bibinfo{author}{\bibnamefont{{Thielemann, F.-K.}}}, \bibnamefont{and}
  \bibinfo{author}{\bibnamefont{{Liebend\"orfer, M.}}},
  \bibinfo{journal}{Astron. \& Astrophys.} \textbf{\bibinfo{volume}{517}},
  \bibinfo{pages}{A80} (\bibinfo{year}{2010}).

\bibitem[{\citenamefont{Woosley and Hoffman}(1992)}]{Woosley.Hoffman:1992}
\bibinfo{author}{\bibfnamefont{S.~E.} \bibnamefont{Woosley}} \bibnamefont{and}
  \bibinfo{author}{\bibfnamefont{R.~D.} \bibnamefont{Hoffman}},
  \bibinfo{journal}{\apj} \textbf{\bibinfo{volume}{395}}, \bibinfo{pages}{202}
  (\bibinfo{year}{1992}).

\bibitem[{\citenamefont{Witti et~al.}(1994)\citenamefont{Witti, Janka, and
  Takahashi}}]{Witti.Janka.Takahashi:1994}
\bibinfo{author}{\bibfnamefont{J.}~\bibnamefont{Witti}},
  \bibinfo{author}{\bibfnamefont{H.-T.} \bibnamefont{Janka}}, \bibnamefont{and}
  \bibinfo{author}{\bibfnamefont{K.}~\bibnamefont{Takahashi}},
  \bibinfo{journal}{Astron. \& Astrophys.} \textbf{\bibinfo{volume}{286}},
  \bibinfo{pages}{841} (\bibinfo{year}{1994}).

\bibitem[{\citenamefont{Fr{\"o}hlich
  et~al.}(2006{\natexlab{a}})\citenamefont{Fr{\"o}hlich, Mart{\'\i}nez-Pinedo,
  Liebend{\"o}rfer, Thielemann, Bravo, Hix, Langanke, and
  Zinner}}]{Froehlich.Martinez-Pinedo.ea:2006}
\bibinfo{author}{\bibfnamefont{C.}~\bibnamefont{Fr{\"o}hlich}},
  \bibinfo{author}{\bibfnamefont{G.}~\bibnamefont{Mart{\'\i}nez-Pinedo}},
  \bibinfo{author}{\bibfnamefont{M.}~\bibnamefont{Liebend{\"o}rfer}},
  \bibinfo{author}{\bibfnamefont{F.-K.} \bibnamefont{Thielemann}},
  \bibinfo{author}{\bibfnamefont{E.}~\bibnamefont{Bravo}},
  \bibinfo{author}{\bibfnamefont{W.~R.} \bibnamefont{Hix}},
  \bibinfo{author}{\bibfnamefont{K.}~\bibnamefont{Langanke}}, \bibnamefont{and}
  \bibinfo{author}{\bibfnamefont{N.~T.} \bibnamefont{Zinner}},
  \bibinfo{journal}{\prl} \textbf{\bibinfo{volume}{96}}, \bibinfo{eid}{142502}
  (\bibinfo{year}{2006}{\natexlab{a}}).

\bibitem[{\citenamefont{Pruet et~al.}(2006)\citenamefont{Pruet, Hoffman,
  Woosley, Janka, and Buras}}]{Pruet.Hoffman.ea:2006}
\bibinfo{author}{\bibfnamefont{J.}~\bibnamefont{Pruet}},
  \bibinfo{author}{\bibfnamefont{R.~D.} \bibnamefont{Hoffman}},
  \bibinfo{author}{\bibfnamefont{S.~E.} \bibnamefont{Woosley}},
  \bibinfo{author}{\bibfnamefont{H.-T.} \bibnamefont{Janka}}, \bibnamefont{and}
  \bibinfo{author}{\bibfnamefont{R.}~\bibnamefont{Buras}},
  \bibinfo{journal}{\apj} \textbf{\bibinfo{volume}{644}}, \bibinfo{pages}{1028}
  (\bibinfo{year}{2006}).

\bibitem[{\citenamefont{Wanajo}(2006)}]{Wanajo:2006}
\bibinfo{author}{\bibfnamefont{S.}~\bibnamefont{Wanajo}},
  \bibinfo{journal}{\apj} \textbf{\bibinfo{volume}{647}}, \bibinfo{pages}{1323}
  (\bibinfo{year}{2006}).

\bibitem[{\citenamefont{Woosley et~al.}(1994)\citenamefont{Woosley, Wilson,
  Mathews, Hoffman, and Meyer}}]{Woosley.Wilson.ea:1994}
\bibinfo{author}{\bibfnamefont{S.~E.} \bibnamefont{Woosley}},
  \bibinfo{author}{\bibfnamefont{J.~R.} \bibnamefont{Wilson}},
  \bibinfo{author}{\bibfnamefont{G.~J.} \bibnamefont{Mathews}},
  \bibinfo{author}{\bibfnamefont{R.~D.} \bibnamefont{Hoffman}},
  \bibnamefont{and} \bibinfo{author}{\bibfnamefont{B.~S.} \bibnamefont{Meyer}},
  \bibinfo{journal}{Astrophys. J.} \textbf{\bibinfo{volume}{433}},
  \bibinfo{pages}{229} (\bibinfo{year}{1994}).

\bibitem[{\citenamefont{Takahashi et~al.}(1994)\citenamefont{Takahashi, Witti,
  and Janka}}]{Takahashi.Witti.Janka:1994}
\bibinfo{author}{\bibfnamefont{K.}~\bibnamefont{Takahashi}},
  \bibinfo{author}{\bibfnamefont{J.}~\bibnamefont{Witti}}, \bibnamefont{and}
  \bibinfo{author}{\bibfnamefont{H.-T.} \bibnamefont{Janka}},
  \bibinfo{journal}{Astron. \& Astrophys.} \textbf{\bibinfo{volume}{286}},
  \bibinfo{pages}{857} (\bibinfo{year}{1994}).

\bibitem[{\citenamefont{{Qian} and {Woosley}}(1996)}]{Qian.Woosley:1996}
\bibinfo{author}{\bibfnamefont{Y.-Z.} \bibnamefont{{Qian}}} \bibnamefont{and}
  \bibinfo{author}{\bibfnamefont{S.~E.} \bibnamefont{{Woosley}}},
  \bibinfo{journal}{Astrophys. J.} \textbf{\bibinfo{volume}{471}},
  \bibinfo{pages}{331} (\bibinfo{year}{1996}).

\bibitem[{\citenamefont{{Hoffman} et~al.}(1997)\citenamefont{{Hoffman},
  {Woosley}, and {Qian}}}]{hoffman.woosley.qian:1997}
\bibinfo{author}{\bibfnamefont{R.~D.} \bibnamefont{{Hoffman}}},
  \bibinfo{author}{\bibfnamefont{S.~E.} \bibnamefont{{Woosley}}},
  \bibnamefont{and} \bibinfo{author}{\bibfnamefont{Y.-Z.}
  \bibnamefont{{Qian}}}, \bibinfo{journal}{Astrophys. J.}
  \textbf{\bibinfo{volume}{482}}, \bibinfo{pages}{951} (\bibinfo{year}{1997}).

\bibitem[{\citenamefont{{Otsuki} et~al.}(2000)\citenamefont{{Otsuki},
  {Tagoshi}, {Kajino}, and {Wanajo}}}]{Otsuki.Tagoshi.ea:2000}
\bibinfo{author}{\bibfnamefont{K.}~\bibnamefont{{Otsuki}}},
  \bibinfo{author}{\bibfnamefont{H.}~\bibnamefont{{Tagoshi}}},
  \bibinfo{author}{\bibfnamefont{T.}~\bibnamefont{{Kajino}}}, \bibnamefont{and}
  \bibinfo{author}{\bibfnamefont{S.}~\bibnamefont{{Wanajo}}},
  \bibinfo{journal}{Astrophys. J.} \textbf{\bibinfo{volume}{533}},
  \bibinfo{pages}{424} (\bibinfo{year}{2000}).

\bibitem[{\citenamefont{Thompson et~al.}(2001)\citenamefont{Thompson, Burrows,
  and Meyer}}]{Thompson.Burrows.Meyer:2001}
\bibinfo{author}{\bibfnamefont{T.~A.} \bibnamefont{Thompson}},
  \bibinfo{author}{\bibfnamefont{A.}~\bibnamefont{Burrows}}, \bibnamefont{and}
  \bibinfo{author}{\bibfnamefont{B.~S.} \bibnamefont{Meyer}},
  \bibinfo{journal}{Astrophys. J.} \textbf{\bibinfo{volume}{562}},
  \bibinfo{pages}{887} (\bibinfo{year}{2001}).

\bibitem[{\citenamefont{{H{\"u}depohl}
  et~al.}(2010)\citenamefont{{H{\"u}depohl}, {M{\"u}ller}, {Janka}, {Marek},
  and {Raffelt}}}]{Huedepohl.ea:2010}
\bibinfo{author}{\bibfnamefont{L.}~\bibnamefont{{H{\"u}depohl}}},
  \bibinfo{author}{\bibfnamefont{B.}~\bibnamefont{{M{\"u}ller}}},
  \bibinfo{author}{\bibfnamefont{H.T.}~\bibnamefont{{Janka}}},
  \bibinfo{author}{\bibfnamefont{A.}~\bibnamefont{{Marek}}}, \bibnamefont{and}
  \bibinfo{author}{\bibfnamefont{G.~G.} \bibnamefont{{Raffelt}}},
  \bibinfo{journal}{Phys. Rev. Lett.} \textbf{\bibinfo{volume}{104}},
  \bibinfo{eid}{251101} (\bibinfo{year}{2010}).

\bibitem[{\citenamefont{{Kratz} et~al.}(1993)\citenamefont{{Kratz}, {Bitouzet},
  {Thielemann}, {Moeller}, and {Pfeiffer}}}]{Kratz.Bitouzet.ea:1993}
\bibinfo{author}{\bibfnamefont{K.}~\bibnamefont{{Kratz}}},
  \bibinfo{author}{\bibfnamefont{J.}~\bibnamefont{{Bitouzet}}},
  \bibinfo{author}{\bibfnamefont{F.}~\bibnamefont{{Thielemann}}},
  \bibinfo{author}{\bibfnamefont{P.}~\bibnamefont{{Moeller}}},
  \bibnamefont{and}
  \bibinfo{author}{\bibfnamefont{B.}~\bibnamefont{{Pfeiffer}}},
  \bibinfo{journal}{Astrophys. J.} \textbf{\bibinfo{volume}{403}},
  \bibinfo{pages}{216} (\bibinfo{year}{1993}).

\bibitem[{\citenamefont{Kratz et~al.}(1998)\citenamefont{Kratz, Pfeiffer, and
  Thielemann}}]{Kratz.Pfeiffer.Thielemann:1998}
\bibinfo{author}{\bibfnamefont{K.-L.} \bibnamefont{Kratz}},
  \bibinfo{author}{\bibfnamefont{B.}~\bibnamefont{Pfeiffer}}, \bibnamefont{and}
  \bibinfo{author}{\bibfnamefont{F.-K.} \bibnamefont{Thielemann}},
  \bibinfo{journal}{Nucl. Phys. A} \textbf{\bibinfo{volume}{630}},
  \bibinfo{pages}{352} (\bibinfo{year}{1998}).

\bibitem[{\citenamefont{Howard et~al.}(1993)\citenamefont{Howard, Goriely,
  Rayet, and Arnould}}]{Howard.Goriely.ea:1993}
\bibinfo{author}{\bibfnamefont{W.~M.} \bibnamefont{Howard}},
  \bibinfo{author}{\bibfnamefont{S.}~\bibnamefont{Goriely}},
  \bibinfo{author}{\bibfnamefont{M.}~\bibnamefont{Rayet}}, \bibnamefont{and}
  \bibinfo{author}{\bibfnamefont{M.}~\bibnamefont{Arnould}},
  \bibinfo{journal}{Astrophys. J.} \textbf{\bibinfo{volume}{417}},
  \bibinfo{pages}{713} (\bibinfo{year}{1993}).

\bibitem[{\citenamefont{{Freiburghaus}
  et~al.}(1999)\citenamefont{{Freiburghaus}, {Rembges}, {Rauscher}, {Kolbe},
  {Thielemann}, {Kratz}, {Pfeiffer}, and
  {Cowan}}}]{Freiburghaus.Rembges.ea:1999}
\bibinfo{author}{\bibfnamefont{C.}~\bibnamefont{{Freiburghaus}}},
  \bibinfo{author}{\bibfnamefont{J.-F.} \bibnamefont{{Rembges}}},
  \bibinfo{author}{\bibfnamefont{T.}~\bibnamefont{{Rauscher}}},
  \bibinfo{author}{\bibfnamefont{E.}~\bibnamefont{{Kolbe}}},
  \bibinfo{author}{\bibfnamefont{F.-K.} \bibnamefont{{Thielemann}}},
  \bibinfo{author}{\bibfnamefont{K.-L.} \bibnamefont{{Kratz}}},
  \bibinfo{author}{\bibfnamefont{B.}~\bibnamefont{{Pfeiffer}}},
  \bibnamefont{and} \bibinfo{author}{\bibfnamefont{J.~J.}
  \bibnamefont{{Cowan}}}, \bibinfo{journal}{\apj}
  \textbf{\bibinfo{volume}{516}}, \bibinfo{pages}{381} (\bibinfo{year}{1999}).

\bibitem[{\citenamefont{{Meyer} and {Brown}}(1997)}]{Meyer.Brown:1997}
\bibinfo{author}{\bibfnamefont{B.~S.} \bibnamefont{{Meyer}}} \bibnamefont{and}
  \bibinfo{author}{\bibfnamefont{J.~S.} \bibnamefont{{Brown}}},
  \bibinfo{journal}{Astrophys. J. Suppl.} \textbf{\bibinfo{volume}{112}},
  \bibinfo{pages}{199} (\bibinfo{year}{1997}).

\bibitem[{\citenamefont{{Wanajo} et~al.}(2004)\citenamefont{{Wanajo},
  {Goriely}, {Samyn}, and {Itoh}}}]{Wanajo.Goriely.ea:2004}
\bibinfo{author}{\bibfnamefont{S.}~\bibnamefont{{Wanajo}}},
  \bibinfo{author}{\bibfnamefont{S.}~\bibnamefont{{Goriely}}},
  \bibinfo{author}{\bibfnamefont{M.}~\bibnamefont{{Samyn}}}, \bibnamefont{and}
  \bibinfo{author}{\bibfnamefont{N.}~\bibnamefont{{Itoh}}},
  \bibinfo{journal}{\apj} \textbf{\bibinfo{volume}{606}}, \bibinfo{pages}{1057}
  (\bibinfo{year}{2004}).

\bibitem[{\citenamefont{{Farouqi} et~al.}(2010)\citenamefont{{Farouqi},
  {Kratz}, {Pfeiffer}, {Rauscher}, {Thielemann}, and
  {Truran}}}]{Farouqi.etal:2010}
\bibinfo{author}{\bibfnamefont{K.}~\bibnamefont{{Farouqi}}},
  \bibinfo{author}{\bibfnamefont{K.}~\bibnamefont{{Kratz}}},
  \bibinfo{author}{\bibfnamefont{B.}~\bibnamefont{{Pfeiffer}}},
  \bibinfo{author}{\bibfnamefont{T.}~\bibnamefont{{Rauscher}}},
  \bibinfo{author}{\bibfnamefont{F.}~\bibnamefont{{Thielemann}}},
  \bibnamefont{and} \bibinfo{author}{\bibfnamefont{J.~W.}
  \bibnamefont{{Truran}}}, \bibinfo{journal}{Astrophys. J.}
  \textbf{\bibinfo{volume}{712}}, \bibinfo{pages}{1359} (\bibinfo{year}{2010}).

\bibitem[{\citenamefont{Surman et~al.}(1997)\citenamefont{Surman, Engel,
  Bennett, and Meyer}}]{Surman.Engel.ea:1997}
\bibinfo{author}{\bibfnamefont{R.}~\bibnamefont{Surman}},
  \bibinfo{author}{\bibfnamefont{J.}~\bibnamefont{Engel}},
  \bibinfo{author}{\bibfnamefont{J.~R.} \bibnamefont{Bennett}},
  \bibnamefont{and} \bibinfo{author}{\bibfnamefont{B.~S.} \bibnamefont{Meyer}},
  \bibinfo{journal}{Phys. Rev. Lett.} \textbf{\bibinfo{volume}{79}},
  \bibinfo{pages}{1809} (\bibinfo{year}{1997}).

\bibitem[{\citenamefont{Surman and Engel}(2001)}]{Surman.Engel:2001}
\bibinfo{author}{\bibfnamefont{R.}~\bibnamefont{Surman}} \bibnamefont{and}
  \bibinfo{author}{\bibfnamefont{J.}~\bibnamefont{Engel}},
  \bibinfo{journal}{\prc} \textbf{\bibinfo{volume}{64}}, \bibinfo{eid}{035801}
  (\bibinfo{year}{2001}).

\bibitem[{\citenamefont{Surman et~al.}(2009)\citenamefont{Surman, Beun,
  McLaughlin, and Hix}}]{Surman.Beun.ea:2009}
\bibinfo{author}{\bibfnamefont{R.}~\bibnamefont{Surman}},
  \bibinfo{author}{\bibfnamefont{J.}~\bibnamefont{Beun}},
  \bibinfo{author}{\bibfnamefont{G.~C.} \bibnamefont{McLaughlin}},
  \bibnamefont{and} \bibinfo{author}{\bibfnamefont{W.~R.} \bibnamefont{Hix}},
  \bibinfo{journal}{Phys. Rev. C} \textbf{\bibinfo{volume}{79}},
  \bibinfo{pages}{045809} (\bibinfo{year}{2009}).

\bibitem[{\citenamefont{Beun et~al.}(2009{\natexlab{a}})\citenamefont{Beun,
  Blackmon, Hix, {McLaughlin}, Smith, and Surman}}]{beun.etal:2009}
\bibinfo{author}{\bibfnamefont{J.}~\bibnamefont{Beun}},
  \bibinfo{author}{\bibfnamefont{J.~C.} \bibnamefont{Blackmon}},
  \bibinfo{author}{\bibfnamefont{W.~R.} \bibnamefont{Hix}},
  \bibinfo{author}{\bibfnamefont{G.~C.} \bibnamefont{{McLaughlin}}},
  \bibinfo{author}{\bibfnamefont{M.~S.} \bibnamefont{Smith}}, \bibnamefont{and}
  \bibinfo{author}{\bibfnamefont{R.}~\bibnamefont{Surman}},
  \bibinfo{journal}{Journal of Physics G Nuclear Physics}
  \textbf{\bibinfo{volume}{36}}, \bibinfo{pages}{5201}
  (\bibinfo{year}{2009}{\natexlab{a}}).

\bibitem[{\citenamefont{Borzov et~al.}(2008)\citenamefont{Borzov,
  Cuenca-Garc{\'i}a, Langanke, Mart{\'i}nez-Pinedo, and
  Montes}}]{Borzov.Cuenca-Garcia.ea:2008}
\bibinfo{author}{\bibfnamefont{I.~N.} \bibnamefont{Borzov}},
  \bibinfo{author}{\bibfnamefont{J.~J.} \bibnamefont{Cuenca-Garc{\'i}a}},
  \bibinfo{author}{\bibfnamefont{K.}~\bibnamefont{Langanke}},
  \bibinfo{author}{\bibfnamefont{G.}~\bibnamefont{Mart{\'i}nez-Pinedo}},
  \bibnamefont{and} \bibinfo{author}{\bibfnamefont{F.}~\bibnamefont{Montes}},
  \bibinfo{journal}{Nucl. Phys. A} \textbf{\bibinfo{volume}{814}},
  \bibinfo{pages}{159} (\bibinfo{year}{2008}).

\bibitem[{\citenamefont{Dillmann et~al.}(2003)\citenamefont{Dillmann, Kratz,
  W{\"o}hr, Arndt, Brown, Hoff, Hjorth-Jensen, K{\"o}ster, Ostrowski, Pfeiffer
  et~al.}}]{Dillmann.Kratz.ea:2003}
\bibinfo{author}{\bibfnamefont{I.}~\bibnamefont{Dillmann}},
  \bibinfo{author}{\bibfnamefont{K.-L.} \bibnamefont{Kratz}},
  \bibinfo{author}{\bibfnamefont{A.}~\bibnamefont{W{\"o}hr}},
  \bibinfo{author}{\bibfnamefont{O.}~\bibnamefont{Arndt}},
  \bibinfo{author}{\bibfnamefont{B.~A.} \bibnamefont{Brown}},
  \bibinfo{author}{\bibfnamefont{P.}~\bibnamefont{Hoff}},
  \bibinfo{author}{\bibfnamefont{M.}~\bibnamefont{Hjorth-Jensen}},
  \bibinfo{author}{\bibfnamefont{U.}~\bibnamefont{K{\"o}ster}},
  \bibinfo{author}{\bibfnamefont{A.~N.} \bibnamefont{Ostrowski}},
  \bibinfo{author}{\bibfnamefont{B.}~\bibnamefont{Pfeiffer}},
  \bibnamefont{et~al.}, \bibinfo{journal}{\prl} \textbf{\bibinfo{volume}{91}},
  \bibinfo{eid}{162503} (\bibinfo{year}{2003}).

\bibitem[{\citenamefont{{Arcones} and {Montes}}(2010)}]{Arcones.Montes:2010}
\bibinfo{author}{\bibfnamefont{A.}~\bibnamefont{{Arcones}}} \bibnamefont{and}
  \bibinfo{author}{\bibfnamefont{F.}~\bibnamefont{{Montes}}}
  (\bibinfo{year}{2010}), \eprint{1007.1275}.

\bibitem[{\citenamefont{{Terasawa} et~al.}(2002)\citenamefont{{Terasawa},
  {Sumiyoshi}, {Yamada}, {Suzuki}, and {Kajino}}}]{Terasawa.Sumiyoshi.ea:2002}
\bibinfo{author}{\bibfnamefont{M.}~\bibnamefont{{Terasawa}}},
  \bibinfo{author}{\bibfnamefont{K.}~\bibnamefont{{Sumiyoshi}}},
  \bibinfo{author}{\bibfnamefont{S.}~\bibnamefont{{Yamada}}},
  \bibinfo{author}{\bibfnamefont{H.}~\bibnamefont{{Suzuki}}}, \bibnamefont{and}
  \bibinfo{author}{\bibfnamefont{T.}~\bibnamefont{{Kajino}}},
  \bibinfo{journal}{Astrophys. J.} \textbf{\bibinfo{volume}{578}},
  \bibinfo{pages}{L137} (\bibinfo{year}{2002}).

\bibitem[{\citenamefont{{Wanajo} et~al.}(2002)\citenamefont{{Wanajo}, {Itoh},
  {Ishimaru}, {Nozawa}, and {Beers}}}]{Wanajo.Itoh.ea:2002}
\bibinfo{author}{\bibfnamefont{S.}~\bibnamefont{{Wanajo}}},
  \bibinfo{author}{\bibfnamefont{N.}~\bibnamefont{{Itoh}}},
  \bibinfo{author}{\bibfnamefont{Y.}~\bibnamefont{{Ishimaru}}},
  \bibinfo{author}{\bibfnamefont{S.}~\bibnamefont{{Nozawa}}}, \bibnamefont{and}
  \bibinfo{author}{\bibfnamefont{T.~C.} \bibnamefont{{Beers}}},
  \bibinfo{journal}{Astrophys. J.} \textbf{\bibinfo{volume}{577}},
  \bibinfo{pages}{853} (\bibinfo{year}{2002}).

\bibitem[{\citenamefont{{Wanajo}}(2007)}]{Wanajo:2007}
\bibinfo{author}{\bibfnamefont{S.}~\bibnamefont{{Wanajo}}},
  \bibinfo{journal}{Astrophys. J.} \textbf{\bibinfo{volume}{666}},
  \bibinfo{pages}{L77} (\bibinfo{year}{2007}), \eprint{arXiv:0706.4360}.

\bibitem[{\citenamefont{{Kuroda} et~al.}(2008)\citenamefont{{Kuroda}, {Wanajo},
  and {Nomoto}}}]{Kuroda.Wanajo.Nomoto:2008}
\bibinfo{author}{\bibfnamefont{T.}~\bibnamefont{{Kuroda}}},
  \bibinfo{author}{\bibfnamefont{S.}~\bibnamefont{{Wanajo}}}, \bibnamefont{and}
  \bibinfo{author}{\bibfnamefont{K.}~\bibnamefont{{Nomoto}}},
  \bibinfo{journal}{Astrophys. J.} \textbf{\bibinfo{volume}{672}},
  \bibinfo{pages}{1068} (\bibinfo{year}{2008}).

\bibitem[{\citenamefont{{Panov} and {Janka}}(2009)}]{Panov.Janka:2009}
\bibinfo{author}{\bibfnamefont{I.~V.} \bibnamefont{{Panov}}} \bibnamefont{and}
  \bibinfo{author}{\bibfnamefont{H.-T.} \bibnamefont{{Janka}}},
  \bibinfo{journal}{Astron. \& Astrophys.} \textbf{\bibinfo{volume}{494}},
  \bibinfo{pages}{829} (\bibinfo{year}{2009}).

\bibitem[{\citenamefont{{Wanajo} et~al.}(2001)\citenamefont{{Wanajo}, {Kajino},
  {Mathews}, and {Otsuki}}}]{Wanajo.Kajino.ea:2001}
\bibinfo{author}{\bibfnamefont{S.}~\bibnamefont{{Wanajo}}},
  \bibinfo{author}{\bibfnamefont{T.}~\bibnamefont{{Kajino}}},
  \bibinfo{author}{\bibfnamefont{G.~J.} \bibnamefont{{Mathews}}},
  \bibnamefont{and} \bibinfo{author}{\bibfnamefont{K.}~\bibnamefont{{Otsuki}}},
  \bibinfo{journal}{Astrophys. J.} \textbf{\bibinfo{volume}{554}},
  \bibinfo{pages}{578} (\bibinfo{year}{2001}).

\bibitem[{\citenamefont{{Scheck} et~al.}(2006)\citenamefont{{Scheck},
  {Kifonidis}, {Janka}, and
  {M{\"u}ller}}}]{Scheck.Kifonidis.Janka.Mueller:2006}
\bibinfo{author}{\bibfnamefont{L.}~\bibnamefont{{Scheck}}},
  \bibinfo{author}{\bibfnamefont{K.}~\bibnamefont{{Kifonidis}}},
  \bibinfo{author}{\bibfnamefont{H.-T.} \bibnamefont{{Janka}}},
  \bibnamefont{and}
  \bibinfo{author}{\bibfnamefont{E.}~\bibnamefont{{M{\"u}ller}}},
  \bibinfo{journal}{Astron. \& Astrophys.} \textbf{\bibinfo{volume}{457}},
  \bibinfo{pages}{963} (\bibinfo{year}{2006}).

\bibitem[{\citenamefont{{Kifonidis} et~al.}(2006)\citenamefont{{Kifonidis},
  {Plewa}, {Scheck}, {Janka}, and {M{\"u}ller}}}]{Kifonidis.Plewa.ea:2006}
\bibinfo{author}{\bibfnamefont{K.}~\bibnamefont{{Kifonidis}}},
  \bibinfo{author}{\bibfnamefont{T.}~\bibnamefont{{Plewa}}},
  \bibinfo{author}{\bibfnamefont{L.}~\bibnamefont{{Scheck}}},
  \bibinfo{author}{\bibfnamefont{H.-T.} \bibnamefont{{Janka}}},
  \bibnamefont{and}
  \bibinfo{author}{\bibfnamefont{E.}~\bibnamefont{{M{\"u}ller}}},
  \bibinfo{journal}{Astron. \& Astrophys.} \textbf{\bibinfo{volume}{453}},
  \bibinfo{pages}{661} (\bibinfo{year}{2006}).

\bibitem[{\citenamefont{{Marek} et~al.}(2006)\citenamefont{{Marek},
  {Dimmelmeier}, {Janka}, {M{\"u}ller}, and
  {Buras}}}]{Marek.Dimmelmeier.ea:2006}
\bibinfo{author}{\bibfnamefont{A.}~\bibnamefont{{Marek}}},
  \bibinfo{author}{\bibfnamefont{H.}~\bibnamefont{{Dimmelmeier}}},
  \bibinfo{author}{\bibfnamefont{H.-T.} \bibnamefont{{Janka}}},
  \bibinfo{author}{\bibfnamefont{E.}~\bibnamefont{{M{\"u}ller}}},
  \bibnamefont{and} \bibinfo{author}{\bibfnamefont{R.}~\bibnamefont{{Buras}}},
  \bibinfo{journal}{Astron. \& Astrophys.} \textbf{\bibinfo{volume}{445}},
  \bibinfo{pages}{273} (\bibinfo{year}{2006}).

\bibitem[{\citenamefont{{Arcones} and {Janka}}(2010)}]{Arcones.Janka:2010}
\bibinfo{author}{\bibfnamefont{A.}~\bibnamefont{{Arcones}}} \bibnamefont{and}
  \bibinfo{author}{\bibfnamefont{H.}~\bibnamefont{{Janka}}}
  (\bibinfo{year}{2010}), \eprint{1008.0882}.

\bibitem[{\citenamefont{{Timmes} and {Arnett}}(1999)}]{Timmes.Arnett:1999}
\bibinfo{author}{\bibfnamefont{F.~X.} \bibnamefont{{Timmes}}} \bibnamefont{and}
  \bibinfo{author}{\bibfnamefont{D.}~\bibnamefont{{Arnett}}},
  \bibinfo{journal}{Astrophys. J. Suppl.} \textbf{\bibinfo{volume}{125}},
  \bibinfo{pages}{277} (\bibinfo{year}{1999}).

\bibitem[{\citenamefont{{Wanajo} et~al.}(2010)\citenamefont{{Wanajo}, {Janka},
  and {Kubono}}}]{Wanajo.etal:2010}
\bibinfo{author}{\bibfnamefont{S.}~\bibnamefont{{Wanajo}}},
  \bibinfo{author}{\bibfnamefont{H.}~\bibnamefont{{Janka}}}, \bibnamefont{and}
  \bibinfo{author}{\bibfnamefont{S.}~\bibnamefont{{Kubono}}}
  (\bibinfo{year}{2010}), \eprint{1004.4487}.

\bibitem[{\citenamefont{Fr{\"o}hlich
  et~al.}(2006{\natexlab{b}})\citenamefont{Fr{\"o}hlich, Hauser,
  Liebend{\"o}rfer, Mart{\'i}nez-Pinedo, Thielemann, Bravo, Zinner, Hix,
  Langanke, Mezzacappa et~al.}}]{Froehlich.Hauser.ea:2006}
\bibinfo{author}{\bibfnamefont{C.}~\bibnamefont{Fr{\"o}hlich}},
  \bibinfo{author}{\bibfnamefont{P.}~\bibnamefont{Hauser}},
  \bibinfo{author}{\bibfnamefont{M.}~\bibnamefont{Liebend{\"o}rfer}},
  \bibinfo{author}{\bibfnamefont{G.}~\bibnamefont{Mart{\'i}nez-Pinedo}},
  \bibinfo{author}{\bibfnamefont{F.-K.} \bibnamefont{Thielemann}},
  \bibinfo{author}{\bibfnamefont{E.}~\bibnamefont{Bravo}},
  \bibinfo{author}{\bibfnamefont{N.~T.} \bibnamefont{Zinner}},
  \bibinfo{author}{\bibfnamefont{W.~R.} \bibnamefont{Hix}},
  \bibinfo{author}{\bibfnamefont{K.}~\bibnamefont{Langanke}},
  \bibinfo{author}{\bibfnamefont{A.}~\bibnamefont{Mezzacappa}},
  \bibnamefont{et~al.}, \bibinfo{journal}{\apj} \textbf{\bibinfo{volume}{637}},
  \bibinfo{pages}{415} (\bibinfo{year}{2006}{\natexlab{b}}).

\bibitem[{\citenamefont{Fuller et~al.}(1982{\natexlab{a}})\citenamefont{Fuller,
  Fowler, and Newman}}]{Fuller.Fowler.Newman:1982b}
\bibinfo{author}{\bibfnamefont{G.~M.} \bibnamefont{Fuller}},
  \bibinfo{author}{\bibfnamefont{W.~A.} \bibnamefont{Fowler}},
  \bibnamefont{and} \bibinfo{author}{\bibfnamefont{M.~J.}
  \bibnamefont{Newman}}, \bibinfo{journal}{\apj}
  \textbf{\bibinfo{volume}{252}}, \bibinfo{pages}{715}
  (\bibinfo{year}{1982}{\natexlab{a}}).

\bibitem[{\citenamefont{Fuller et~al.}(1982{\natexlab{b}})\citenamefont{Fuller,
  Fowler, and Newman}}]{Fuller.Fowler.Newman:1982a}
\bibinfo{author}{\bibfnamefont{G.~M.} \bibnamefont{Fuller}},
  \bibinfo{author}{\bibfnamefont{W.~A.} \bibnamefont{Fowler}},
  \bibnamefont{and} \bibinfo{author}{\bibfnamefont{M.~J.}
  \bibnamefont{Newman}}, \bibinfo{journal}{Astrophys. J. Suppl.}
  \textbf{\bibinfo{volume}{48}}, \bibinfo{pages}{279}
  (\bibinfo{year}{1982}{\natexlab{b}}).

\bibitem[{\citenamefont{Langanke and
  Mart{\'\i}nez-Pinedo}(2000)}]{Langanke.Martinez-Pinedo:2000}
\bibinfo{author}{\bibfnamefont{K.}~\bibnamefont{Langanke}} \bibnamefont{and}
  \bibinfo{author}{\bibfnamefont{G.}~\bibnamefont{Mart{\'\i}nez-Pinedo}},
  \bibinfo{journal}{Nucl. Phys. A} \textbf{\bibinfo{volume}{673}},
  \bibinfo{pages}{481} (\bibinfo{year}{2000}).

\bibitem[{\citenamefont{{Langanke} and
  {Mart{\'\i}nez-Pinedo}}(2001)}]{Langanke.Martinez-Pinedo:2001}
\bibinfo{author}{\bibfnamefont{K.}~\bibnamefont{{Langanke}}} \bibnamefont{and}
  \bibinfo{author}{\bibfnamefont{G.}~\bibnamefont{{Mart{\'\i}nez-Pinedo}}},
  \bibinfo{journal}{At. Data. Nucl. Data Tables} \textbf{\bibinfo{volume}{79}},
  \bibinfo{pages}{1} (\bibinfo{year}{2001}).

\bibitem[{\citenamefont{Zinner}(2007)}]{Zinner:2007}
\bibinfo{author}{\bibfnamefont{N.~T.} \bibnamefont{Zinner}}, Ph.D. thesis,
  \bibinfo{school}{University of Aarhus, Denmark} (\bibinfo{year}{2007}).

\bibitem[{\citenamefont{Mocelj}(2007)}]{Mocelj:2007}
\bibinfo{author}{\bibfnamefont{D.}~\bibnamefont{Mocelj}}, Ph.D. thesis,
  \bibinfo{school}{Basel University} (\bibinfo{year}{2007}).

\bibitem[{\citenamefont{{Cowan} et~al.}(1983)\citenamefont{{Cowan}, {Cameron},
  and {Truran}}}]{Cowan.Cameron.Truran:1983}
\bibinfo{author}{\bibfnamefont{J.~J.} \bibnamefont{{Cowan}}},
  \bibinfo{author}{\bibfnamefont{A.~G.~W.} \bibnamefont{{Cameron}}},
  \bibnamefont{and} \bibinfo{author}{\bibfnamefont{J.~W.}
  \bibnamefont{{Truran}}}, \bibinfo{journal}{Astrophys. J.}
  \textbf{\bibinfo{volume}{265}}, \bibinfo{pages}{429} (\bibinfo{year}{1983}).

\bibitem[{\citenamefont{Hix and Thielemann}(1999)}]{Hix.Thielemann:1999b}
\bibinfo{author}{\bibfnamefont{W.~R.} \bibnamefont{Hix}} \bibnamefont{and}
  \bibinfo{author}{\bibfnamefont{F.-K.} \bibnamefont{Thielemann}},
  \bibinfo{journal}{J. Comput. Appl. Math.} \textbf{\bibinfo{volume}{109}},
  \bibinfo{pages}{321} (\bibinfo{year}{1999}).

\bibitem[{\citenamefont{Hix and Meyer}(2006)}]{Hix.Meyer:2006}
\bibinfo{author}{\bibfnamefont{W.~R.} \bibnamefont{Hix}} \bibnamefont{and}
  \bibinfo{author}{\bibfnamefont{B.~S.} \bibnamefont{Meyer}},
  \bibinfo{journal}{Nucl. Phys. A} \textbf{\bibinfo{volume}{777}},
  \bibinfo{pages}{188} (\bibinfo{year}{2006}).

\bibitem[{\citenamefont{Schenk and G{\"a}rtner}(2004)}]{Schenk.Gaertner:2004}
\bibinfo{author}{\bibfnamefont{O.}~\bibnamefont{Schenk}} \bibnamefont{and}
  \bibinfo{author}{\bibfnamefont{K.}~\bibnamefont{G{\"a}rtner}},
  \bibinfo{journal}{Future Generation Computer Systems}
  \textbf{\bibinfo{volume}{20}}, \bibinfo{pages}{475} (\bibinfo{year}{2004}).

\bibitem[{\citenamefont{{M{\"o}ller} et~al.}(1995)\citenamefont{{M{\"o}ller},
  {Nix}, {Myers}, and {Swiatecki}}}]{Moeller.Nix.ea:1995}
\bibinfo{author}{\bibfnamefont{P.}~\bibnamefont{{M{\"o}ller}}},
  \bibinfo{author}{\bibfnamefont{J.~R.} \bibnamefont{{Nix}}},
  \bibinfo{author}{\bibfnamefont{W.~D.} \bibnamefont{{Myers}}},
  \bibnamefont{and} \bibinfo{author}{\bibfnamefont{W.~J.}
  \bibnamefont{{Swiatecki}}}, \bibinfo{journal}{At. Data Nucl. Data Tables}
  \textbf{\bibinfo{volume}{59}}, \bibinfo{pages}{185} (\bibinfo{year}{1995}).

\bibitem[{\citenamefont{Pearson et~al.}(1996)\citenamefont{Pearson, Nayak, and
  Goriely}}]{Pearson.Nayak.Goriely:1996}
\bibinfo{author}{\bibfnamefont{J.~M.} \bibnamefont{Pearson}},
  \bibinfo{author}{\bibfnamefont{R.~C.} \bibnamefont{Nayak}}, \bibnamefont{and}
  \bibinfo{author}{\bibfnamefont{S.}~\bibnamefont{Goriely}},
  \bibinfo{journal}{Phys. Lett. B} \textbf{\bibinfo{volume}{387}},
  \bibinfo{pages}{455} (\bibinfo{year}{1996}).

\bibitem[{\citenamefont{Goriely et~al.}(2009)\citenamefont{Goriely, Chamel, and
  Pearson}}]{Goriely.Chamel.Pearson:2009}
\bibinfo{author}{\bibfnamefont{S.}~\bibnamefont{Goriely}},
  \bibinfo{author}{\bibfnamefont{N.}~\bibnamefont{Chamel}}, \bibnamefont{and}
  \bibinfo{author}{\bibfnamefont{J.~M.} \bibnamefont{Pearson}},
  \bibinfo{journal}{Phys. Rev. Lett.} \textbf{\bibinfo{volume}{102}},
  \bibinfo{eid}{152503} (\bibinfo{year}{2009}).

\bibitem[{\citenamefont{Duflo and Zuker}(1995)}]{Duflo.Zuker:1995}
\bibinfo{author}{\bibfnamefont{J.}~\bibnamefont{Duflo}} \bibnamefont{and}
  \bibinfo{author}{\bibfnamefont{A.~P.} \bibnamefont{Zuker}},
  \bibinfo{journal}{\prc} \textbf{\bibinfo{volume}{52}}, \bibinfo{pages}{R23}
  (\bibinfo{year}{1995}).

\bibitem[{\citenamefont{Rauscher and
  Thielemann}(2000)}]{Rauscher.Thielemann:2000}
\bibinfo{author}{\bibfnamefont{T.}~\bibnamefont{Rauscher}} \bibnamefont{and}
  \bibinfo{author}{\bibfnamefont{F.-K.} \bibnamefont{Thielemann}},
  \bibinfo{journal}{At. Data Nucl. Data Tables} \textbf{\bibinfo{volume}{75}},
  \bibinfo{pages}{1} (\bibinfo{year}{2000}).

\bibitem[{\citenamefont{Rauscher and
  Thielemann}(1998)}]{Rauscher.Thielemann:1998}
\bibinfo{author}{\bibfnamefont{T.}~\bibnamefont{Rauscher}} \bibnamefont{and}
  \bibinfo{author}{\bibfnamefont{F.-K.} \bibnamefont{Thielemann}}, in
  \emph{\bibinfo{booktitle}{Stellar Evolution, Stellar Explosions and Galactic
  Chemical Evolution}}, edited by
  \bibinfo{editor}{\bibfnamefont{A.}~\bibnamefont{Mezzacappa}}
  (\bibinfo{publisher}{IOP, Bristol}, \bibinfo{year}{1998}), p.
  \bibinfo{pages}{519}.

\bibitem[{\citenamefont{{Goriely} et~al.}(2008)\citenamefont{{Goriely},
  {Hilaire}, and {Koning}}}]{Goriely.Hilaire.Koning:2008}
\bibinfo{author}{\bibfnamefont{S.}~\bibnamefont{{Goriely}}},
  \bibinfo{author}{\bibfnamefont{S.}~\bibnamefont{{Hilaire}}},
  \bibnamefont{and} \bibinfo{author}{\bibfnamefont{A.~J.}
  \bibnamefont{{Koning}}}, \bibinfo{journal}{Astron. \& Astrophys.}
  \textbf{\bibinfo{volume}{487}}, \bibinfo{pages}{767} (\bibinfo{year}{2008}).

\bibitem[{\citenamefont{Michaud and Fowler}(1970)}]{Michaud.Fowler:1970}
\bibinfo{author}{\bibfnamefont{G.}~\bibnamefont{Michaud}} \bibnamefont{and}
  \bibinfo{author}{\bibfnamefont{W.~A.} \bibnamefont{Fowler}},
  \bibinfo{journal}{Phys. Rev. C} \textbf{\bibinfo{volume}{2}},
  \bibinfo{pages}{2041} (\bibinfo{year}{1970}).

\bibitem[{\citenamefont{Woosley et~al.}(1975)\citenamefont{Woosley, Fowler,
  Holmes, and Zimmerman}}]{Woosley.Fowler.ea:1975}
\bibinfo{author}{\bibfnamefont{S.~E.} \bibnamefont{Woosley}},
  \bibinfo{author}{\bibfnamefont{W.~A.} \bibnamefont{Fowler}},
  \bibinfo{author}{\bibfnamefont{J.~A.} \bibnamefont{Holmes}},
  \bibnamefont{and} \bibinfo{author}{\bibfnamefont{B.~A.}
  \bibnamefont{Zimmerman}}, \bibinfo{type}{Preprint} \bibinfo{number}{OAP-422},
  \bibinfo{institution}{California Institute of Technology, W. K. Kellogg
  Radiation Laboratory} (\bibinfo{year}{1975}).

\bibitem[{\citenamefont{{Mathews} et~al.}(1983)\citenamefont{{Mathews},
  {Mengoni}, {Thielemann}, and {Fowler}}}]{mathews.mengoni.ea:1983}
\bibinfo{author}{\bibfnamefont{G.~J.} \bibnamefont{{Mathews}}},
  \bibinfo{author}{\bibfnamefont{A.}~\bibnamefont{{Mengoni}}},
  \bibinfo{author}{\bibfnamefont{F.-K.} \bibnamefont{{Thielemann}}},
  \bibnamefont{and} \bibinfo{author}{\bibfnamefont{W.~A.}
  \bibnamefont{{Fowler}}}, \bibinfo{journal}{Astrophys. J.}
  \textbf{\bibinfo{volume}{270}}, \bibinfo{pages}{740} (\bibinfo{year}{1983}).

\bibitem[{\citenamefont{M\"oller et~al.}(2003)\citenamefont{M\"oller, Pfeiffer,
  and Kratz}}]{Moeller.Pfeiffer.Kratz:2003}
\bibinfo{author}{\bibfnamefont{P.}~\bibnamefont{M\"oller}},
  \bibinfo{author}{\bibfnamefont{B.}~\bibnamefont{Pfeiffer}}, \bibnamefont{and}
  \bibinfo{author}{\bibfnamefont{K.-L.} \bibnamefont{Kratz}},
  \bibinfo{journal}{\prc} \textbf{\bibinfo{volume}{67}}, \bibinfo{eid}{055802}
  (\bibinfo{year}{2003}).

\bibitem[{\citenamefont{NuDat2}(2009)}]{NuDat2}
\bibinfo{author}{\bibnamefont{NuDat2}} (\bibinfo{year}{2009}),
  \bibinfo{note}{{National Nuclear Data Center, information extracted from the
  NuDat 2 database}}, \urlprefix\url{http://www.nndc.bnl.gov/nudat2/}.

\bibitem[{\citenamefont{{Sneden} et~al.}(2008)\citenamefont{{Sneden}, {Cowan},
  and {Gallino}}}]{Sneden.etal:2008}
\bibinfo{author}{\bibfnamefont{C.}~\bibnamefont{{Sneden}}},
  \bibinfo{author}{\bibfnamefont{J.~J.} \bibnamefont{{Cowan}}},
  \bibnamefont{and}
  \bibinfo{author}{\bibfnamefont{R.}~\bibnamefont{{Gallino}}},
  \bibinfo{journal}{Ann. Rev. Astron. \& Astrop.}
  \textbf{\bibinfo{volume}{46}}, \bibinfo{pages}{241} (\bibinfo{year}{2008}).

\bibitem[{\citenamefont{{Blake} and {Schramm}}(1976)}]{Blake.Schramm:1976}
\bibinfo{author}{\bibfnamefont{J.~B.} \bibnamefont{{Blake}}} \bibnamefont{and}
  \bibinfo{author}{\bibfnamefont{D.~N.} \bibnamefont{{Schramm}}},
  \bibinfo{journal}{Astrophys. J.} \textbf{\bibinfo{volume}{209}},
  \bibinfo{pages}{846} (\bibinfo{year}{1976}).

\bibitem[{\citenamefont{{Schatz} et~al.}(2002)\citenamefont{{Schatz},
  {Toenjes}, {Pfeiffer}, {Beers}, {Cowan}, {Hill}, and
  {Kratz}}}]{Schatz.Toenjes.ea:2002}
\bibinfo{author}{\bibfnamefont{H.}~\bibnamefont{{Schatz}}},
  \bibinfo{author}{\bibfnamefont{R.}~\bibnamefont{{Toenjes}}},
  \bibinfo{author}{\bibfnamefont{B.}~\bibnamefont{{Pfeiffer}}},
  \bibinfo{author}{\bibfnamefont{T.~C.} \bibnamefont{{Beers}}},
  \bibinfo{author}{\bibfnamefont{J.~J.} \bibnamefont{{Cowan}}},
  \bibinfo{author}{\bibfnamefont{V.}~\bibnamefont{{Hill}}}, \bibnamefont{and}
  \bibinfo{author}{\bibfnamefont{K.-L.} \bibnamefont{{Kratz}}},
  \bibinfo{journal}{\apj} \textbf{\bibinfo{volume}{579}}, \bibinfo{pages}{626}
  (\bibinfo{year}{2002}).

\bibitem[{\citenamefont{{Roederer} et~al.}(2009)\citenamefont{{Roederer},
  {Kratz}, {Frebel}, {Christlieb}, {Pfeiffer}, {Cowan}, and
  {Sneden}}}]{Roederer.Kratz.ea:2009}
\bibinfo{author}{\bibfnamefont{I.~U.} \bibnamefont{{Roederer}}},
  \bibinfo{author}{\bibfnamefont{K.}~\bibnamefont{{Kratz}}},
  \bibinfo{author}{\bibfnamefont{A.}~\bibnamefont{{Frebel}}},
  \bibinfo{author}{\bibfnamefont{N.}~\bibnamefont{{Christlieb}}},
  \bibinfo{author}{\bibfnamefont{B.}~\bibnamefont{{Pfeiffer}}},
  \bibinfo{author}{\bibfnamefont{J.~J.} \bibnamefont{{Cowan}}},
  \bibnamefont{and} \bibinfo{author}{\bibfnamefont{C.}~\bibnamefont{{Sneden}}},
  \bibinfo{journal}{Astrophys. J.} \textbf{\bibinfo{volume}{698}},
  \bibinfo{pages}{1963} (\bibinfo{year}{2009}).

\bibitem[{\citenamefont{K{\"a}ppeler}(1999)}]{Kaeppeler:1999}
\bibinfo{author}{\bibfnamefont{F.}~\bibnamefont{K{\"a}ppeler}},
  \bibinfo{journal}{Prog. Part. Nucl. Phys.} \textbf{\bibinfo{volume}{43}},
  \bibinfo{pages}{419} (\bibinfo{year}{1999}).

\bibitem[{\citenamefont{Loens et~al.}(2008)\citenamefont{Loens, Langanke,
  Mart{\'i}nez-Pinedo, Rauscher, and Thielemann}}]{Loens.Langanke.ea:2008}
\bibinfo{author}{\bibfnamefont{H.}~\bibnamefont{Loens}},
  \bibinfo{author}{\bibfnamefont{K.}~\bibnamefont{Langanke}},
  \bibinfo{author}{\bibfnamefont{G.}~\bibnamefont{Mart{\'i}nez-Pinedo}},
  \bibinfo{author}{\bibfnamefont{T.}~\bibnamefont{Rauscher}}, \bibnamefont{and}
  \bibinfo{author}{\bibfnamefont{F.-K.} \bibnamefont{Thielemann}},
  \bibinfo{journal}{Phys. Lett. B} \textbf{\bibinfo{volume}{666}},
  \bibinfo{pages}{395} (\bibinfo{year}{2008}).

\bibitem[{\citenamefont{Litvinova et~al.}(2009)\citenamefont{Litvinova, Loens,
  Langanke, {Mart{\'i}nez-Pinedo}, Rauscher, Ring, Thielemann, and
  Tselyaev}}]{Litvinova.Loens.ea:2009}
\bibinfo{author}{\bibfnamefont{E.}~\bibnamefont{Litvinova}},
  \bibinfo{author}{\bibfnamefont{H.}~\bibnamefont{Loens}},
  \bibinfo{author}{\bibfnamefont{K.}~\bibnamefont{Langanke}},
  \bibinfo{author}{\bibfnamefont{G.}~\bibnamefont{{Mart{\'i}nez-Pinedo}}},
  \bibinfo{author}{\bibfnamefont{T.}~\bibnamefont{Rauscher}},
  \bibinfo{author}{\bibfnamefont{P.}~\bibnamefont{Ring}},
  \bibinfo{author}{\bibfnamefont{F.}~\bibnamefont{Thielemann}},
  \bibnamefont{and} \bibinfo{author}{\bibfnamefont{V.}~\bibnamefont{Tselyaev}},
  \bibinfo{journal}{Nucl. Phys. A} \textbf{\bibinfo{volume}{823}},
  \bibinfo{pages}{26} (\bibinfo{year}{2009}).

\bibitem[{\citenamefont{Rauscher et~al.}(1997)\citenamefont{Rauscher,
  Thielemann, and Kratz}}]{Rauscher.Thielemann.Kratz:1997}
\bibinfo{author}{\bibfnamefont{T.}~\bibnamefont{Rauscher}},
  \bibinfo{author}{\bibfnamefont{F.-K.} \bibnamefont{Thielemann}},
  \bibnamefont{and} \bibinfo{author}{\bibfnamefont{K.-L.} \bibnamefont{Kratz}},
  \bibinfo{journal}{Phys. Rev. C} \textbf{\bibinfo{volume}{56}},
  \bibinfo{pages}{1613} (\bibinfo{year}{1997}).

\bibitem[{\citenamefont{{Goriely}}(1997)}]{Goriely:1997}
\bibinfo{author}{\bibfnamefont{S.}~\bibnamefont{{Goriely}}},
  \bibinfo{journal}{Astron. \& Astrophys.} \textbf{\bibinfo{volume}{325}},
  \bibinfo{pages}{414} (\bibinfo{year}{1997}).

\bibitem[{\citenamefont{{Rauscher}}(2005)}]{Rauscher:2005}
\bibinfo{author}{\bibfnamefont{T.}~\bibnamefont{{Rauscher}}},
  \bibinfo{journal}{Nucl. Phys. A} \textbf{\bibinfo{volume}{758}},
  \bibinfo{pages}{655c} (\bibinfo{year}{2005}).

\bibitem[{\citenamefont{Beun et~al.}(2009{\natexlab{b}})\citenamefont{Beun,
  Blackmon, Hix, McLaughlin, Smith, and Surman}}]{Beun.Blackmon.ea:2009}
\bibinfo{author}{\bibfnamefont{J.}~\bibnamefont{Beun}},
  \bibinfo{author}{\bibfnamefont{J.~C.} \bibnamefont{Blackmon}},
  \bibinfo{author}{\bibfnamefont{W.~R.} \bibnamefont{Hix}},
  \bibinfo{author}{\bibfnamefont{G.~C.} \bibnamefont{McLaughlin}},
  \bibinfo{author}{\bibfnamefont{M.~S.} \bibnamefont{Smith}}, \bibnamefont{and}
  \bibinfo{author}{\bibfnamefont{R.}~\bibnamefont{Surman}},
  \bibinfo{journal}{J. Phys. G: Nucl. Part. Phys.}
  \textbf{\bibinfo{volume}{36}}, \bibinfo{eid}{025201}
  (\bibinfo{year}{2009}{\natexlab{b}}).

\bibitem[{\citenamefont{Meyer et~al.}(1992)\citenamefont{Meyer, Mathews,
  Howard, Woosley, and Hoffman}}]{Meyer.Mathews.ea:1992}
\bibinfo{author}{\bibfnamefont{B.~S.} \bibnamefont{Meyer}},
  \bibinfo{author}{\bibfnamefont{G.~J.} \bibnamefont{Mathews}},
  \bibinfo{author}{\bibfnamefont{W.~M.} \bibnamefont{Howard}},
  \bibinfo{author}{\bibfnamefont{S.~E.} \bibnamefont{Woosley}},
  \bibnamefont{and} \bibinfo{author}{\bibfnamefont{R.~D.}
  \bibnamefont{Hoffman}}, \bibinfo{journal}{\apj}
  \textbf{\bibinfo{volume}{399}}, \bibinfo{pages}{656} (\bibinfo{year}{1992}).

\bibitem[{\citenamefont{Meyer}(1993)}]{Meyer:1993b}
\bibinfo{author}{\bibfnamefont{B.~S.} \bibnamefont{Meyer}}, in
  \emph{\bibinfo{booktitle}{Origin and Evolution of the Elements}}, edited by
  \bibinfo{editor}{\bibfnamefont{N.}~\bibnamefont{Prantzos}},
  \bibinfo{editor}{\bibfnamefont{E.}~\bibnamefont{Vangioni-Flam}},
  \bibnamefont{and} \bibinfo{editor}{\bibfnamefont{M.}~\bibnamefont{Casse}}
  (\bibinfo{publisher}{Cambridge University Press, Cambridge},
  \bibinfo{year}{1993}), pp. \bibinfo{pages}{444--448}.

\bibitem[{\citenamefont{{Burbidge} et~al.}(1957)\citenamefont{{Burbidge},
  {Burbidge}, {Fowler}, and {Hoyle}}}]{Burbidge.Burbidge.ea:1957}
\bibinfo{author}{\bibfnamefont{E.~M.} \bibnamefont{{Burbidge}}},
  \bibinfo{author}{\bibfnamefont{G.~R.} \bibnamefont{{Burbidge}}},
  \bibinfo{author}{\bibfnamefont{W.~A.} \bibnamefont{{Fowler}}},
  \bibnamefont{and} \bibinfo{author}{\bibfnamefont{F.}~\bibnamefont{{Hoyle}}},
  \bibinfo{journal}{Rev. Mod. Phys.} \textbf{\bibinfo{volume}{29}},
  \bibinfo{pages}{547} (\bibinfo{year}{1957}).

\bibitem[{\citenamefont{Cameron}(1957)}]{Cameron:1957}
\bibinfo{author}{\bibfnamefont{A.~G.~W.} \bibnamefont{Cameron}},
  \bibinfo{type}{Report} \bibinfo{number}{CRL-41}, \bibinfo{institution}{Chalk
  River} (\bibinfo{year}{1957}).

\bibitem[{\citenamefont{Audi et~al.}(2003)\citenamefont{Audi, Bersillon,
  Blachot, and Wapstra}}]{Audi.Bersillon.ea:2003}
\bibinfo{author}{\bibfnamefont{G.}~\bibnamefont{Audi}},
  \bibinfo{author}{\bibfnamefont{O.}~\bibnamefont{Bersillon}},
  \bibinfo{author}{\bibfnamefont{J.}~\bibnamefont{Blachot}}, \bibnamefont{and}
  \bibinfo{author}{\bibfnamefont{A.~H.} \bibnamefont{Wapstra}},
  \bibinfo{journal}{Nucl. Phys. A} \textbf{\bibinfo{volume}{729}},
  \bibinfo{pages}{3} (\bibinfo{year}{2003}).

\bibitem[{\citenamefont{{Cameron} et~al.}(1983)\citenamefont{{Cameron},
  {Cowan}, and {Truran}}}]{Cameron.Cowan.Truran:1983}
\bibinfo{author}{\bibfnamefont{A.~G.~W.} \bibnamefont{{Cameron}}},
  \bibinfo{author}{\bibfnamefont{J.~J.} \bibnamefont{{Cowan}}},
  \bibnamefont{and} \bibinfo{author}{\bibfnamefont{J.~W.}
  \bibnamefont{{Truran}}}, \bibinfo{journal}{Astrophys. Space Sci.}
  \textbf{\bibinfo{volume}{91}}, \bibinfo{pages}{235} (\bibinfo{year}{1983}).

\bibitem[{\citenamefont{{Goriely} and {Arnould}}(1996)}]{Goriely.Arnould:1996}
\bibinfo{author}{\bibfnamefont{S.}~\bibnamefont{{Goriely}}} \bibnamefont{and}
  \bibinfo{author}{\bibfnamefont{M.}~\bibnamefont{{Arnould}}},
  \bibinfo{journal}{Astron. \& Astrophys.} \textbf{\bibinfo{volume}{312}},
  \bibinfo{pages}{327} (\bibinfo{year}{1996}).

\end{thebibliography}

\end{document}